\documentclass[preprint,3p]{elsarticle}

\usepackage{graphicx}
\usepackage{amsmath}
\usepackage{array}
\usepackage{amssymb}
\usepackage{multirow}
\usepackage{subfig}

\usepackage{amssymb}

\journal{Journal of Computational Physics}

\begin{document}

\begin{frontmatter}



\title{Stable pseudoanalytical computation of electromagnetic fields from arbitrarily-oriented dipoles in cylindrically stratified media}


\author[rvt]{H. Moon\corref{cor1}}
\ead{moon.173@osu.edu}

\author[rvt]{F. L. Teixeira}
\ead{teixeira@ece.osu.edu}

\author[rvt2]{B. Donderici}
\ead{burkay.donderici@halliburton.com}

\cortext[cor1]{Corresponding author}

\address[rvt]{ElectroScience Laboratory, The Ohio State University, Columbus, OH 43212, USA}

\address[rvt2]{Sensor Physics \& Technology, Halliburton Energy Services, Houston, TX 77032, USA}

\begin{abstract}
Computation of electromagnetic fields due to point sources (Hertzian dipoles) in cylindrically stratified media is a classical problem
for which analytical expressions of the associated tensor Green's function have been long known.
However, under finite-precision arithmetic, direct numerical computations
based on the application of such analytical (canonical) expressions
invariably lead to underflow and overflow problems related to the poor scaling of the eigenfunctions (cylindrical Bessel and Hankel functions) for
extreme arguments and/or high-order, as well as convergence problems related to the numerical integration over the spectral wavenumber and to the truncation of the infinite series over the azimuth mode number. These problems are exacerbated when a disparate range of values is to be considered for the layers' thicknesses and
material properties (resistivities, permittivities, and permeabilities), the transverse and longitudinal distances between source and observation points, as well as the source frequency.
To overcome these challenges in a systematic fashion, we introduce herein different sets of range-conditioned, modified cylindrical functions (in lieu of standard cylindrical eigenfunctions),
each associated with
nonoverlapped subdomains of (numerical) evaluation
to allow for stable computations
under any range of physical parameters. In addition, adaptively-chosen integration contours are employed in the complex spectral wavenumber plane to ensure convergent numerical integration in all cases. We illustrate the application of the algorithm to problems of geophysical interest involving layer resistivities ranging from 1,000 $\Omega\cdot m$ to $10^{-8}$ $\Omega\cdot m$, frequencies of operation ranging from 10 MHz down to the low magnetotelluric range of 0.01 Hz, and for various combinations of layer thicknesses.
\end{abstract}


\end{frontmatter}


%
\newcommand{\Jn}{J_n}
\newcommand{\Jnd}{J'_n}
\newcommand{\Hn}{H^{(1)}_n}
\newcommand{\Hnd}{H'^{(1)}_n}
\newcommand{\hJn}{\hat{J}_n}
\newcommand{\hJnd}{\hat{J}'_n}
\newcommand{\hHn}{\hat{H}^{(1)}_n}
\newcommand{\hHnd}{\hat{H}'^{(1)}_n}
%
\newcommand{\bJn}{\overline{\mathbf{J}}\,_n}
\newcommand{\hbJn}{\hat{\overline{\mathbf{J}}}\,_n}
\newcommand{\bHn}{\overline{\mathbf{H}}\,^{(1)}_n}
\newcommand{\hbHn}{\hat{\overline{\mathbf{H}}}\,^{(1)}_n}
\newcommand{\bD}{\overline{\mathbf{D}}\,}
\newcommand{\hbD}{\hat{\overline{\mathbf{D}}}\,}
\newcommand{\bR}{\overline{\mathbf{R}}\,}
\newcommand{\tbR}{\widetilde{\overline{\mathbf{R}}}\,}
\newcommand{\hbR}{\hat{\overline{\mathbf{R}}}\,}
\newcommand{\htbR}{\hat{\widetilde{\overline{\mathbf{R}}}}\,}
\newcommand{\bRn}{\overline{\mathbf{R}}_n}
\newcommand{\bT}{\overline{\mathbf{T}}\,}
\newcommand{\tbT}{\widetilde{\overline{\mathbf{T}}}\,}
\newcommand{\hbT}{\hat{\overline{\mathbf{T}}}\,}
\newcommand{\htbT}{\hat{\widetilde{\overline{\mathbf{T}}}}\,}
\newcommand{\bS}{\overline{\mathbf{S}}\,}
\newcommand{\hbS}{\hat{\overline{\mathbf{S}}}\,}
\newcommand{\bI}{\overline{\mathbf{I}}}
\newcommand{\bM}{\overline{\mathbf{M}}\,}
\newcommand{\tbM}{\widetilde{\overline{\mathbf{M}}}\,}
\newcommand{\hbM}{\hat{\overline{\mathbf{M}}}\,}
\newcommand{\bMn}{\overline{\mathbf{M}}_n}
\newcommand{\bN}{\overline{\mathbf{N}}\,}
\newcommand{\tbN}{\widetilde{\overline{\mathbf{N}}}\,}
\newcommand{\hbN}{\hat{\overline{\mathbf{N}}}\,}
\newcommand{\bF}{\overline{\mathbf{F}}}
\newcommand{\bW}{\overline{\mathbf{W}}}
\newcommand{\bX}{\overline{\mathbf{X}}}
\newcommand{\obX}{\ddot{\overline{\mathbf{X}}}}
\newcommand{\bY}{\overline{\mathbf{Y}}}
\newcommand{\obY}{\ddot{\overline{\mathbf{Y}}}}
\newcommand{\bZ}{\overline{\mathbf{Z}}}
\newcommand{\obZ}{\ddot{\overline{\mathbf{Z}}}}
\newcommand{\bBn}{\overline{\mathbf{B}}_n}
\newcommand{\bCn}{\overline{\mathbf{C}}_n}
\newcommand{\bLn}{\overline{\mathbf{L}}_n}
\newcommand{\rr}{\mathbf{r}}
\newcommand{\rp}{\mathbf{r'}}
\newcommand{\suma}{\sum_{n=-\infty}^{\infty}}
\newcommand{\sumb}{\sum_{n=1}^{\infty}}
\newcommand{\intmp}{\int_{-\infty}^{\infty}}
\newcommand{\Fn}{\overline{\mathbf{F}}_n(\rho,\rho')}
\newcommand{\Dj}{\overleftarrow{\mathbf{D}}'_{j}}
\newcommand{\Dja}{\overleftarrow{\mathbf{D}}'_{j1}}
\newcommand{\Djb}{\overleftarrow{\mathbf{D}}'_{j2}}
\newcommand{\Djc}{\overleftarrow{\mathbf{D}}'_{j3}}
\newcommand{\pa}{\partial}
\newcommand{\iu}{\mathrm{i}}
\section{Introduction}
\label{sec.1.intro}
Computation of electromagnetic fields due to arbitrarily-oriented elementary (Hertzian) dipoles in cylindrically stratified media is of interest in a wide range of scenarios, including geophysical exploration, fiber optics,
and radar cross-section analysis.
Assuming the $z$-axis to be the symmetry axis, the analytical formulation of this problem is predicated on the knowledge of the cylindrical eigenfunctions (Bessel and Hankel functions) in the domain transverse to $z$ and their modal amplitudes. The derivation of reflection and transmission coefficients at each cylindrical boundary is then ascertained through the use of the proper boundary conditions. Since the eigenfunctions comprise a continuum spectrum in an unbounded domain, a Fourier-type integral along the spectral wavenumber $k_z$ is subsequently necessary to determine the fields (tensor Green's function)~\cite{Chew83:Singularities, Lovell87:Response, Lovell90:Effect, Tokgoz00:Closed, Hue06:Analysis, Wang08:Numerical, Liu12:Analysis}.
Unfortunately, numerical computations based on direct use of such (canonical) analytical expressions often lead to underflow and/or overflow problems under finite-precision arithmetic. These problems are related to the poor scaling of cylindrical eigenfunctions for extreme arguments and/or high-order, as well as convergence problems related to the numerical evaluation of the spectral integral on $k_z$ and truncation of the infinite series over the azimuth mode number $n$. Underflow and overflow problems become especially acute when a disparate range of values needs to be considered for the physical parameters, viz., the layers' constitutive properties (resistivities, permittivities, and permeabilities) and thicknesses, the transverse and longitudinal distance between source and observation points, as well as the source frequency~\cite{Yousif92:Fortran}.
In order to stabilize the numerical computation in the case of plane-wave scattering by highly absorbing layers,
Swathi and Tong \cite{Swathi88:New} developed an algorithm based upon scaled cylindrical functions. A stabilization procedure to deal with a very large number of cylindrical layers and disparate radii was proposed in~\cite{Jiang06:Improved}. Similar issues appear when computing Mie scattering from multilayered spheres~\cite{Kaiser93:Stable},\cite{Pena09:Scattering}, where continued fractions~\cite{Lentz76:Generating} or logarithmic derivatives~\cite{Toon81:Algorithms}, for example, can be used to circumvent the recurrence instability of Bessel functions of very large order. When the overall computation cost is not an issue, a more extreme strategy to circumvent this problem is to use arbitrary-precision arithmetic~\cite{Suzuki12:Mie}.

In this work, we extend those efforts by considering the stable computation of electromagnetic fields due to points sources in cylindrically stratified media. A salient feature of our work is that we do not limit ourselves to a particular regime of interest (that is very small or very large radii, or highly absorbing layers) but instead
develop a systematic algorithm to enable stable computations in any scenario. Note that, as opposed to Mie scattering or plane-wave scattering from cylinders considered in the above, the required spectral integration over $k_z$ produces, {\it per se}, a large variation on the integrand function arguments. The numerical convergence of such spectral integral, which depends among other factors on the separation
between source and observation points, need to be considered in tandem with the numerical stabilization procedure.
Our methodology is based on the use of various sets of
range-conditioned, modified cylindrical functions (in lieu of standard cylindrical eigenfunctions), each evaluated in nonoverlapped subdomains
to yield stable computations under double-precision floating-point format
for any range of physical parameters. This is combined with different numerically-robust integration contours that are adaptively chosen in the complex $k_z$ plane to yield fast convergence.
We illustrate the algorithm in problems of geophysical interest involving layer resistivities ranging from 1,000 $\Omega\cdot m$ to about $10^{-8}$ $\Omega\cdot m$, frequencies ranging from 10 MHz to as low as 0.01 Hz, and for various layer thicknesses.

\section{Range-Conditioned Formulation}
\label{sec.2.form}
\begin{figure}[!htbp]
    \centering
    \includegraphics[height=2.5in]{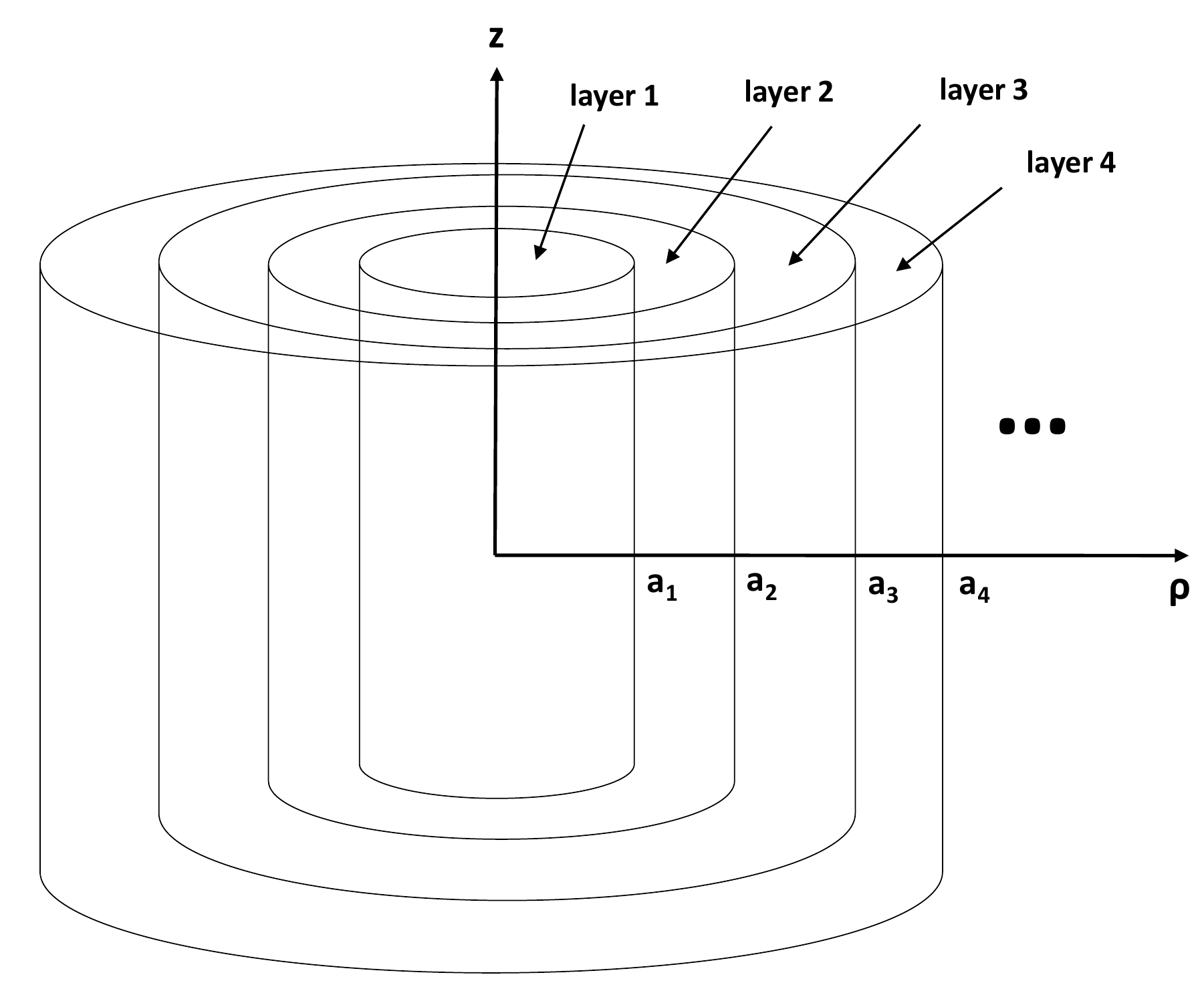}
    \caption{Basic geometry of a cylindrically stratified medium.}
    \label{S2.F.basic.geometry}
\end{figure}

Figure \ref{S2.F.basic.geometry} shows the geometry of a cylindrically stratified medium. Hereinafter we shall refer to the layer where the point source is present as layer $j$ and the region where the fields are computed as layer $i$. As indicated in Fig. \ref{S2.F.basic.geometry},
each successive layer radius is denoted by $a_i$.

We use nonprimed coordinates $(\rho,\phi,z)$ for the observation point and primed coordinates  $(\rho',\phi',z')$ for the source location, $k_{i\rho}=(k_i^2-k_z^2)^{1/2}$ denotes the transverse wavenumber in layer $i$ and $k_z$ is the longitudinal wavenumber, where $k_i^2=\omega^2 \mu_i \epsilon_i$.
As usual, $\omega$ represents the angular frequency, $\mu_i=\mu_{r,i}+\iu \sigma_{m,i}/\omega$ the complex permeability, and $\epsilon_i = \epsilon_{r,i}+\iu \sigma_i/\omega$ the complex permittivity, with $\epsilon_{r,i}$, $\mu_{r,i}$ denoting the real-valued permittivity and permeability, resp., and $\sigma_i$,  $\sigma_{m,i}$ denoting the electric and magnetic conductivities, resp.\footnote{Note that we allow for magnetic conductivities here so that problems involving magnetic dipole sources may be
computed simply by using a corresponding solution for electric dipoles after proper application of the duality theorem. Magnetic dipole sources are used to represent loop antennas at low frequencies, which are widely used in geophysical exploration.} Following~\cite[ch. ~3]{Chew:Inhomogeneous}, the $z$-components of the electric and magnetic fields can be written using the following generic expression

\begin{flalign}
    \begin{bmatrix}
    E_z \\ H_z
    \end{bmatrix}
=\frac{\iu Il}{4\pi\omega\epsilon_{j}}\suma e^{\iu n(\phi-\phi')}\intmp dk_z e^{\iu k_{z}(z-z')}
    \Fn\cdot\Dj, \label{S2.eq.EzHz.general}
\end{flalign}
where $Il$ is the dipole moment, and
\begin{flalign}
\Dj=\frac{\iu}{2}
    \begin{bmatrix}
    (\hat{z}k_j^2-\iu k_z\nabla')\cdot\hat{\alpha}' \\
    \iu\omega\epsilon_j\hat{\alpha}'\cdot\hat{z}\times\nabla'
    \end{bmatrix} \label{S2.eq.Dj.multi}
\end{flalign}
is an operator acting on the primed variables at the left, with $\hat{\alpha}'$ being a unit vector corresponding to the dipole orientation. The transverse field components can be easily derived from the the knowledge of the $z$-components.
Four distinct expressions for the generic factor $\Fn$ in the integrand of (\ref{S2.eq.EzHz.general}) exist depending on the relative position of the source and observation points, as follows~\cite[p.~176]{Chew:Inhomogeneous}:
\\
\\
\begin{subequations}
\text{{\it Case 1}: $\rho$ and $\rho'$ are in the same layer and $\rho\geq\rho'$}
\begin{flalign}
&\Fn=
    \left[\Hn(k_{j\rho}\rho)\bI+\Jn(k_{j\rho}\rho)\tbR_{j,j+1}\right]\cdot
    \tbM_{j+}\cdot
    \left[\Jn(k_{j\rho}\rho')\bI+\Hn(k_{j\rho}\rho')\tbR_{j,j-1}\right],&
    \label{S2.eq.Fn.case1.orig}
\end{flalign}
\\
\text{{\it Case 2}: $\rho$ and $\rho'$ are in the same layer and $\rho<\rho'$}
\begin{flalign}
&\Fn=
    \left[\Jn(k_{j\rho}\rho)\bI+\Hn(k_{j\rho}\rho)\tbR_{j,j-1}\right]\cdot
    \tbM_{j-}\cdot
    \left[\Hn(k_{j\rho}\rho')\bI+\Jn(k_{j\rho}\rho')\tbR_{j,j+1}\right],&
    \label{S2.eq.Fn.case2.orig}
\end{flalign}
\\
\text{{\it Case 3}: $\rho$ and $\rho'$ are in different layers and $\rho>\rho'$}
\begin{flalign}
&\Fn=
    \left[\Hn(k_{i\rho}\rho)\bI+\Jn(k_{i\rho}\rho)\tbR_{i,i+1}\right]\cdot
    \bN_{i+}\cdot\tbT_{ji}\cdot\tbM_{j+}\cdot
    \left[\Jn(k_{j\rho}\rho')\bI+\Hn(k_{j\rho}\rho')\tbR_{j,j-1}\right],&
    \label{S2.eq.Fn.case3.orig}
\end{flalign}
\\
\text{{\it Case 4}: $\rho$ and $\rho'$ are in different layers and $\rho<\rho'$}
\begin{flalign}
&\Fn=
    \left[\Jn(k_{i\rho}\rho)\bI+\Hn(k_{i\rho}\rho)\tbR_{i,i-1}\right]\cdot
    \bN_{i-}\cdot\tbT_{ji}\cdot\tbM_{j-}\cdot
    \left[\Hn(k_{j\rho}\rho')\bI+\Jn(k_{j\rho}\rho')\tbR_{j,j+1}\right],&
    \label{S2.eq.Fn.case4.orig}
\end{flalign}
\end{subequations}
where $\Jn$ and $\Hn$ are cylindrical Bessel and Hankel functions of the first kind and order $n$, $\bI$ is the $2 \times 2$ identity matrix,
$\tbM_{j\pm} = \left[\bI - \tbR_{j,j \mp 1} \cdot \tbR_{j,j \pm 1} \right]^{-1}$, and
$\bN_{i\pm}=\left[\bI-\bR_{i,i\mp 1}\cdot\tbR_{i,i\pm 1}\right]^{-1}
$ with $\bR_{j,j\pm 1}$ representing the $2 \times 2$ {\it local} reflection matrix between two adjacent cylindrical layers,
$\tbR_{j,j\pm 1}$ representing the $2 \times 2$ {\it generalized} reflection matrix between two adjacent cylindrical layers, and $\tbT_{ji}$ representing $2 \times 2$ {\it generalized} transmission matrix, respectively. The reflection and transmission coefficients between cylindrical layers are 2$\times$2 matrices because both TE$_z$ and TM$_z$ waves are in general needed to match the boundary conditions at the interfaces. Expressions for
$\bR_{j,j\pm 1}$,
$\tbR_{j,j\pm 1}$, $\tbT_{ji}$ are found further down below and are also provided in
~\cite[ch. ~3]{Chew:Inhomogeneous}.
A note should be made that insofar as the expressions for Cases 3 and 4 are concerned, the leftmost factors $\tbM_{i\pm}$ found in \cite[p.~176]{Chew:Inhomogeneous}
are incorrect and should instead be replaced by
$\bN_{i\pm}$,
as shown above.

As alluded before, even though \eqref{S2.eq.EzHz.general} provides an exact analytical expression, in many instances the wildly disparate behavior in magnitude of the many different factors that comprise the integrand
makes it difficult to obtain accurate results numerically. Furthermore, care should be exercised in choosing the integration path in the complex $k_z$ plane so that a convergent numerical integration is obtained, in a robust fashion. These two challenges are tackled in the remainder of this paper.

\subsection{Range-conditioned cylindrical functions}
\label{sec.2.1}
The evaluation of products of $\Jn$ and $\Hn$ (and their derivatives) is needed for the computation of the integrals in \eqref{S2.eq.Fn.case1.orig} -- \eqref{S2.eq.Fn.case4.orig} (and also in similar integrals for the computation of the transverse field components). When $|z| \ll 1$, $\Hn(z)$ has a very large value whereas $\Jn(z)$ has a very small value, and this disparity becomes even more extreme for larger order. On the other hand, when $\Im m[z] \gg 1$ (a condition that arises, for example, at low frequencies in layers with small resistivity), $\Hn(z)$ has a very small value while $\Jn(z)$ has a very large value. In cylindrically stratified media, these disparities in magnitude can coexist, to a varying extent, in many different layers. In addition, since the magnitude of the arguments for $\Jn$ and $\Hn$ also depends on $k_z$, large variations occur in the course of numerical integration over $k_z$ as well. As a result, significant round-off errors can accumulate unless the inherent characteristics of such functions are taken into consideration beforehand.

\subsubsection{Definitions}
\label{sec.2.1.1}
When $|k_{i\rho}a_i| \ll 1$, $\Jn(k_{i\rho}a_i)$ and $\Hn(k_{i\rho}a_i)$ can be expressed through the following small argument approximations for $n>0$ \cite[p.~15]{Chew:Inhomogeneous}.
\begin{subequations}
\begin{flalign}
\Jn(k_{i\rho}a_i)&=\frac{1}{n!}\left(\frac{k_{i\rho}a_i}{2}\right)^n=
    \frac{1}{n!}\left(\frac{k_{i\rho}}{2}\right)^n\cdot a_i^n\cdot 1=
    G_i a_i^n\hJn(k_{i\rho}a_i), \label{S2.eq.small.approx.Jn}\\
\Jnd(k_{i\rho}a_i)&=\frac{1}{2(n-1)!}\left(\frac{k_{i\rho}a_i}{2}\right)^{n-1}=
    \frac{1}{n!}\left(\frac{k_{i\rho}}{2}\right)^n\cdot a_i^n\cdot \frac{n}{k_{i\rho}a_i}=
    G_i a_i^n\hJnd(k_{i\rho}a_i), \label{S2.eq.small.approx.Jnd}\\
\Hn(k_{i\rho}a_i)&=-\frac{\iu(n-1)!}{\pi}\left(\frac{2}{k_{i\rho}a_i}\right)^n=
    n!\left(\frac{2}{k_{i\rho}}\right)^n\cdot a_i^{-n}\cdot \left(-\frac{\iu}{n\pi}\right)=
    G_i^{-1}a_i^{-n}\hHn(k_{i\rho}a_i), \label{S2.eq.small.approx.Hn}\\
\Hnd(k_{i\rho}a_i)&=\frac{\iu(n!)}{2\pi}\left(\frac{2}{k_{i\rho}a_i}\right)^{n+1}=
    n!\left(\frac{2}{k_{i\rho}}\right)^n\cdot a_i^{-n}\cdot \left(\frac{\iu}{\pi k_{i\rho}a_i}\right)=
    G_i^{-1}a_i^{-n}\hHnd(k_{i\rho}a_i), \label{S2.eq.small.approx.Hnd}
\end{flalign}
\end{subequations}
where $G_i=(k_{i\rho}/2)^n/n!$. In \eqref{S2.eq.small.approx.Jn} -- \eqref{S2.eq.small.approx.Hnd}, $\hJn$, $\hJnd$, $\hHn$, and $\hHnd$ are the range-conditioned cylindrical functions for small arguments. Note that the definition is such that $(1)$ the multiplicative factors $G_i$ and $a_i^n$ associated with a given cylindrical function and its derivative are the same and $(2)$ the multiplicative factors for $\hJn(k_{i\rho}a_i)$ and $\hJnd(k_{i\rho}a_i)$ are reciprocal to ones for $\hHn(k_{i\rho}a_i)$ and $\hHnd(k_{i\rho}a_i)$. These multiplicative factors play a determinant role in producing the extreme values for the cylindrical eigenfunctions and, as we will see below, should be analytically manipulated (reduced) in an appropriate fashion {\it before} any numerical evaluations to stabilize the computation. As seen below, these definitions facilitate subsequent computations.

On the other hand, when $|k_{i\rho}a_i| \gg 1$, large argument approximations can be used for the cylindrical eigenfunctions~\cite[p.~131, 236]{Jin:SpecialFunctions}. Likewise, range-conditioned cylindrical functions for large arguments can be defined as
\begin{flalign}
\Jn(k_{i\rho}a_i)&=\sqrt{\frac{2}{\pi k_{i\rho}a_i}}
    \left[P(n,k_{i\rho}a_i)\cos\chi-Q(n,k_{i\rho}a_i)\sin\chi\right]\notag\\
&=e^{\chi''}\sqrt{\frac{2}{\pi k_{i\rho}a_i}}\left[P(n,k_{i\rho}a_i)\frac{e^{-\iu\chi'}}{2}
    -Q(n,k_{i\rho}a_i)\frac{\iu e^{-\iu\chi'}}{2}\right]\notag\\
&=e^{k''_{i\rho}a_i}\sqrt{\frac{2}{\pi k_{i\rho}a_i}}\left[P(n,k_{i\rho}a_i)\frac{e^{-\iu\chi'}}{2}
    -Q(n,k_{i\rho}a_i)\frac{\iu e^{-\iu\chi'}}{2}\right]\notag\\
&=e^{k''_{i\rho}a_i}\hJn(k_{i\rho}a_i), \label{S2.eq.large.approx.Jn}
\end{flalign}
\begin{flalign}
\Hn(k_{i\rho}a_i)&=\frac{2}{\pi}(-\iu)^{n+1}K_n(-\iu k_{i\rho}a_i)\notag\\
&=\frac{2}{\pi}(-\iu)^{n+1}\sqrt{\frac{\pi}{2z}}e^{-k''_{i\rho}a_i}e^{\iu k'_{i\rho}a_i}
    \left[1+\frac{(\mu-1)}{1!(8z)}+\frac{(\mu-1)(\mu-9)}{2!(8z)^2}+\dotsb \right]\notag\\
&=e^{-k''_{i\rho}a_i}(-\iu)^{n+1}\sqrt{\frac{2}{\pi z}}e^{\iu k'_{i\rho}a_i}
    \left[1+\frac{(\mu-1)}{1!(8z)}+\frac{(\mu-1)(\mu-9)}{2!(8z)^2}+\dotsb \right]\notag\\
&=e^{-k''_{i\rho}a_i}\hHn(k_{i\rho}a_i), \label{S2.eq.large.approx.Hn}
\end{flalign}
where $\chi=k_{i\rho}a_i-\frac{n\pi}{2}-\frac{\pi}{4}$, $\mu=4n^2$, $K_n(\cdot)$ is the modified Bessel function of the second kind, and $P(\cdot)$ and $Q(\cdot)$ are phase and quadrature polynomial functions with tabulated expressions. In the above, we used the fact that, for $\chi = \chi'+\iu\chi''$, $\sin \chi$ and $\cos \chi$ in \eqref{S2.eq.large.approx.Jn} can be decomposed into two terms and one of them can be ignored when $\chi''$ is large enough such that
\begin{subequations}
\begin{flalign}
\cos\chi&=\cos(\chi'+\iu\chi'')
\cong\frac{1}{2}e^{\chi''}e^{-\iu\chi'}, \label{S2.eq.cos.chi}\\
\sin\chi&=\sin(\chi'+\iu\chi'')
\cong\frac{\iu}{2}e^{\chi''}e^{-\iu\chi'}. \label{S2.eq.sin.chi}
\end{flalign}
\end{subequations}
Derivatives of range-conditioned cylindrical functions with large arguments can be obtained through the recursive formulas below
\begin{flalign}
\Jnd(k_{i\rho}a_i)&=J_{n-1}(k_{i\rho}a_i)-\frac{n}{k_{i\rho}a_i}\Jn(k_{i\rho}a_i) \notag\\
&=e^{k''_{i\rho}a_i}\hat{J}_{n-1}(k_{i\rho}a_i)
    -e^{k''_{i\rho}a_i}\frac{n}{k_{i\rho}a_i}\hJn(k_{i\rho}a_i)
=e^{k''_{i\rho}a_i}\hJnd(k_{i\rho}a_i), \label{S2.eq.large.approx.Jnd}\\
\Hnd(k_{i\rho}a_i)&=H^{(1)}_{n-1}(k_{i\rho}a_i)-\frac{n}{k_{i\rho}a_i}\Hn(k_{i\rho}a_i) \notag\\
&=e^{-k''_{i\rho}a_i}\hat{H}^{(1)}_{n-1}(k_{i\rho}a_i)
    -e^{-k''_{i\rho}a_i}\frac{n}{k_{i\rho}a_i}\hHn(k_{i\rho}a_i)
=e^{-k''_{i\rho}a_i}\hHnd(k_{i\rho}a_i). \label{S2.eq.large.approx.Hnd}
\end{flalign}
Note again that the associated multiplicative factors in \eqref{S2.eq.large.approx.Jn} and
\eqref{S2.eq.large.approx.Hn}, as well as in \eqref{S2.eq.large.approx.Jnd} and \eqref{S2.eq.large.approx.Hnd}
are chosen to be reciprocal to each other.

When argument is neither too small nor too large, range-conditioned cylindrical functions are introduced akin to those of small and large arguments, i.e.,
\begin{subequations}
\begin{flalign}
\Jn(k_{i\rho}a_i)&=P_{ii}\hJn(k_{i\rho}a_i), \label{S2.eq.moderate.Jn}\\
\Jnd(k_{i\rho}a_i)&=P_{ii}\hJnd(k_{i\rho}a_i), \label{S2.eq.moderate.Jnd}\\
\Hn(k_{i\rho}a_i)&=P_{ii}^{-1}\hHn(k_{i\rho}a_i), \label{S2.eq.moderate.Hn}\\
\Hnd(k_{i\rho}a_i)&=P_{ii}^{-1}\hHnd(k_{i\rho}a_i), \label{S2.eq.moderate.Hnd}
\end{flalign}
\end{subequations}
where $P_{ii}$ is determined in \ref{app.b}, and its first and second subscript correspond to the radial wavenumber and layer radius, respectively.

In summary, the argument for range-conditioned cylindrical functions can be classified into three types according to its magnitude: small, moderate, and large. The multiplicative factors that define the
respective range-conditioned functions for each type of argument are summarized in Table \ref{S2.T.def.RCCF}.
\begin{table}[!htbp]
\begin{center}
\renewcommand{\arraystretch}{1.6}
\setlength{\tabcolsep}{4pt}
\caption{Definition of range-conditioned cylindrical functions for all types of arguments. The associated multiplicative factors are different according to the argument type.}
\begin{tabular}{ccc}
    \hline
    small arguments & moderate arguments & large arguments \\
    \hline
    $\Jn(k_{i\rho}a_i)=G_ia_i^n\hJn(k_{i\rho}a_i)$ &
    $\Jn(k_{i\rho}a_i)=P_{ii}\hJn(k_{i\rho}a_i)$ &
    $\Jn(k_{i\rho}a_i)=e^{k''_{i\rho}a_i}\hJn(k_{i\rho}a_i)$ \\

    $\Jnd(k_{i\rho}a_i)=G_ia_i^n\hJnd(k_{i\rho}a_i)$ &
    $\Jnd(k_{i\rho}a_i)=P_{ii}\hJnd(k_{i\rho}a_i)$ &
    $\Jnd(k_{i\rho}a_i)=e^{k''_{i\rho}a_i}\hJnd(k_{i\rho}a_i)$ \\

    $\Hn(k_{i\rho}a_i)=G^{-1}_ia_i^{-n}\hHn(k_{i\rho}a_i)$ &
    $\Hn(k_{i\rho}a_i)=P_{ii}^{-1}\hHn(k_{i\rho}a_i)$ &
    $\Hn(k_{i\rho}a_i)=e^{-k''_{i\rho}a_i}\hHn(k_{i\rho}a_i)$ \\

    $\Hnd(k_{i\rho}a_i)=G^{-1}_ia_i^{-n}\hHnd(k_{i\rho}a_i)$ &
    $\Hnd(k_{i\rho}a_i)=P_{ii}^{-1}\hHnd(k_{i\rho}a_i)$ &
    $\Hnd(k_{i\rho}a_i)=e^{-k''_{i\rho}a_i}\hHnd(k_{i\rho}a_i)$ \\
    \hline
\end{tabular}
\label{S2.T.def.RCCF}
\end{center}
\end{table}

As shown in \ref{app.a},
the local reflection and transmission matrices $\bR_{j,j\pm 1}$ and $\bT_{j,j\pm 1}$ are written in terms of $\bJn$ and $\bHn$ matrices.
The latter are defined in terms of cylindrical functions so that range-conditioned versions thereof can be constructed as well.
After some algebra, it can be shown that the multiplicative factors associated with the $\bJn$ and $\bHn$ matrices are the same as for the range-conditioned cylindrical functions. That is, for small arguments we have
\begin{subequations}
\begin{flalign}
\bJn(k_{i\rho}a_i)&=G_i a_i^n\frac{1}{k^2_{i\rho}a_i}
    \begin{bmatrix}
    \iu\omega\epsilon_i k_{i\rho}a_i\hJnd(k_{i\rho}a_i) & -nk_z\hJn(k_{i\rho}a_i)\\
    -nk_z\hJn(k_{i\rho}a_i) & -\iu\omega\mu_i k_{i\rho}a_i\hJnd(k_{i\rho}a_i)
    \end{bmatrix}
=G_i a_i^n\hbJn(k_{i\rho}a_i), \label{S2.eq.small.matrixJ}\\
\bHn(k_{i\rho}a_i)&=G^{-1}_i a_i^{-n}\frac{1}{k^2_{i\rho}a_i}
    \begin{bmatrix}
    \iu\omega\epsilon_i k_{i\rho}a_i\hHnd(k_{i\rho}a_i) & -nk_z\hHn(k_{i\rho}a_i)\\
    -nk_z\hHn(k_{i\rho}a_i) & -\iu\omega\mu_{i}k_{i\rho}a_i\hHnd(k_{i\rho}a_i)
    \end{bmatrix}
=G^{-1}_i a_i^{-n}\hbHn(k_{i\rho}a_i), \label{S2.eq.small.matrixH}
\end{flalign}
\end{subequations}
for large arguments we have
\begin{subequations}
\begin{flalign}
\bJn(k_{i\rho}a_i)&=e^{|k''_{i\rho}a_i|}\frac{1}{k^2_{i\rho}a_i}
    \begin{bmatrix}
    \iu\omega\epsilon_i k_{i\rho}a_i\hJnd(k_{i\rho}a_i) & -nk_z\hJn(k_{i\rho}a_i)\\
    -nk_z\hJn(k_{i\rho}a_i) & -\iu\omega\mu_i k_{i\rho}a_i\hJnd(k_{i\rho}a_i)
    \end{bmatrix}
=e^{|k''_{i\rho}a_i|}\hbJn(k_{i\rho}a_i), \label{S2.eq.large.matrixJ} \\
\bHn(k_{i\rho}a_i)&=e^{-k''_{i\rho}a_i}\frac{1}{k^2_{i\rho}a_i}
    \begin{bmatrix}
    \iu\omega\epsilon_i k_{i\rho}a_i\hHnd(k_{i\rho}a_i) & -nk_z\hHn(k_{i\rho}a_i)\\
    -nk_z\hHn(k_{i\rho}a_i) & -\iu\omega\mu_{i}k_{i\rho}a_i\hHnd(k_{i\rho}a_i)
    \end{bmatrix}
=e^{-k''_{i\rho}a_i}\hbHn(k_{i\rho}a_i), \label{S2.eq.large.matrixH}
\end{flalign}
\end{subequations}
and for moderate arguments we have
\begin{subequations}
\begin{flalign}
\bJn(k_{i\rho}a_i)&=P_{ii}\hbJn(k_{i\rho}a_i), \label{S2.eq.moderate.matrixJ}\\
\bHn(k_{i\rho}a_i)&=P_{ii}^{-1}\hbHn(k_{i\rho}a_i). \label{S2.eq.moderate.matrixH}
\end{flalign}
\end{subequations}

For a numerical computation, the actual threshold values among the three argument types above need to be determined. This is done in \ref{app.b} assuming standard double-precision arithmetics.

\subsection{Reflection and transmission coefficients for two cylindrical layers}
\label{sec.2.2}
When a medium consists of two cylindrical layers, the expressions for the reflection $\bR_{12}$, $\bR_{21}$
and transmission $\bT_{12}$, $\bT_{21}$ coefficients at boundary $\rho=a_1$ are given in
~\cite[ch.~3]{Chew:Inhomogeneous} and, for convenience, also presented in~\ref{app.a}.
When one or two of the involved layers are associated with small or large arguments, those coefficients need to be rewritten using range-conditioned cylindrical functions. Therefore, there are total of nine cases to be considered according to the argument types as listed in Tables \ref{S2.T.1st.gr} -- \ref{S2.T.3rd.gr}. The first group (Cases 1, 2, 3) only incorporates small and moderate arguments; the second group (Cases 4, 5, 6) only incorporates large and moderate arguments; and the third group (Cases 7, 8, 9) consists of remaining combinations.
\begin{table}[!htbp]
\begin{center}
\renewcommand{\arraystretch}{1.6}
\setlength{\tabcolsep}{3pt}
\begin{minipage}[b]{0.3\linewidth}\centering
\caption{First group.}
\begin{tabular}{ccc}
    \hline
     & $k_{1\rho}a_1$ & $k_{2\rho}a_1$ \\
    \hline
    Case 1 & small & small \\
    Case 2 & small & moderate \\
    Case 3 & moderate & small \\
    \hline
\end{tabular}
\label{S2.T.1st.gr}
\end{minipage}
\hspace{0.5cm}
\begin{minipage}[b]{0.3\linewidth}\centering
\caption{Second group.}
\begin{tabular}{ccc}
    \hline
     & $k_{1\rho}a_1$ & $k_{2\rho}a_1$ \\
    \hline
    Case 4 & large & large \\
    Case 5 & large & moderate \\
    Case 6 & moderate & large \\
    \hline
\end{tabular}
\label{S2.T.2nd.gr}
\end{minipage}
\hspace{0.5cm}
\begin{minipage}[b]{0.3\linewidth}\centering
\caption{Third group.}
\begin{tabular}{ccc}
    \hline
     & $k_{1\rho}a_1$ & $k_{2\rho}a_1$ \\
    \hline
    Case 7 & small & large \\
    Case 8 & large & small \\
    Case 9 & moderate & moderate \\
    \hline
\end{tabular}
\label{S2.T.3rd.gr}
\end{minipage}
\end{center}
\end{table}

The canonical expressions for the (local) reflection and transmission coefficients in~\ref{app.a}
 can be rewritten using range-conditioned cylindrical functions for all the above nine cases. The redefined (conditioned) coefficients for all three groups are summarized in Table \ref{S2.T.redefined.R.T}. Note that for the second (Cases 4, 5, 6) and third groups (Cases 7, 8, 9), the redefinition of the coefficients is basically similar to the first group but the associated multiplicative factors are different.
\begin{table}[!htbp]
\begin{center}
\renewcommand{\arraystretch}{1.8}
\setlength{\tabcolsep}{10pt}
\caption{Redefined (local) reflection and transmission coefficients using the range-conditioned cylindrical functions for two cylindrical layers.}
\begin{tabular}{ccccc}
    \hline
     & $\bR_{12}$ & $\bR_{21}$ & $\bT_{12}$ & $\bT_{21}$ \\
    \hline
    Case 1 & $G^{-2}_1 a^{-2n}_1\hbR_{12}$ & $G^2_2 a^{2n}_1\hbR_{21}$ &
		$G^{-1}_1 a^{-n}_1 G_2 a^n_1\hbT_{12}$ & $G^{-1}_1 a^{-n}_1 G_2 a^n_1\hbT_{21}$ \\
	Case 2 & $G^{-2}_1 a^{-2n}_1\hbR_{12}$ & $P_{21}^{2}\hbR_{21}$ &
		$G^{-1}_1 a^{-n}_1 P_{21}\hbT_{12}$ & $G^{-1}_1 a^{-n}_1 P_{21}\hbT_{21}$ \\
	Case 3 & $P_{11}^{-2}\hbR_{12}$ & $G^2_2 a^{2n}_1\hbR_{21}$ &
		$G_2 a^n_1 P_{11}^{-1}\hbT_{12}$ & $G_2 a^n_1 P_{11}^{-1}\hbT_{21}$ \\

	Case 4 & $e^{-2k''_{1\rho}a_1}\hbR_{12}$ & $e^{2k''_{2\rho}a_1}\hbR_{21}$ &
		$e^{-k''_{1\rho}a_1}e^{k''_{2\rho}a_1}\hbT_{12}$ & $e^{-k''_{1\rho}a_1}e^{k''_{2\rho}a_1}\hbT_{21}$ \\
	Case 5 & $e^{-2k''_{1\rho}a_1}\hbR_{12}$ & $P_{21}^{2}\hbR_{21}$ &
		$e^{-k''_{1\rho}a_1} P_{21}\hbT_{12}$ & $e^{-k''_{1\rho}a_1} P_{21}\hbT_{21}$ \\
	Case 6 & $P_{11}^{-2}\hbR_{12}$ & $e^{2k''_{2\rho}a_1}\hbR_{21}$ &
		$e^{k''_{2\rho}a_1} P_{11}^{-1}\hbT_{12}$ & $e^{k''_{2\rho}a_1} P_{11}^{-1}\hbT_{21}$ \\

	Case 7 & $G^{-2}_1 a^{-2n}_1\hbR_{12}$ & $e^{2k''_{2\rho}a_1}\hbR_{21}$ &
		$G^{-1}_1 a^{-n}_1 e^{k''_{2\rho}a_1}\hbT_{12}$ & $G^{-1}_1 a^{-n}_1 e^{k''_{2\rho}a_1}\hbT_{21}$ \\
	Case 8 & $e^{-2k''_{1\rho}a_1}\hbR_{12}$ & $G^{2}_{2}a^{2n}_1\hbR_{21}$ &
		$e^{-k''_{1\rho}a_1} G_2 a^n_1\hbT_{12}$ & $e^{-k''_{1\rho}a_1} G_2 a^n_1\hbT_{21}$ \\
	Case 9 & $P_{11}^{-2}\hbR_{12}$ & $P_{21}^{2}\hbR_{21}$ &
		$P_{11}^{-1}P_{21}\hbT_{12}$ & $P_{11}^{-1}P_{21}\hbT_{21}$ \\
    \hline
\end{tabular}
\label{S2.T.redefined.R.T}
\end{center}
\end{table}
Explicit expressions for $\hbR_{12}$, $\hbR_{21}$, $\hbT_{12}$, and $\hbT_{21}$ are also provided in \ref{app.a}. It should be noted that there is a simple relationship between the original coefficients and the range-conditioned coefficients in all cases. For any two arbitrarily-indexed cylindrical layers, the relationship can be succinctly expressed as
\begin{subequations}
\begin{flalign}
\bR_{i,i+1}&=\alpha^2_{ii}\hbR_{i,i+1}, \label{S2.eq.local.Rij}\\
\bR_{i+1,i}&=\beta^2_{i+1,i}\hbR_{i+1,i}, \label{S2.eq.local.Rji}\\
\bT_{i,i+1}&=\alpha_{ii}\beta_{i+1,i}\hbT_{i,i+1}, \label{S2.eq.local.Tij}\\
\bT_{i+1,i}&=\alpha_{ii}\beta_{i+1,i}\hbT_{i+1,i}, \label{S2.eq.local.Tji}
\end{flalign}
\end{subequations}
where $\alpha_{ii}$ is the function of $k_{i\rho}a_i$ and $\beta_{i+1,i}$ is the function of $k_{i+1,\rho}a_i$. The first subscript of $\alpha$ and $\beta$ stands for the radial wavenumber and second subscript stands for the radial distance. The values for $\alpha_{ii}$ and $\beta_{i+1,i}$ are summarized in Table \ref{S2.T.alpha.beta} for the three types of arguments. The coefficients $\alpha$ and $\beta$ obey two important properties that will be exploited later on. (1) {\it Reciprocity}: $\alpha$ and $\beta$ are reciprocal to each other for identical subscripts, i.e.
$\alpha_{ii}=1/\beta_{ii}$. (2) {\it Boundness}: when the radial wavenumbers of $\alpha$ and $\beta$ are the same but the radial distance for $\alpha$ is larger than that for $\beta$, the absolute value of the product of $\alpha$ and $\beta$ is always less than or equal to unity, i.e. $
|\beta_{ii}\,\alpha_{ij}|\leq 1$ for $a_i < a_j$.

\begin{table}[t]
\begin{center}
\renewcommand{\arraystretch}{1.6}
\setlength{\tabcolsep}{10pt}
\caption{Definition of $\alpha_{ii}$ and $\beta_{i+1,i}$.}
\begin{tabular}{ccc}
    \hline
    argument type & $\alpha_{ii}$ & $\beta_{i+1,i}$ \\
    \hline
    small  & $G^{-1}_i a^{-n}_i$ & $G_{i+1} a^n_i$ \\
    moderate & $P_{ii}^{-1}$ & $P_{i+1,i}$\\
    large  & $e^{-k''_{i\rho}a_i}$ & $e^{k''_{i+1,\rho}a_i}$\\
    \hline
\end{tabular}
\label{S2.T.alpha.beta}
\end{center}
\end{table}

\subsection{Generalized reflection and transmission coefficients for three or more cylindrical layers}
\label{sec.2.3}
\begin{figure}[t]
    \begin{minipage}[b]{0.45\linewidth}
    \centering
    \includegraphics[width=2.3in]{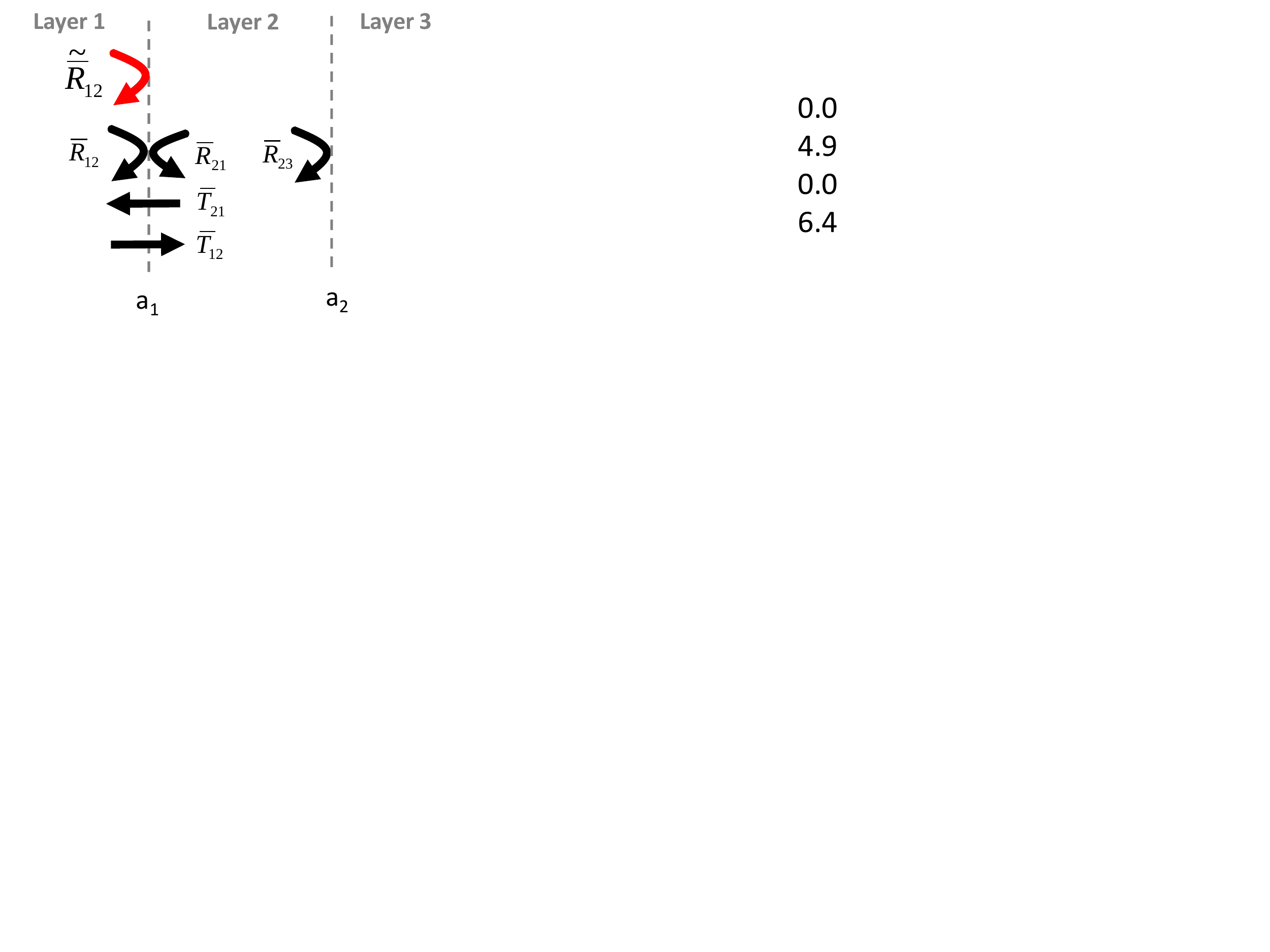}
    \caption{Generalized reflection coefficient for the outgoing-wave case for three cylindrical layers in the $\rho z$-plane.}
    \label{S2.F.RF.Out}
    \end{minipage}
    \hspace{1.0cm}
    \begin{minipage}[b]{0.45\linewidth}
    \centering
    \includegraphics[width=2.3in]{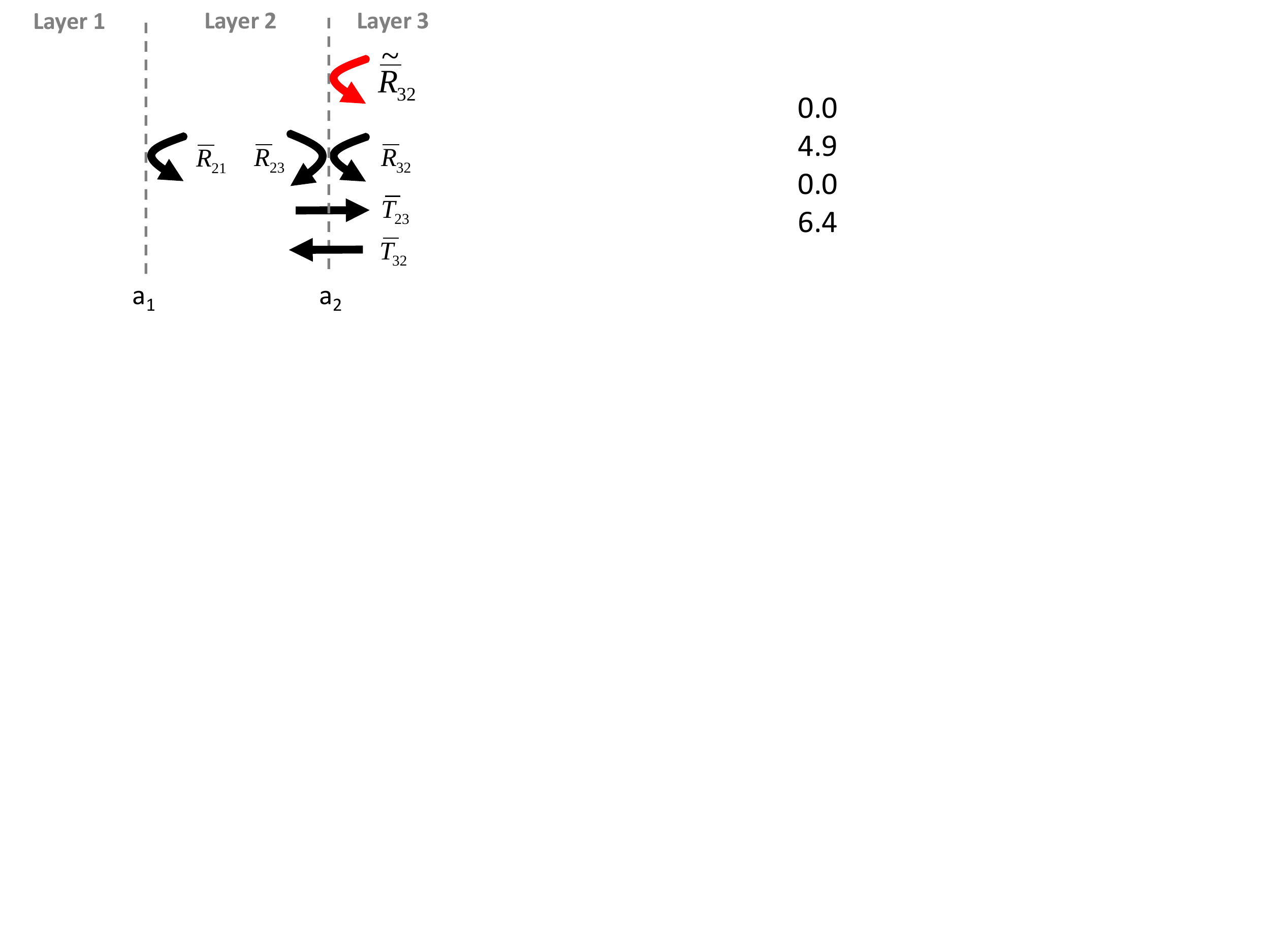}
    \caption{Generalized reflection coefficient for the standing-wave case for three cylindrical layers in the $\rho z$-plane.}
    \label{S2.F.RF.Stand}
    \end{minipage}
\end{figure}

When more than two cylindrical layers are present, {\it generalized} reflection and transmission coefficients can be defined recursively from the local reflection and transmission coefficients between each two layers
~\cite[ch. ~3]{Chew:Inhomogeneous}. These generalized coefficients incorporate multiple reflections and transmissions. Using the redefined local reflection and transmission coefficients in the previous section, \eqref{S2.eq.local.Rij} -- \eqref{S2.eq.local.Tji}, generalized reflection and transmission coefficients can be redefined accordingly. Figure \ref{S2.F.RF.Out} and \ref{S2.F.RF.Stand} illustrate the local coefficients used to define generalized reflection coefficients for the outgoing-wave and standing-wave cases, in a three-layer example. In this case,
the generalized reflection coefficient at $a_1$ for the outgoing-wave case is
\begin{flalign}
\tbR_{12}=
    \bR_{12}+\bT_{21}\cdot\bR_{23}\cdot
    \left[\bI-\bR_{21}\cdot\bR_{23}\right]^{-1}\cdot\bT_{12}. \label{S2.eq.general.R12}
\end{flalign}
Assuming all coefficients in \eqref{S2.eq.general.R12} are replaced with their range-conditioned-versions, we have
\begin{flalign}
\tbR_{12}&=
    \bR_{12}+\bT_{21}\cdot\bR_{23}\cdot
    \left[\bI-\bR_{21}\cdot\bR_{23}\right]^{-1}\cdot\bT_{12} \notag\\
&=\alpha^2_{11}\hbR_{12}+\left(\alpha_{11}\beta_{21}\hbT_{21}\right)\cdot
    \left(\alpha^2_{22}\hbR_{23}\right)\cdot
    \left[\bI-\left(\beta^2_{21}\hbR_{21}\right)\cdot
        \left(\alpha^2_{22}\hbR_{23}\right)\right]^{-1}\cdot
    \left(\alpha_{11}\beta_{21}\hbT_{12}\right) \notag\\
&=\alpha^2_{11}\left\{
    \hbR_{12}+\beta^2_{21}\alpha^2_{22}\hbT_{21}\cdot
    \hbR_{23}\cdot
        \left[\bI-\beta^2_{21}\alpha^2_{22}\hbR_{21}\cdot\hbR_{23}\right]^{-1}\cdot
    \hbT_{12}\right\} \notag\\
&=\alpha^2_{11}\htbR_{12}. \label{S2.eq.general.R12.re}
\end{flalign}
Note that $|\beta_{21}\alpha_{22}|\leq1$ for any medium properties of Layer 1, 2, and 3 (refer to Table \ref{S2.T.alpha.beta} for definitions). This condition makes the evaluation of the redefined generalized reflection coefficient numerically stable. Similarly, the generalized reflection coefficient at $a_2$ for the standing-wave case is
\begin{flalign}
\tbR_{32}=
    \bR_{32}+\bT_{23}\cdot\bR_{21}\cdot
    \left[\bI-\bR_{23}\cdot\bR_{21}\right]^{-1}\cdot\bT_{32}. \label{S2.eq.general.R32}
\end{flalign}
Hence, the generalized reflection coefficient for the standing-wave case can be likewise redefined as
\begin{flalign}
\tbR_{32}&=
    \bR_{32}+\bT_{23}\cdot\bR_{21}\cdot
    \left[\bI-\bR_{23}\cdot\bR_{21}\right]^{-1}\cdot\bT_{32} \notag\\
&=\beta^2_{32}\hbR_{32}+\left(\alpha_{22}\beta_{32}\hbT_{23}\right)\cdot
    \left(\beta^2_{21}\hbR_{21}\right)\cdot
    \left[\bI-\left(\alpha^2_{22}\hbR_{23}\right)\cdot
        \left(\beta^2_{21}\hbR_{21}\right)\right]^{-1}\cdot
    \left(\alpha_{22}\beta_{32}\hbT_{32}\right) \notag\\
&=\beta^2_{32}\left\{
    \hbR_{32}+\beta^2_{21}\alpha^2_{22}\hbT_{23}\cdot
    \hbR_{21}\cdot
        \left[\bI-\beta^2_{21}\alpha^2_{22}\hbR_{23}\cdot\hbR_{21}\right]^{-1}\cdot
    \hbT_{32}\right\} \notag\\
&=\beta^2_{32}\htbR_{32}, \label{S2.eq.general.R32.re}
\end{flalign}
where again the condition of $|\beta_{21}\alpha_{22}|\leq1$ stabilizes the evaluation of the redefined generalized reflection coefficient \eqref{S2.eq.general.R32.re}.
Note that the coefficients associated with the redefined generalized reflection coefficients are identical to those associated with the redefined local reflection coefficients (see \eqref{S2.eq.local.Rij}, \eqref{S2.eq.local.Rji} and \eqref{S2.eq.general.R12.re}, \eqref{S2.eq.general.R32.re}). This property makes it straightforward to redefine the generalized (conditioned) reflection coefficients for a generic cylindrically stratified medium (i.e. with any number of layers) as
\begin{subequations}
\begin{flalign}
\tbR_{i,i+1}&=\alpha^2_{ii}\htbR_{i,i+1}, \label{S2.eq.general.Rij}\\
\tbR_{i+1,i}&=\beta^2_{i+1,i}\htbR_{i+1,i}. \label{S2.eq.general.Rji}
\end{flalign}
\end{subequations}

\begin{figure}[t]
    \begin{minipage}[b]{0.45\linewidth}
    \centering
    \includegraphics[width=2.3in]{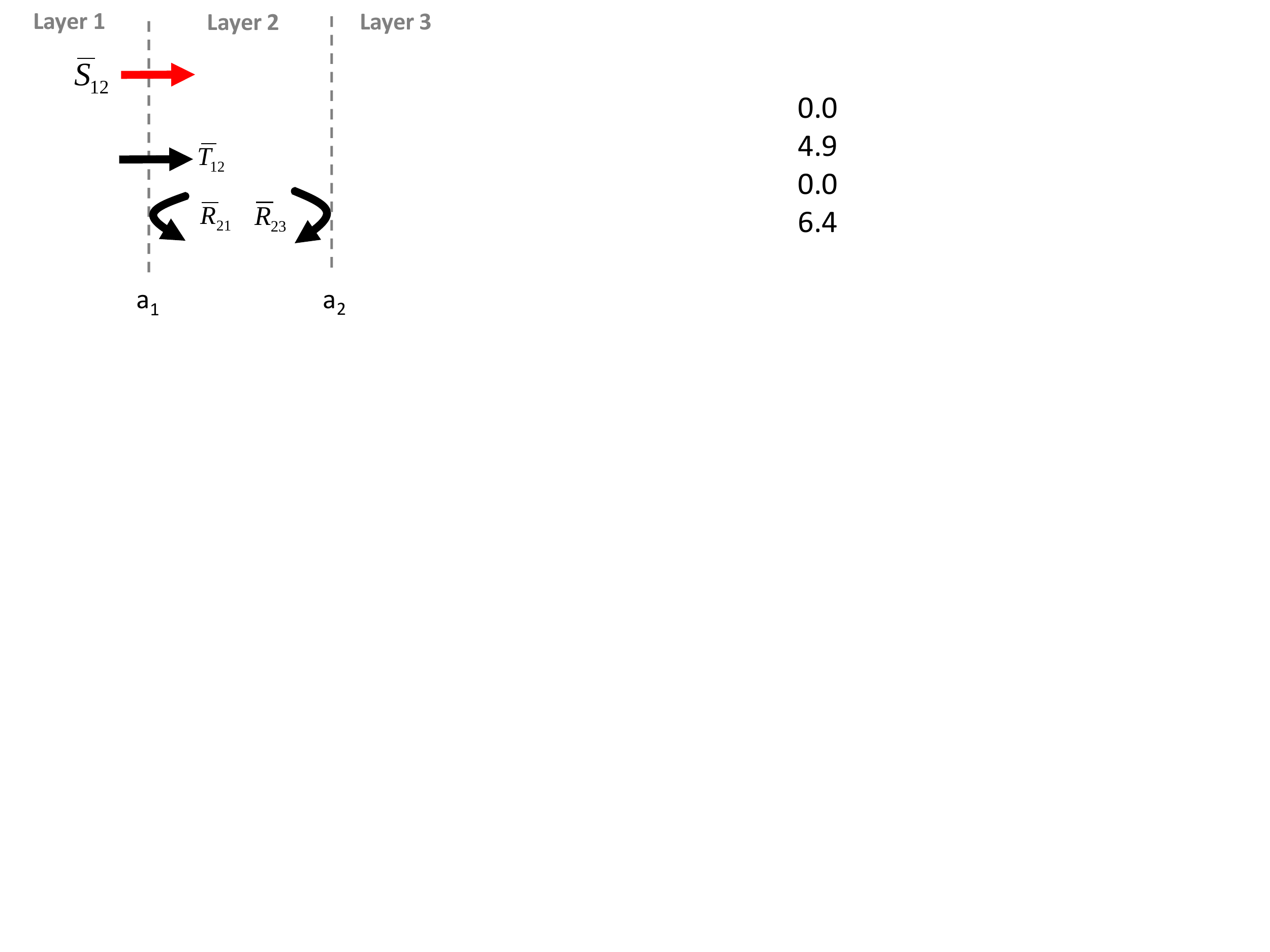}
    \caption{$S$-coefficient for the outgoing-wave case for three cylindrical layers in the $\rho z$-plane.}
    \label{S2.F.S.Out}
    \end{minipage}
    \hspace{1.0cm}
    \begin{minipage}[b]{0.45\linewidth}
    \centering
    \includegraphics[width=2.3in]{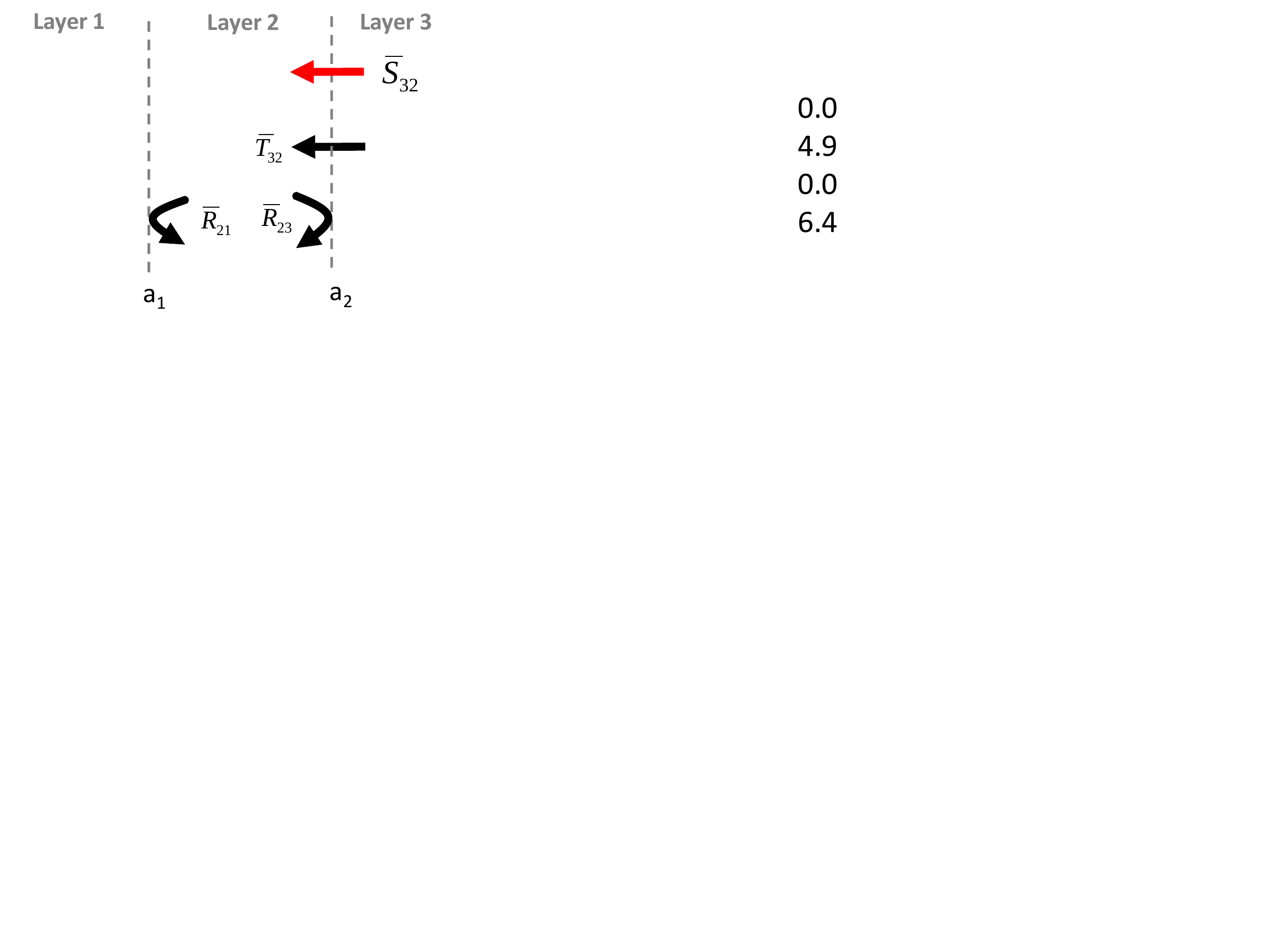}
    \caption{$S$-coefficient for the standing-wave case for three cylindrical layers in the $\rho z$-plane.}
    \label{S2.F.S.Stand}
    \end{minipage}
\end{figure}

Before redefining the generalized transmission coefficients, we first need to consider the so-called $\bS$ coefficients \cite[p.~168, 171]{Chew:Inhomogeneous}, which are used in the definition of the transmission coefficients. The $\bS$ coefficients can be regarded as auxiliary factors incorporating multiple reflections and transmissions in the relevant layers. The two types of $\bS$ coefficients are depicted in Figure \ref{S2.F.S.Out} and \ref{S2.F.S.Stand}, in a three-layer example.
For the outgoing-wave case, the $\bS$ coefficient is
\begin{flalign}
\bS_{12}=\left[\bI-\bR_{21}\cdot\bR_{23}\right]^{-1}\cdot\bT_{12}. \label{S2.eq.S12}
\end{flalign}
Using the range-conditioned cylindrical functions, \eqref{S2.eq.S12} can be rewritten as
\begin{flalign}
\bS_{12}&=
    \left[\bI-\left(\beta^2_{21}\hbR_{21}\right)\cdot
        \left(\alpha^2_{22}\hbR_{23}\right)\right]^{-1}\cdot
    \left(\alpha_{11}\beta_{21}\hbT_{12}\right) \notag\\
&=\alpha_{11}\beta_{21}
    \left[\bI-\beta^2_{21}\alpha^2_{22}\hbR_{21}\cdot\hbR_{23}\right]^{-1}\cdot
    \hbT_{12}
=\alpha_{11}\beta_{21}\hbS_{12}. \label{S2.eq.S12.re}
\end{flalign}
Similarly, for the standing-wave case, the $\bS$ coefficient is
\begin{flalign}
\bS_{32}=\left[\bI-\bR_{23}\cdot\bR_{21}\right]^{-1}\cdot\bT_{32}, \label{S2.eq.S32}
\end{flalign}
which can be rewritten using the range-conditioned cylindrical functions as
\begin{flalign}
\bS_{32}&=
    \left[\bI-\left(\alpha^2_{22}\hbR_{23}\right)\cdot
        \left(\beta^2_{21}\hbR_{21}\right)\right]^{-1}\cdot
    \left(\alpha_{22}\beta_{32}\hbT_{32}\right) \notag\\
&=\alpha_{22}\beta_{32}
    \left[\bI-\beta^2_{21}\alpha^2_{22}\hbR_{23}\cdot\hbR_{21}\right]^{-1}\cdot
    \hbT_{32}
	=\alpha_{22}\beta_{32}\hbS_{32}. \label{S2.eq.S32.re}
\end{flalign}

When more than three layers are present, the reflection coefficients $\bR_{23}$ in \eqref{S2.eq.S12} and $\bR_{21}$ in \eqref{S2.eq.S32} should be replaced with generalized ones such as $\tbR_{23}$ and $\tbR_{21}$. The above redefinition of $\bS$ coefficients is fully consistent with such generalization. As a result, the redefined coefficients for a generic cylindrically stratified medium are written as
\begin{subequations}
\begin{flalign}
\hbS_{i,i+1}&=\bN_{(i+1)+}\cdot\hbT_{i,i+1}, \label{S2.eq.general.Sij}\\
\hbS_{i+1,i}&=\bN_{i-}\cdot\hbT_{i+1,1}. \label{S2.eq.general.Sji}
\end{flalign}
\end{subequations}
Using \eqref{S2.eq.general.Sij} and \eqref{S2.eq.general.Sji}, generalized transmission coefficients can be obtained next. For the outgoing-wave case, we have
\begin{flalign}
\tbT_{ji}&=\bT_{i-1,i}\cdot\bS_{i-2,i-1}\cdots\bS_{j,j+1}, \label{S2.eq.Tji.Out}
\end{flalign}
where $i>j$. The above equation can be rewritten as
\begin{flalign}
\tbT_{ji}&=\bT_{i-1,i}\cdot\bS_{i-2,i-1}\cdots\bS_{j,j+1}
	=\bT_{i-1,i}\cdot\bX_{j,i-1}\cdot\bI \notag\\
&=\left(\beta_{i,i-1}\hbT_{i-1,i}\right)\cdot
    \left(\prod_{k=j+1}^{i-1}\beta_{k,k-1}\alpha_{kk}
        \bN_{k+}\cdot\hbT_{k-1,k}\right)\cdot\left(\alpha_{jj}\bI\right)
=\beta_{i,i-1}\htbT_{ji}\cdot\left(\alpha_{jj}\bI\right), \label{S2.eq.Tji.Out.re}
\end{flalign}
where
\begin{flalign}
\bN_{k+}=
    \left[\bI-\beta^2_{k,k-1}\alpha^2_{kk}\hbR_{k,k-1}\cdot\htbR_{k,k+1}\right]^{-1}.
    \label{S2.eq.Nk.plus}
\end{flalign}
It should be noted that the product in \eqref{S2.eq.Tji.Out.re} is indeed the product of a number of 2$\times$2 matrices, so the order of the product should be made clear. The 2$\times$2 matrix for $k=j+1$ and 2$\times$2 matrix for $k=i-1$ should be placed in the rightmost and leftmost in the matrix product, respectively. Furthermore, when $i=j+1$, the matrix product is not defined. In this case, $\bX_{j,i-1}=\bI$.

Generalized transmission coefficient for the standing-wave case is defined in a slightly different way:
\begin{flalign}
\tbT_{ji}&=\bT_{i+1,i}\cdot\bS_{i+2,i+1}\cdots\bS_{j,j-1}, \label{S2.eq.Tji.Stand}
\end{flalign}
where $i<j$. Using the range-conditioned cylindrical functions, \eqref{S2.eq.Tji.Stand} can be rewritten as
\begin{flalign}
\tbT_{ji}&=\bT_{i+1,i}\cdot\bS_{i+2,i+1}\cdots\bS_{j,j-1}
	=\bT_{i+1,i}\cdot\bX_{j,i+1}\cdot\bI \notag\\
&=\left(\alpha_{ii}\hbT_{i+1,i}\right)\cdot
    \left(\prod_{k=i+1}^{j-1}\beta_{k,k-1}\alpha_{kk}
        \bN_{k-}\cdot\hbT_{k+1,k}\right)\cdot
    \left(\beta_{j,j-1}\bI\right)
	=\alpha_{ii}\htbT_{ji}\cdot\left(\beta_{j,j-1}\bI\right), \label{S2.eq.Tji.Stand.re}
\end{flalign}
where
\begin{flalign}
\bN_{k-}=
    \left[\bI-\beta^2_{k,k-1}\alpha^2_{kk}\hbR_{k,k+1}\cdot\htbR_{k,k-1}\right]^{-1}.
    \label{S2.eq.Nk.minus}
\end{flalign}
Again, the product in \eqref{S2.eq.Tji.Stand.re} is indeed the product of 2$\times$2 matrices. In contrast to \eqref{S2.eq.Tji.Out.re}, the order of the matrix product is just the opposite. The 2$\times$2 matrix for $k=i+1$ and 2$\times$2 matrix for $k=j-1$ should be placed in the leftmost and rightmost in the matrix product, respectively. Furthermore, when $i=j-1$ the matrix product is again not defined, so $\bX_{j,i+1}=\bI$.

\subsection{Conditioned integrands for all argument types}
\label{sec.2.4}

Recall that there are four integrand types depending on the relative positions of $\rho$ and $\rho'$ (see \eqref{S2.eq.Fn.case1.orig} -- \eqref{S2.eq.Fn.case4.orig}). For Case 1, there are four arguments of interest: $k_{j\rho}a_{j-1}$, $k_{j\rho}\rho'$, $k_{j\rho}\rho$, and $k_{j\rho}a_j$. For convenience, we let $a_{j-1}=a_1$, $\rho'=a_2$, $\rho=a_3$, and $a_j=a_4$ so that $a_1<a_2<a_3<a_4$. The conditioned integrand factor $\Fn$
then becomes in this case
\begin{flalign}
\Fn&=
    \left[\Hn(k_{j\rho}\rho)\bI+\Jn(k_{j\rho}\rho)\tbR_{j,j+1}\right]\cdot\tbM_{j+}\cdot
    \left[\Jn(k_{j\rho}\rho')\bI+\Hn(k_{j\rho}\rho')\tbR_{j,j-1}\right] \notag\\
    &=\left[\alpha_{j3}\hHn(k_{j\rho}\rho)\bI+
		\beta_{j3}\hJn(k_{j\rho}\rho)\alpha_{j4}^2\htbR_{j,j+1}\right]\cdot
    \tbM_{j+} \notag\\
	&\qquad\qquad\qquad\qquad\cdot\left[\beta_{j2}\hJn(k_{j\rho}\rho')\bI+
		\alpha_{j2}\hHn(k_{j\rho}\rho')\beta_{j1}^2\htbR_{j,j-1}\right] \notag\\
	&=\left[\beta_{j2}\alpha_{j3}\hHn(k_{j\rho}\rho)\bI+
		(\beta_{j2}\alpha_{j4})(\beta_{j3}\alpha_{j4})\hJn(k_{j\rho}\rho)\htbR_{j,j+1}\right]\cdot
    \tbM_{j+} \notag\\
	&\qquad\qquad\qquad\qquad\cdot\left[\hJn(k_{j\rho}\rho')\bI+
		(\beta_{j1}\alpha_{j2})^2\hHn(k_{j\rho}\rho')\htbR_{j,j-1}\right] \notag\\
	&=\left[A_1\hHn(k_{j\rho}\rho)\bI+A_2\hJn(k_{j\rho}\rho)\htbR_{j,j+1}\right]\cdot\tbM_{j+}
	\left[A_3\hJn(k_{j\rho}\rho')\bI+A_4\hHn(k_{j\rho}\rho')\htbR_{j,j-1}\right].
\label{S2.eq.Fn.case1}
\end{flalign}
Note that the reciprocity property $\beta_{j2}^{-1}=\alpha_{j2}$ has been used in the above, and that the magnitudes of the multiplicative factors $A_1$, $A_2$, $A_3$, and $A_4$ are never greater than unity due to the boundness property.

For Case 2, the four arguments of interest are: $k_{j\rho}a_{j-1}$, $k_{j\rho}\rho$, $k_{j\rho}\rho'$, and $k_{j\rho}a_j$. Again, we let $a_{j-1}=a_1$, $\rho=a_2$, $\rho'=a_3$, and $a_j=a_4$ so that $a_1<a_2<a_3<a_4$.
Similarly, the conditioned integrand becomes
\begin{flalign}
\Fn&=
    \left[\Jn(k_{j\rho}\rho)\bI+\Hn(k_{j\rho}\rho)\tbR_{j,j-1}\right]\cdot\tbM_{j-}\cdot
    \left[\Hn(k_{j\rho}\rho')\bI+\Jn(k_{j\rho}\rho')\tbR_{j,j+1}\right] \notag\\
    &=\left[\beta_{j2}\hJn(k_{j\rho}\rho)\bI+
		\alpha_{j2}\hHn(k_{j\rho}\rho)\beta_{j1}^2\htbR_{j,j-1}\right]\cdot
    \tbM_{j-} \notag\\
	&\qquad\qquad\qquad\qquad\cdot\left[\alpha_{j3}\hHn(k_{j\rho}\rho')\bI+
		\beta_{j3}\hJn(k_{j\rho}\rho')\alpha_{j4}^2\htbR_{j,j+1}\right] \notag\\
    &=\left[\beta_{j2}\alpha_{j3}\hJn(k_{j\rho}\rho)\bI+
		(\beta_{j1}\alpha_{j2})(\beta_{j1}\alpha_{j3})\hHn(k_{j\rho}\rho)\htbR_{j,j-1}\right]\cdot
    \tbM_{j-} \notag\\
	&\qquad\qquad\qquad\qquad\cdot\left[\hHn(k_{j\rho}\rho')\bI+
		(\beta_{j3}\alpha_{j4})^2\hJn(k_{j\rho}\rho')\htbR_{j,j+1}\right] \notag\\
    &=\left[B_1\hJn(k_{j\rho}\rho)\bI+B_2\hHn(k_{j\rho}\rho)\htbR_{j,j-1}\right]\cdot\tbM_{j-}
	\cdot\left[B_3\hHn(k_{j\rho}\rho')\bI+B_4\hJn(k_{j\rho}\rho')\htbR_{j,j+1}\right],
\label{S2.eq.Fn.case2}
\end{flalign}
where $\alpha_{j3}^{-1}=\beta_{j3}$ has been used. Again, the magnitudes of the multiplicative factors $B_1$, $B_2$, $B_3$, and $B_4$ are never greater than unity due to the boundness property.

For Case 3, there are six arguments of interest: $k_{i\rho}a_{i-1}$, $k_{i\rho}\rho$, $k_{i\rho}a_i$, $k_{j\rho}a_{j-1}$, $k_{j\rho}\rho'$, and $k_{j\rho}a_j$. We let $a_{i-1}=a_1$, $\rho=a_2$, $a_i=a_3$, $a_{j-1}=b_1$, $\rho'=b_2$, and $a_j=b_3$ so that $a_1<a_2<a_3$ and $b_1<b_2<b_3$. The conditioned integrand then writes as
\begin{flalign}
\Fn&=
    \left[\Hn(k_{i\rho}\rho)\bI+\Jn(k_{i\rho}\rho)\tbR_{i,i+1}\right]\cdot
    \tbN_{i+}\cdot\tbT_{ji}\cdot\tbM_{j+} \notag\\
    &\qquad\qquad\qquad\cdot
    \left[\Jn(k_{j\rho}\rho')\bI+\Hn(k_{j\rho}\rho')
        \tbR_{j,j-1}\right] \notag\\
&=\left[\Hn(k_{i\rho}\rho)\bI+\Jn(k_{i\rho}\rho)\tbR_{i,i+1}\right]\cdot
    \tbN_{i+}\cdot\bT_{i-1,i}\cdot\bX_{j,i-1} \notag\\
    &\qquad\qquad\qquad\cdot\bI\cdot\tbM_{j+}\cdot
    \left[\Jn(k_{j\rho}\rho')\bI+\Hn(k_{j\rho}\rho')
        \tbR_{j,j-1}\right] \notag\\
&=\left[\beta_{i1}\alpha_{i2}\hHn(k_{i\rho}\rho)\bI+
	(\beta_{i1}\alpha_{i3})(\beta_{i2}\alpha_{i3})\hJn(k_{i\rho}\rho)\htbR_{i,i+1}\right]\cdot
    \tbN_{i+}\cdot\htbT_{ji} \notag\\
&\qquad\qquad\qquad\cdot\tbM_{j+}\cdot
    \left[\beta_{j2}\alpha_{j3}\hJn(k_{j\rho}\rho')\bI+
	(\beta_{j1}\alpha_{j2})(\beta_{j1}\alpha_{j3})\hHn(k_{j\rho}\rho')\htbR_{j,j-1}\right] \notag\\
&=\left[C_1\hHn(k_{i\rho}\rho)\bI+C_2\hJn(k_{i\rho}\rho)\htbR_{i,i+1}\right]\cdot
    \tbN_{i+}\cdot\htbT_{ji}\cdot\tbM_{j+} \notag\\
&\qquad\qquad\qquad\cdot
    \left[C_3\hJn(k_{j\rho}\rho')\bI+C_4\hHn(k_{j\rho}\rho')\htbR_{j,j-1}\right].
\label{S2.eq.Fn.case3}
\end{flalign}
The magnitudes of all multiplicative factors $C_1$, $C_2$, $C_3$, and $C_4$ are again never greater than unity.

For Case 4, the arguments of interest are the same as those for Case 3, and the integrand becomes
\begin{flalign}
\Fn&=
    \left[\Jn(k_{i\rho}\rho)\bI+\Hn(k_{i\rho}\rho)\tbR_{i,i-1}\right]\cdot
    \tbN_{i-}\cdot\tbT_{ji}\cdot\tbM_{j-} \notag\\
    &\qquad\qquad\qquad\cdot
    \left[\Hn(k_{j\rho}\rho')\bI+\Jn(k_{j\rho}\rho')\tbR_{j,j+1}\right] \notag\\
&=\left[\Jn(k_{i\rho}\rho)\bI+\Hn(k_{i\rho}\rho)\tbR_{i,i-1}\right]\cdot
    \tbN_{i-}\cdot\bT_{i+1,i}\cdot\bX_{j,i+1} \notag\\
    &\qquad\qquad\qquad\cdot\bI\cdot\tbM_{j-}\cdot
    \left[\Hn(k_{j\rho}\rho')\bI+\Jn(k_{j\rho}\rho')\tbR_{j,j+1}\right] \notag\\
&=\left[\beta_{i2}\alpha_{i3}\hJn(k_{i\rho}\rho)\bI+
	(\beta_{i1}\alpha_{i2})(\beta_{i1}\alpha_{i3})\hHn(k_{i\rho}\rho)\htbR_{i,i-1}\right]\cdot
    \tbN_{i-}\cdot\htbT_{ji} \notag\\
    &\qquad\qquad\qquad\cdot\tbM_{j-}\cdot
    \left[\beta_{j1}\alpha_{j2}\hHn(k_{j\rho}\rho')\bI+
	(\beta_{j1}\alpha_{j3})(\beta_{j2}\alpha_{j3})\hJn(k_{j\rho}\rho')\htbR_{j,j+1}\right] \notag\\	
&=\left[D_1\hJn(k_{i\rho}\rho)\bI+D_2\hHn(k_{i\rho}\rho)\htbR_{i,i-1}\right]\cdot
    \tbN_{i-}\cdot\htbT_{ji}\cdot\tbM_{j-} \notag\\
    &\qquad\qquad\qquad\cdot
    \left[D_3\hHn(k_{j\rho}\rho')\bI+D_4\hJn(k_{j\rho}\rho')\htbR_{j,j+1}\right].
\label{S2.eq.Fn.case4}
\end{flalign}
Once again, the magnitudes of the multiplicative factors $D_1$, $D_2$, $D_3$, and $D_4$ are never greater than unity. It should be noted that $D_1$ and $D_2$ for Case 4 have precisely the same form as $C_3$ and $C_4$ for Case 3, respectively. Also, $D_3$ and $D_4$ for Case 4 have precisely the same form as $C_1$ and $C_2$ for Case 3, respectively.

\subsection{Source factor for an arbitrarily-oriented dipole}
\label{sec.2.5}

In addition to $\Fn$, the source factor $\Dj$ should be determined in order to compute \eqref{S2.eq.EzHz.general} for an arbitrarily-oriented electric dipole.
An elementary electric dipole along the unit vector $\hat{\alpha}'$ direction can be represented as $\mathbf{J}(\rr)=Il\hat{\alpha}'\delta(\rr-\rp)$. We write again the source factor here as
\begin{flalign}
\Dj=\frac{\iu}{2}
    \begin{bmatrix}
    \left(\hat{z}k^2_j-\iu k_z\nabla'\right)\cdot\hat{\alpha}' \\
    \iu\omega\epsilon_j\hat{\alpha}'\cdot\hat{z}\times\nabla'
    \end{bmatrix},
\label{S2.eq.Dj}
\end{flalign}
where $\nabla'=\hat{\rho}'\frac{\pa}{\pa\rho'}-\hat{\phi}'\frac{\iu n}{\rho'}-\hat{z}\iu k_z$. Note that the dipole is located at $\rp$, hence the primed components for $\hat{\alpha}'$. In more explicit form, \eqref{S2.eq.Dj} can be written as
\begin{flalign}
\Dj&=\frac{\iu}{2}
    \begin{bmatrix}
    (k^2_{j\rho})\alpha_{z'} \\ 0
    \end{bmatrix}
+\frac{\iu}{2}
    \begin{bmatrix}
    \frac{nk_z}{\rho'}\left(\alpha_{x'}\sin\phi'-\alpha_{y'}\cos\phi'\right) \\
    -\frac{n\omega\epsilon_j}{\rho'}\left(\alpha_{x'}\cos\phi'+\alpha_{y'}\sin\phi'\right)
    \end{bmatrix}
+\frac{\iu}{2}
    \begin{bmatrix}
    -\iu k_z\left(\alpha_{x'}\cos\phi'+\alpha_{y'}\sin\phi'\right) \\
    \iu\omega\epsilon_j\left(-\alpha_{x'}\sin\phi'+\alpha_{y'}\cos\phi'\right)
    \end{bmatrix}\frac{\pa}{\pa\rho'} \notag\\
&=\frac{\iu}{2}\left(\Dja+\Djb+\Djc\frac{\pa}{\pa\rho'}\right).
\label{S2.eq.Dj.Cartesian.b}
\end{flalign}
In \eqref{S2.eq.Dj.Cartesian.b} for a fixed $k_z$, $\Dja$ is constant, $\Djb$ depends on the order $n$, and $\Djc$ is associated with the partial derivative with respect to $\rho'$. To convert \eqref{S2.eq.Dj.Cartesian.b} into cylindrical coordinates, we let $\hat{\alpha}'$=$\hat{x}'\alpha_{x'}+\hat{y}'\alpha_{y'}+\hat{z}'\alpha_{z'}$=
$\hat{\rho}'\alpha_{\rho'}+\hat{\phi}'\alpha_{\phi'}+\hat{z}'\alpha_{z'}$. After some algebra, we can write the source factor in cylindrical coordinates as
\begin{flalign}
\Dj=
    \frac{\iu}{2}
    \left(\Dja+\Djb+\Djc\frac{\pa}{\pa\rho'}\right)
=\frac{\iu}{2}
\left(
    \begin{bmatrix}
    (k^2_{j\rho})\alpha_{z'} \\ 0
    \end{bmatrix}
    +
    \begin{bmatrix}
    -\frac{nk_z}{\rho'}\alpha_{\phi'} \\
    -\frac{n\omega\epsilon_j}{\rho'}\alpha_{\rho'}
    \end{bmatrix}
    +
    \begin{bmatrix}
    -\iu k_z\alpha_{\rho'} \\
    \iu\omega\epsilon_j\alpha_{\phi'}
    \end{bmatrix}\frac{\pa}{\pa\rho'}
\right). \label{S2.eq.Dj.cylin}
\end{flalign}

\subsection{Azimuth modal summation}
\label{sec.2.6}

For the azimuthal mode summation, we use the basic property~\cite{Abramowitz:Handbook}
$B_{-n}(z)=(-1)^n B_n(z)$,
where $n$ is a positive integer, and $B_{n}$ stands for either $J_{n}(z)$, $J'_{n}(z)$, $H^{(1)}_{n}(z)$, or $H'^{(1)}_{n}(z)$ to fold the sum, as usual.
Reflection and transmission coefficients are 2$\times$2 matrices in cylindrical coordinates and include the matrices in \eqref{S2.eq.small.matrixJ} -- \eqref{S2.eq.large.matrixH}.
Therefore, only off-diagonal elements of the reflection and transmission coefficients as well as the auxiliary coefficients ($\tbM_{j\pm}$ and $\bN_{i\pm}$) in $\Fn$ change sign for negative integer order modes. Consequently, only off-diagonal elements of $\Fn$ change sign. Since the order of summation and integration can be interchanged, \eqref{S2.eq.EzHz.general} becomes
\begin{flalign}
    \begin{bmatrix}
    E_z \\ H_z
    \end{bmatrix}
&=\frac{\iu Il}{4\pi\omega\epsilon_j} \intmp dk_ze^{\iu k_z(z-z')}
    \left[\suma e^{\iu n(\phi-\phi')}\Fn\cdot\Dj\right].
\label{S2.eq.EHz.modified.integrand}
\end{flalign}
Using \eqref{S2.eq.Dj.cylin}, the integrand factor inside the brackets in \eqref{S2.eq.EHz.modified.integrand} can be expanded as
\begin{flalign}
\left[\suma e^{\iu n(\phi-\phi')}\Fn\cdot\Dj\right]
&=\frac{\iu}{2}
    \left[\suma e^{\iu n(\phi-\phi')}\Fn\right]\cdot\Dja
+\frac{\iu}{2}
    \left[\suma e^{\iu n(\phi-\phi')}\Fn\cdot\Djb\right] \notag\\
&\qquad+\frac{\iu}{2}\frac{\pa}{\pa\rho'}
    \left[\suma e^{\iu n(\phi-\phi')}\Fn\right]\cdot\Djc.
\label{S2.eq.bracket.z.expanded}
\end{flalign}
Note that neither $\Dja$ nor $\Djc$ depend on the azimuth mode number $n$, so they both can be factored out from the sum. The three sums that appear in the right hand side of \eqref{S2.eq.bracket.z.expanded} can be folded in such a way that only non-negative modes are involved.

For completeness, the conditioned integral expressions for the transverse field components are detailed in \ref{app.c}.

\section{Numerically-robust integration paths}
\label{sec.3.path}

In this section, two numerically-robust integration paths for the integration along $k_z$ are considered. As discussed below, the optimal choice depends on the longitudinal distance between the source and observation points.
\begin{figure}[!htbp]
    \begin{minipage}[b]{0.47\linewidth}
	\centering
	\includegraphics[height=2.2in]{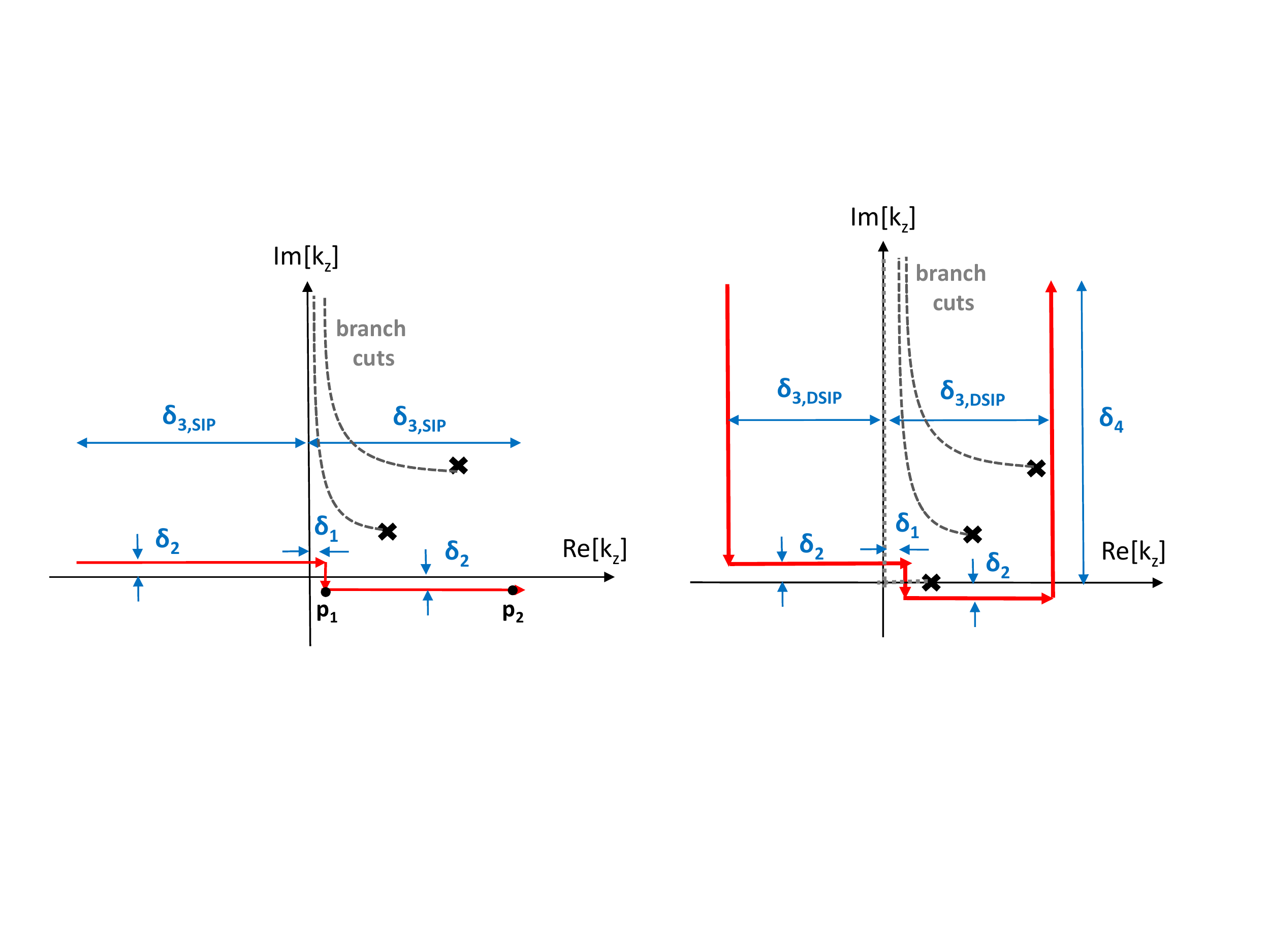}
	\caption{Sommerfeld integration path (SIP) and associated free-parameters.}
	\label{S3.F.SIP}
	\end{minipage}
    \hspace{0.5cm}
    \begin{minipage}[b]{0.47\linewidth}
	\centering
	\includegraphics[height=2.3in]{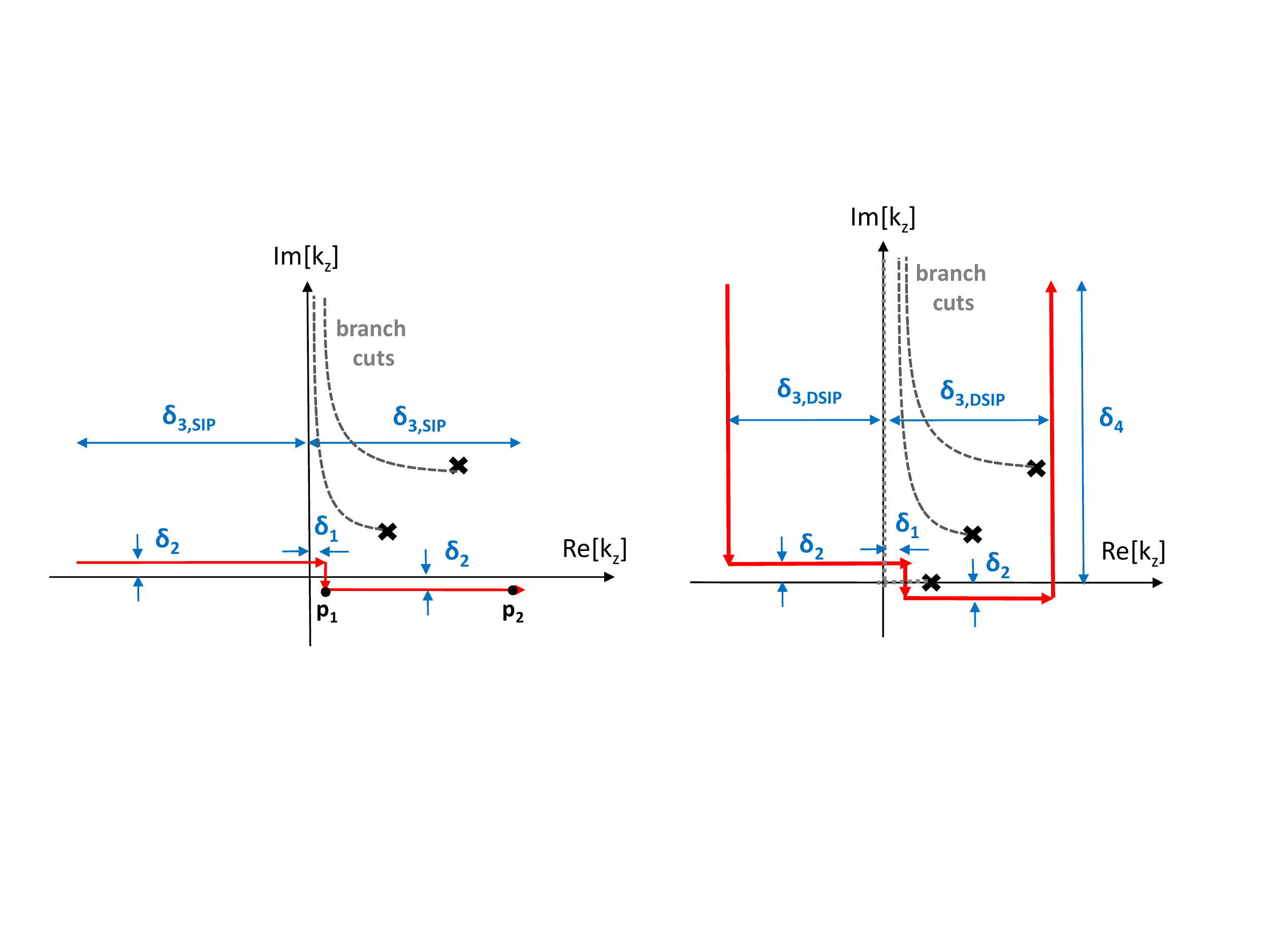}
    \caption{Deformed Sommerfeld integration path (DSIP) and associated free-parameters.}
    \label{S3.F.DSIP}
	\end{minipage}
\end{figure}

\subsection{Sommerfeld integration path (SIP)}
\label{sec.3.1}
The original integration path along the real $k_z$ axis can be deformed into a numerically robust Sommerfeld integration path (SIP) as depicted in Figure \ref{S3.F.SIP}. The SIP is usually employed to avoid integrand singularities in the complex $k_z$ plane due to poles and various branch points/cuts associated with $k_\rho=\sqrt{k^2-k_z^2}$, as well as the singularity of the Hankel function at the origin. Note in particular, that the SIP is chosen above the logarithmic branch-point singularity at the origin from $H^{(1)}_0(z)$.
The SIP is easy to implement numerically since it does not require detailed individual tracking of the singularities. This is in contrast, for example, to an integration along (or near to) the steepest-descent path, where detailed pole-tracking need to be performed for each physical scenario.  As seen from Figure \ref{S3.F.SIP}, only three free-parameters are used to define the SIP: $\delta_1$, $\delta_2$, and $\delta_{3,SIP}$.
Here, the parameters $\delta_1$ and $\delta_2$ are chosen as $\delta_1= \Re e[(k_i)_{\text{min}}]/5$ and
$\delta_2=\Im m[(k_i)_{\text{min}}]/5$, where
$(k_i)_{\text{min}} = \omega (\mu_i \epsilon_i)^{1/2}_{\text{min}}$ refers to the branch point closest to the origin.
The factor $1/5$ was verified to produce accurate results, but it can be modified as long as the path does not cross branch points. If the smallest branch point is purely real because the associated layer is lossless, $\delta_2$ is set equal to $\delta_1$. In this case branch cut crossing is inevitable and a proper tracking of the correct value of $k_\rho$ as the integration path enters into the bottom Riemann sheet and emerges back is required.

The SIP is the most appropriate when $|z-z'|$ is small. Otherwise, the integrand factor $e^{\iu k_z(z-z')}$ is rapidly oscillating along the SIP, which requires a high sampling rate and hence costly numerical integration. When the SIP is appropriate, the integrand exponentially decreases away from the origin and $\delta_{3,SIP}$ can be taken to be a point at which the ratio of the magnitudes of the integrand evaluated at $k_z=p_2$ and $k_z=p_1$, as indicated in Figure \ref{S3.F.SIP}, is below some threshold. For the numerical results shown here, we choose this threshold to be $10^{-20}$. When the SIP is not appropriate, a deformed SIP (DSIP) as depicted in Figure \ref{S3.F.DSIP} should be used, as discussed next.

\subsection{Deformed Sommerfeld integration path (DSIP)}
\label{sec.3.2}
Assuming $z-z'>0$, the DSIP is constructed by deforming the SIP such that the two horizontal paths are bent upwards  while making sure that all branch cuts and singularities are enclosed~\cite{Ebihara03:Calculation}, as depicted in Figure \ref{S3.F.DSIP}. The DSIP exploits the fact that, along its two vertical branches, the factor
\begin{flalign}
e^{\iu k_z(z-z')}=e^{\iu(k'_z+ik''_z)(z-z')}=
    e^{-k''_z(z-z')}e^{\iu k'_z(z-z')}
\label{S3.eq.exp.part}
\end{flalign}
decays exponentially when $z-z'>0$, since $k''_z >0$. If $z-z'<0$, one can instead bend the SIP downwards so that $k''_z<0$ for $e^{\iu k_z(z-z')}$ to decay exponentially.

\begin{figure}[!htbp]
	\centering
	\subfloat[\label{S3.F.delat3.case1}]{%
      \includegraphics[width=3.0in]{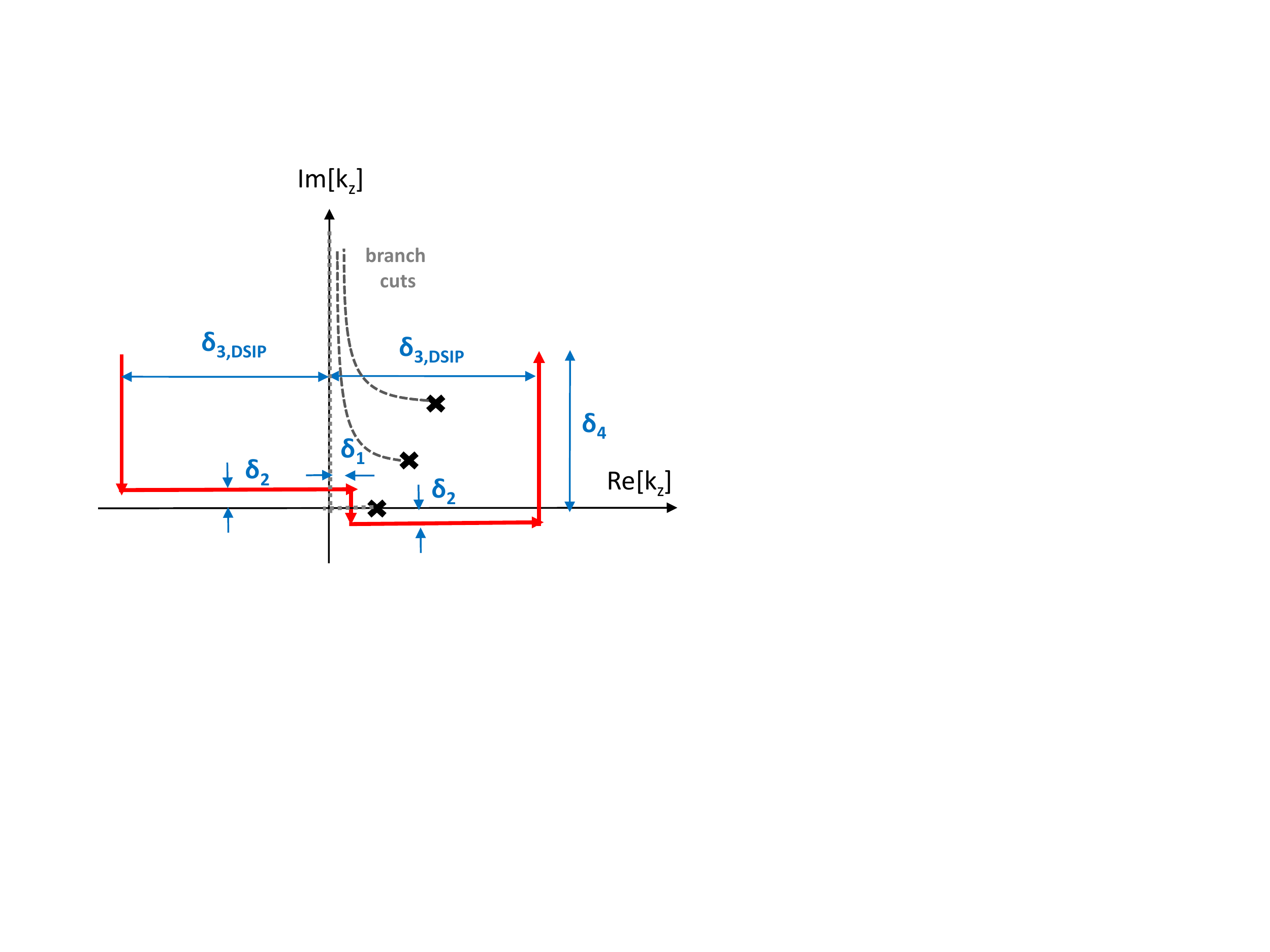}
    }
    \hfill
    \subfloat[\label{S3.F.delat3.case2}]{%
      \includegraphics[width=3.0in]{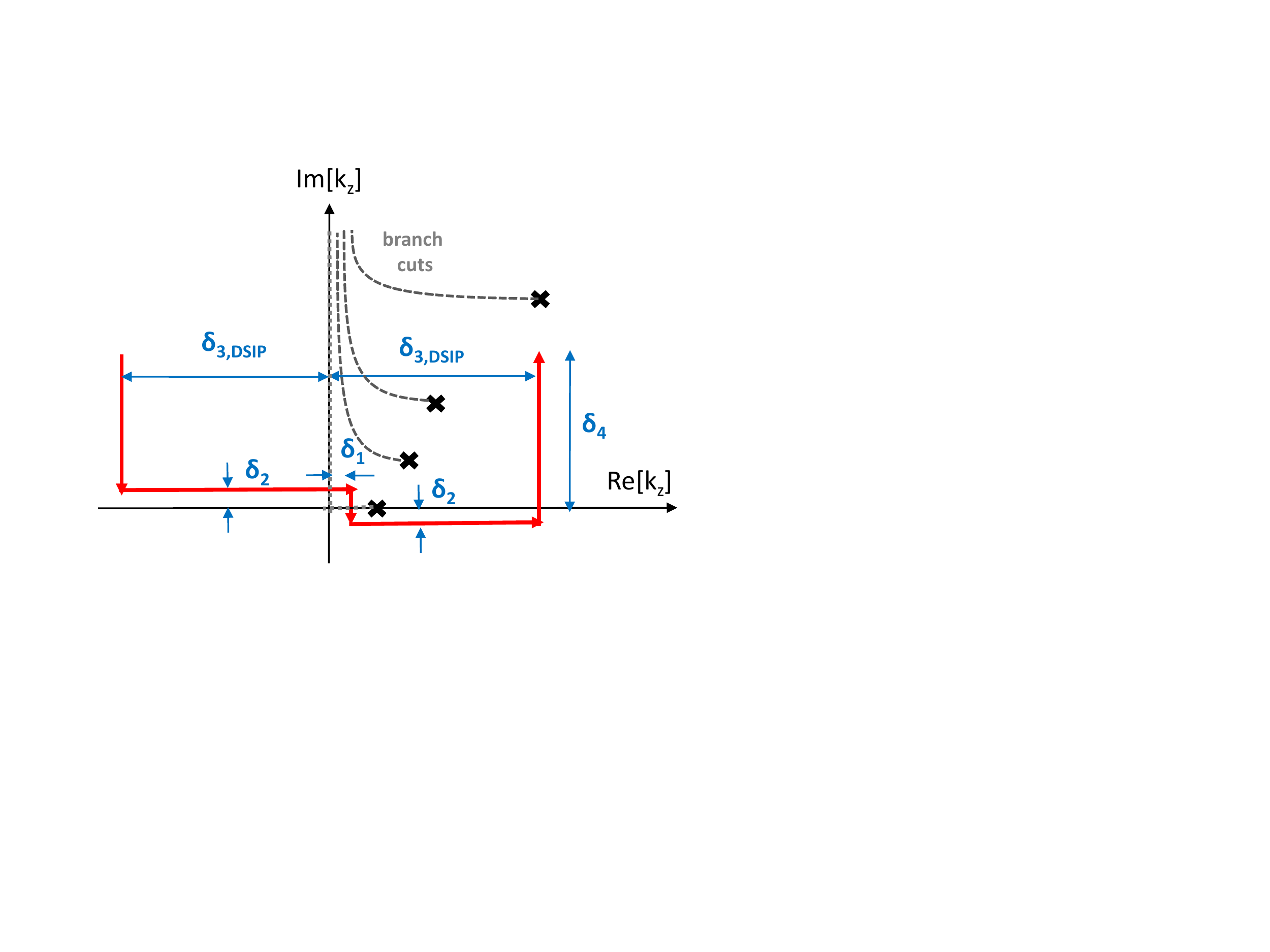}
    }
    \caption{Determination of $\delta_{3,DSIP}$: (a) Case 1 and (b) Case 2.}
    \label{S3.F.delat3.case12}
\end{figure}

There are four free-parameters for the numerical integration along the DSIP: $\delta_1$, $\delta_2$, $\delta_{3,DSIP}$, and $\delta_4$. The parameters $\delta_1$ and $\delta_2$ are defined in the same manner as for the SIP. The truncation parameter $\delta_4$ is determined so that the factor $e^{-\delta_4|z-z'|}$ is below some tolerance $\gamma$, i.e.,
$\delta_4=-\ln\gamma/|z-z'|$. As for $\delta_{3,DSIP}$,
Figure \ref{S3.F.delat3.case1} and \ref{S3.F.delat3.case2} show two different cases to help
clarify its determination. Branch points with {\it imaginary} part greater than $\delta_4$ are excluded from the determination of $\delta_{3,DSIP}$ since integrand values around those points are very small and their contribution to the integral is negligible. Consequently, $\delta_{3,DSIP}$ is defined as some multiplicative factor $f$ times the value of the largest {\it real} part of the remaining (relevant) branch points, i.e.
$\delta_{3,DSIP}=fRe[k_i^{\text{rel}}]_{\text{max}}$,
where $k_i^{\text{rel}}$ represents the group of relevant branch points. We have verified that $f=2$ gives good results. This factor can be increased or decreased under certain limits but it should not be made too close to unity otherwise the integration path would be too close to a singularity.

\subsection{Convergence analysis versus source/field separation}
\label{sec.3.3}
\begin{figure}[!htbp]
  \centering
  \includegraphics[height=2.0in]{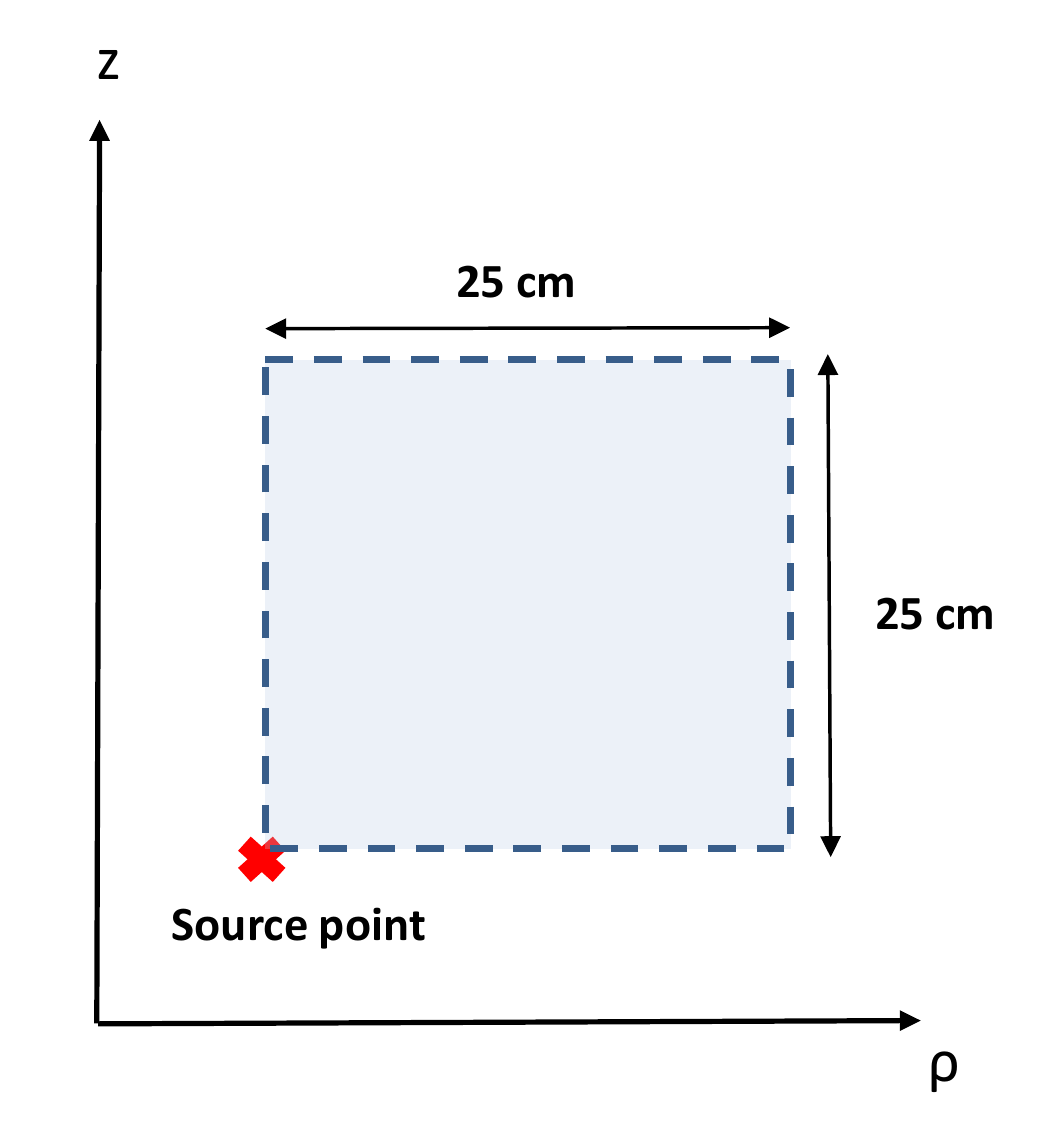}
  \caption{Region of fields points with an arbitrarily situated source in the $\rho z$-plane.}
  \label{S3.F.Field.region}
\end{figure}

\begin{figure}[!htbp]
	\centering
	\subfloat[\label{S3.F.DSIP.10.2000}]{%
      \includegraphics[width=2.5in]{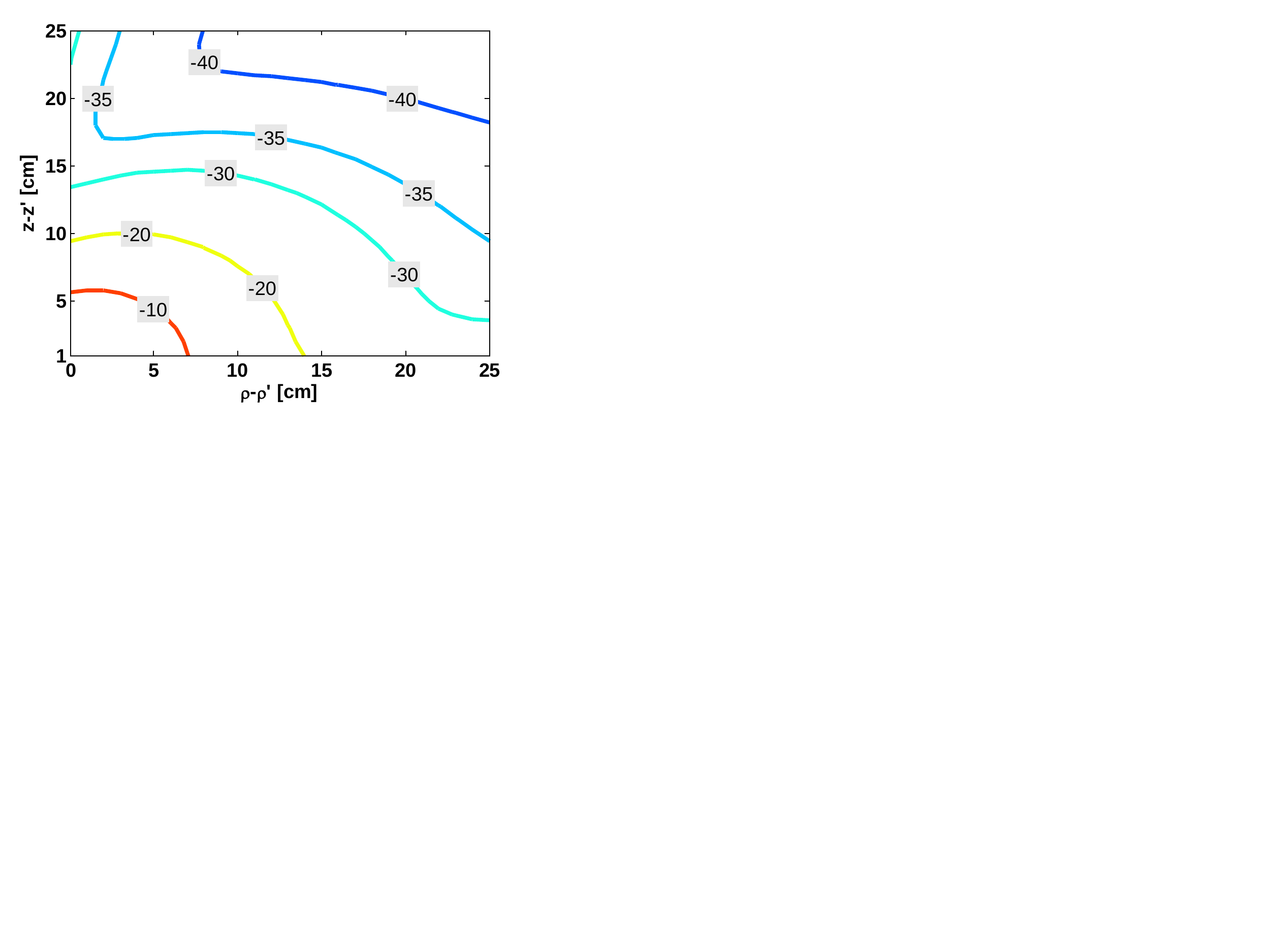}
    }
    \qquad\qquad
    \subfloat[\label{S3.F.DSIP.20.2000}]{%
      \includegraphics[width=2.5in]{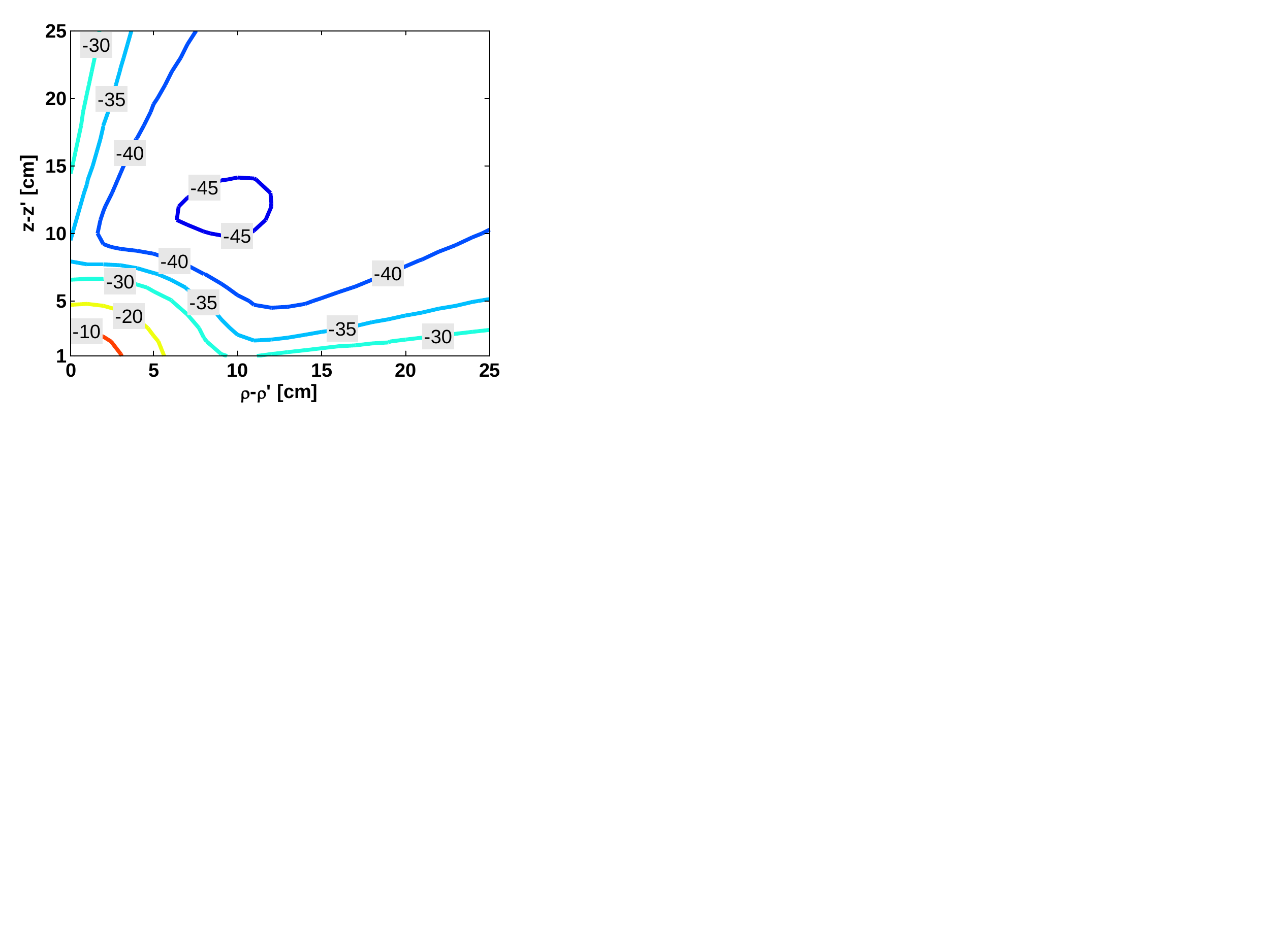}
    }

    \subfloat[\label{S3.F.DSIP.10.4000}]{%
      \includegraphics[width=2.5in]{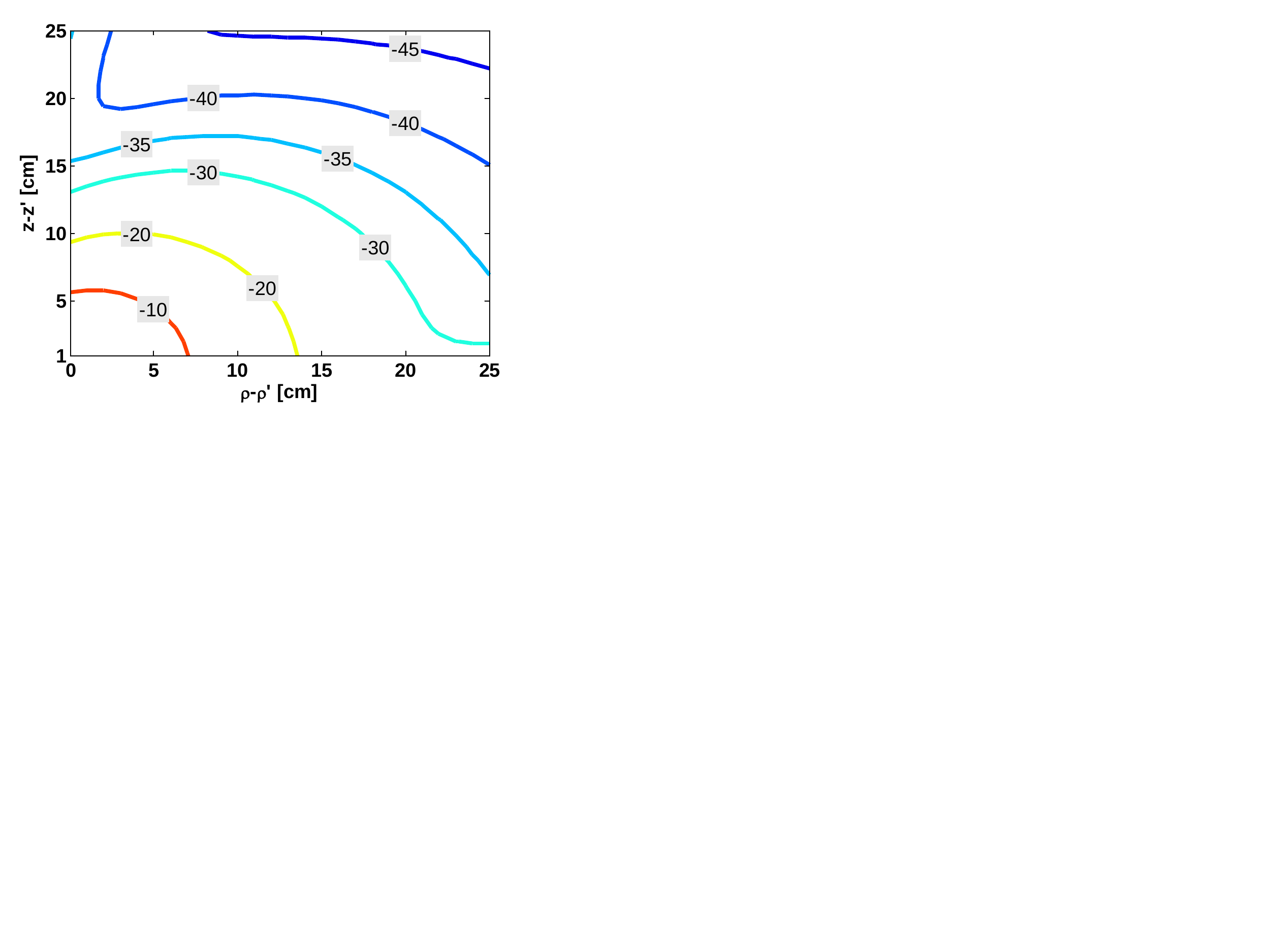}
    }
    \qquad\qquad
    \subfloat[\label{S3.F.DSIP.20.4000}]{%
      \includegraphics[width=2.5in]{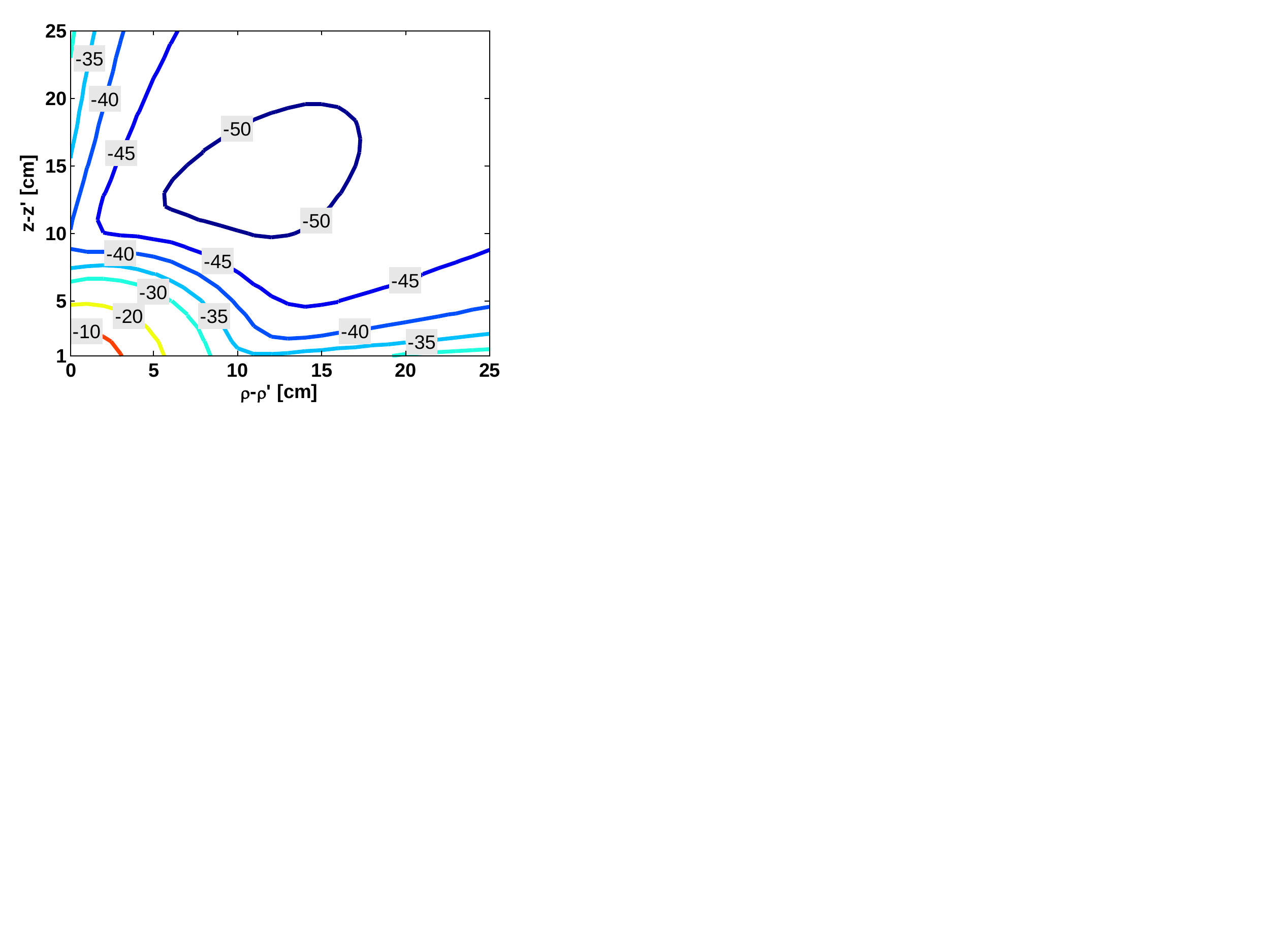}
    }
    \caption{Relative error distributions of the DSIP for various maximum orders $n_{max}$ and integration (quadrature) points $n_{int}$: (a) $n_{max}=10$, $n_{int}=2000$, (b) $n_{max}=20$, $n_{int}=2000$, (c) $n_{max}=10$, $n_{int}=4000$, and (d) $n_{max}=20$, $n_{int}=4000$.}
    \label{S3.F.DSIP.4cases}
\end{figure}

In order to examine the convergence of the numerical integrals along the SIP and DSIP in more detail, consider the square domain shown in Figure \ref{S3.F.Field.region} corresponding to a homogeneous region with $\epsilon_r=1$, $\mu_r=1$, and $\sigma=1$. We assume a $\phi$-directed magnetic dipole operating at 36 kHz and compute the $\phi$-component
of the magnetic field in this square region using (a) the numerical integral for varying number of integration (quadrature) points and azimuthal mode numbers and (b) the closed-form analytical solution (which is available for this simple geometry). Figure \ref{S3.F.DSIP.10.2000} -- \ref{S3.F.DSIP.20.4000} show the relative error distribution using the DSIP. The relative error $\varepsilon_{dB}$ is computed as
\begin{flalign}
\varepsilon_{dB} = 10\log_{10}\frac{|H_{\phi,a}-H_{\phi,n}|}{|H_{\phi,a}|},
\label{S3.eq.relative.error}
\end{flalign}
where $H_{\phi,a}$ is the closed-form (exact) value and $H_{\phi,n}$ is the numerical integration value. As the maximum order and the number of integration points increase, a smaller relative error is obtained, as expected.

\begin{figure}[!htbp]
	\centering
	\subfloat[\label{S3.F.SIP.10.2000}]{%
      \includegraphics[width=2.5in]{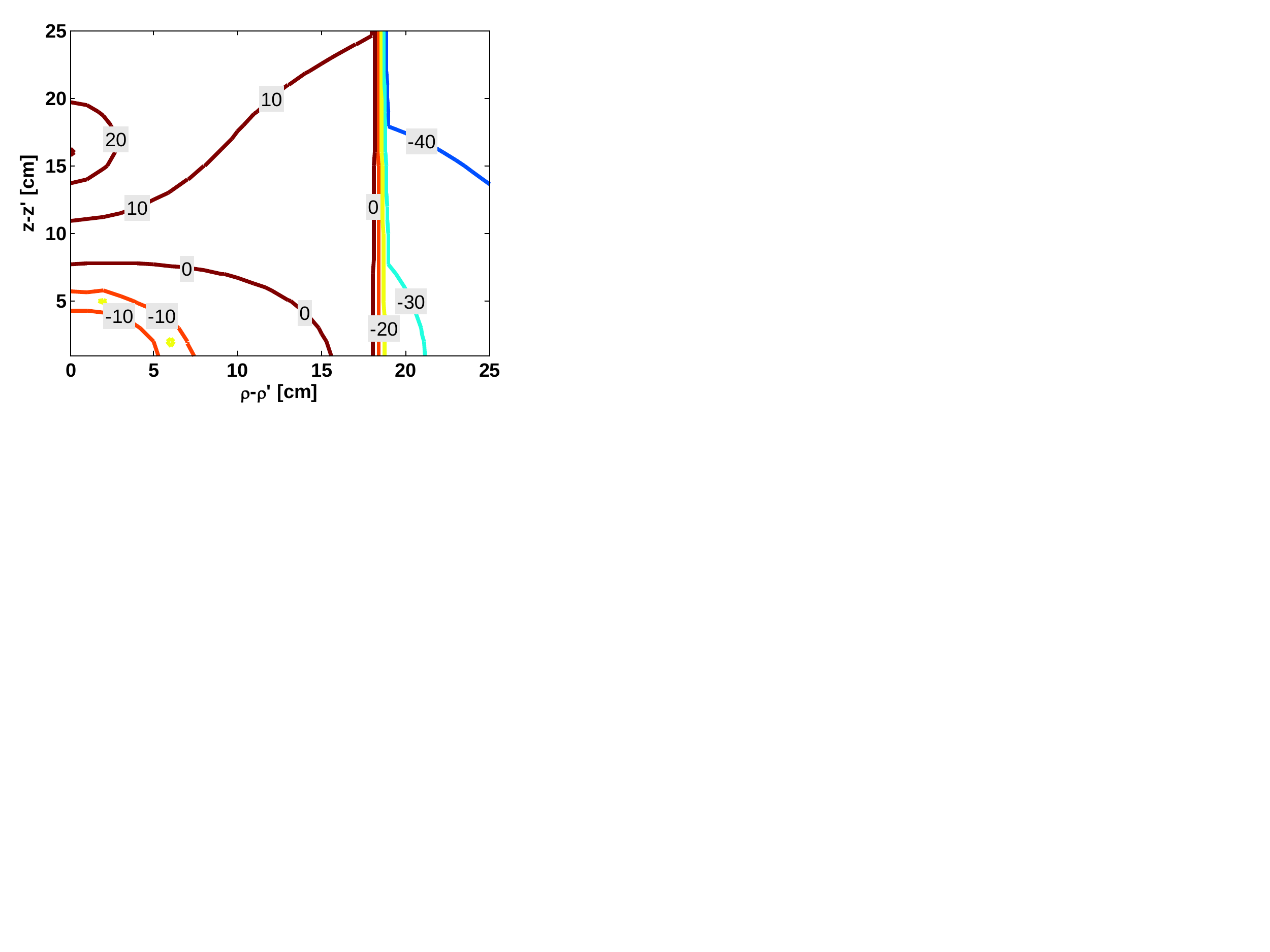}
    }
    \qquad\qquad
    \subfloat[\label{S3.F.SIP.20.4000}]{%
      \includegraphics[width=2.5in]{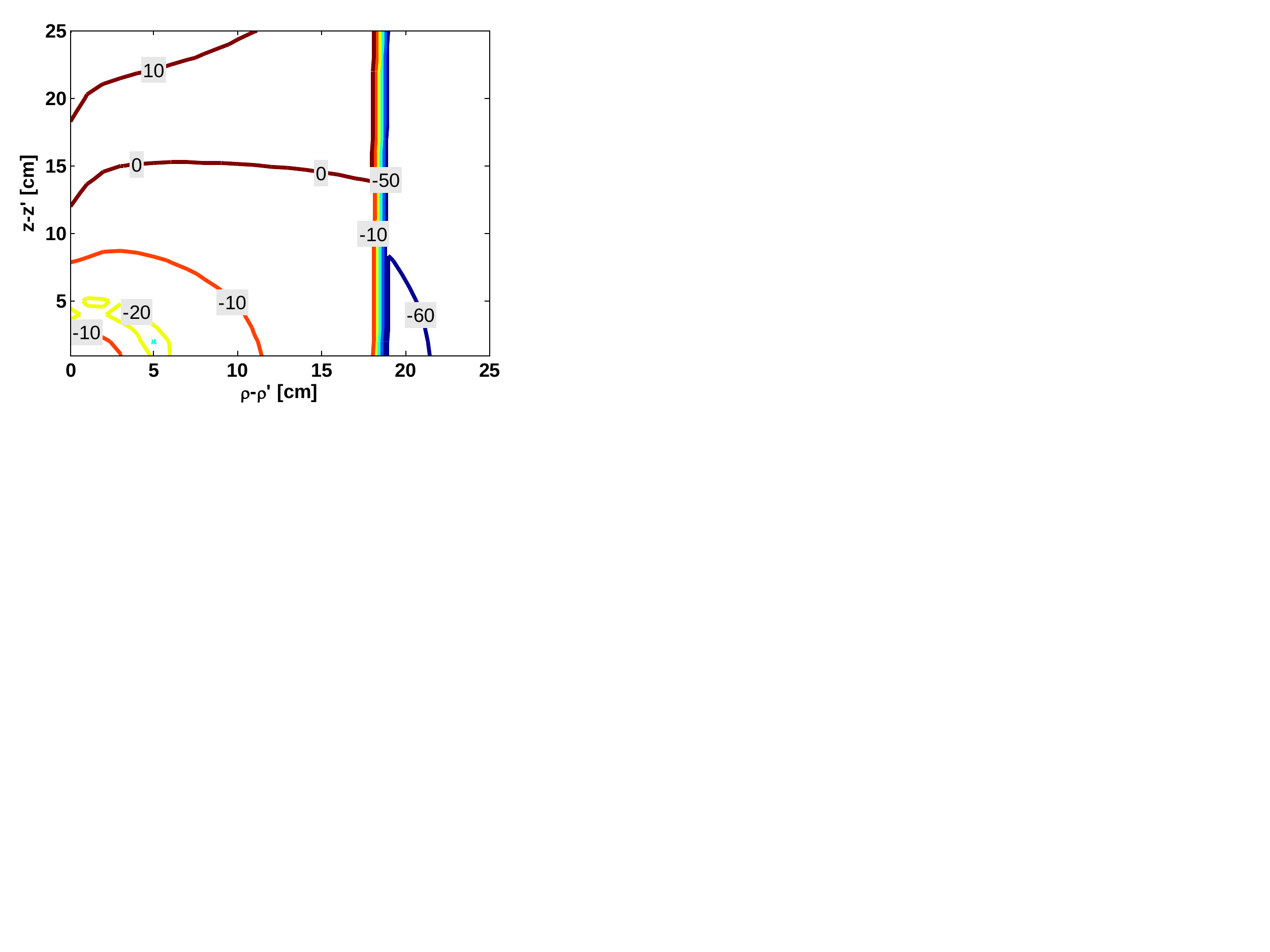}
    }
    \caption{Relative error distributions of the SIP for various maximum orders $n_{max}$ and integration (quadrature) points $n_{int}$: (a) $n_{max}=10$, $n_{int}=2000$ and (b) $n_{max}=20$, $n_{int}=4000$.}
    \label{S3.F.SIP.2cases}
\end{figure}

Figure \ref{S3.F.SIP.10.2000} and \ref{S3.F.SIP.20.4000} show a similar distribution of relative errors now using the SIP. The SIP yields a very small relative error in the region beyond $\rho-\rho'=18$ cm,
where the integrand provides sufficient decay along the SIP regardless of the exponential factor
$e^{\iu k_z(z-z')}$. In this region, an accurate numerical integration can be obtained with small $\delta_{3,SIP}$; otherwise, the DSIP should be used.
It is seen from Figures \ref{S3.F.DSIP.20.4000} and \ref{S3.F.SIP.20.4000} that the DSIP provides in general more robust results and hence can be considered as the primary integration path. Still, the DSIP results show that the convergence gradually degrades as $z \rightarrow z'$ or $\rho\rightarrow\rho'$: strategies to address these two cases are considered next.

\subsubsection{Small longitudinal separation: $z \approx z'$}
\label{sec.3.3.1}

\begin{figure}[!htbp]
	\centering
	\subfloat[\label{S3.F.DSIP.zz.30.4000}]{%
      \includegraphics[width=2.5in]{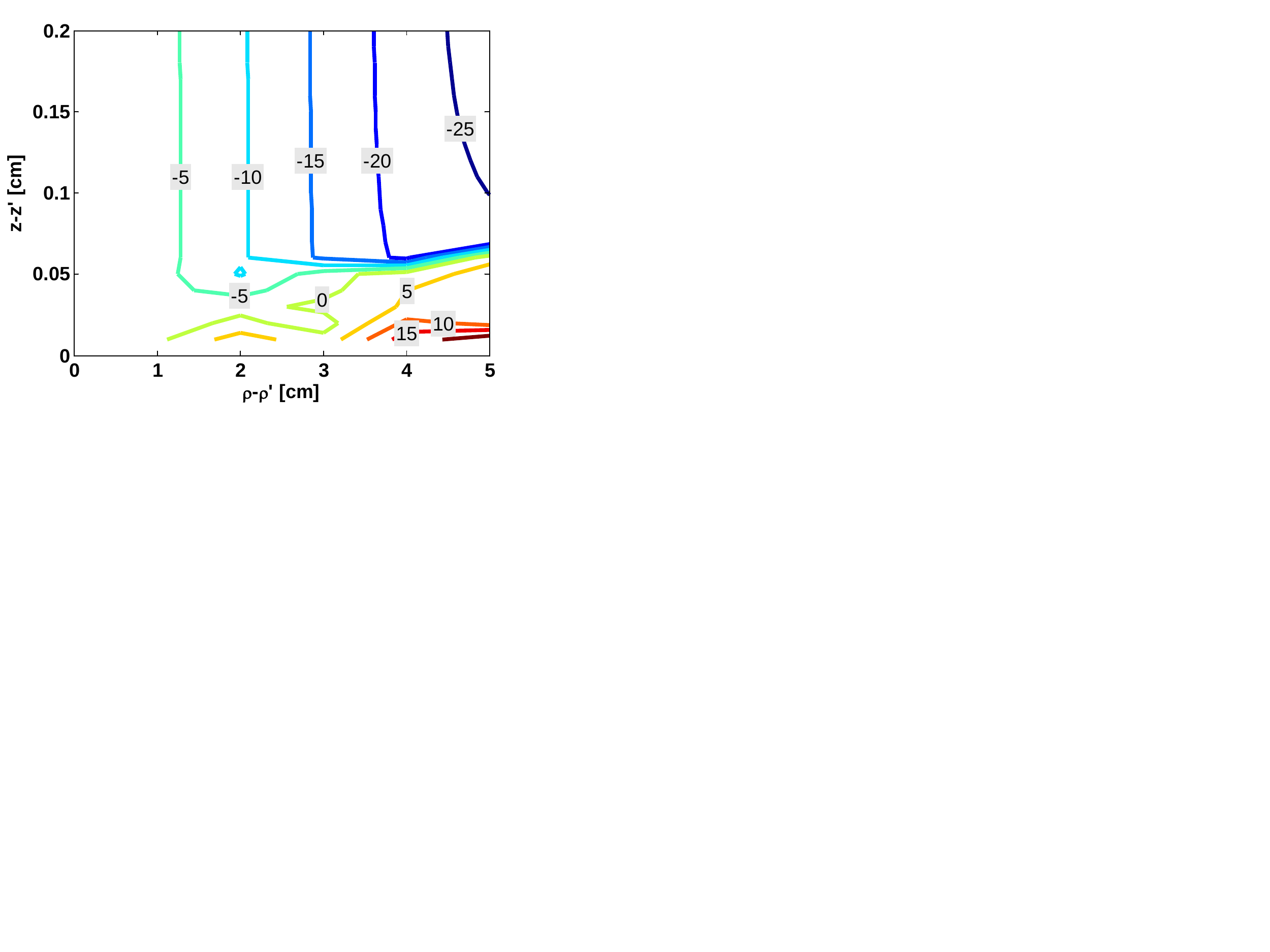}
    }
    \qquad\qquad
    \subfloat[\label{S3.F.SIP.zz.30.4000}]{%
      \includegraphics[width=2.5in]{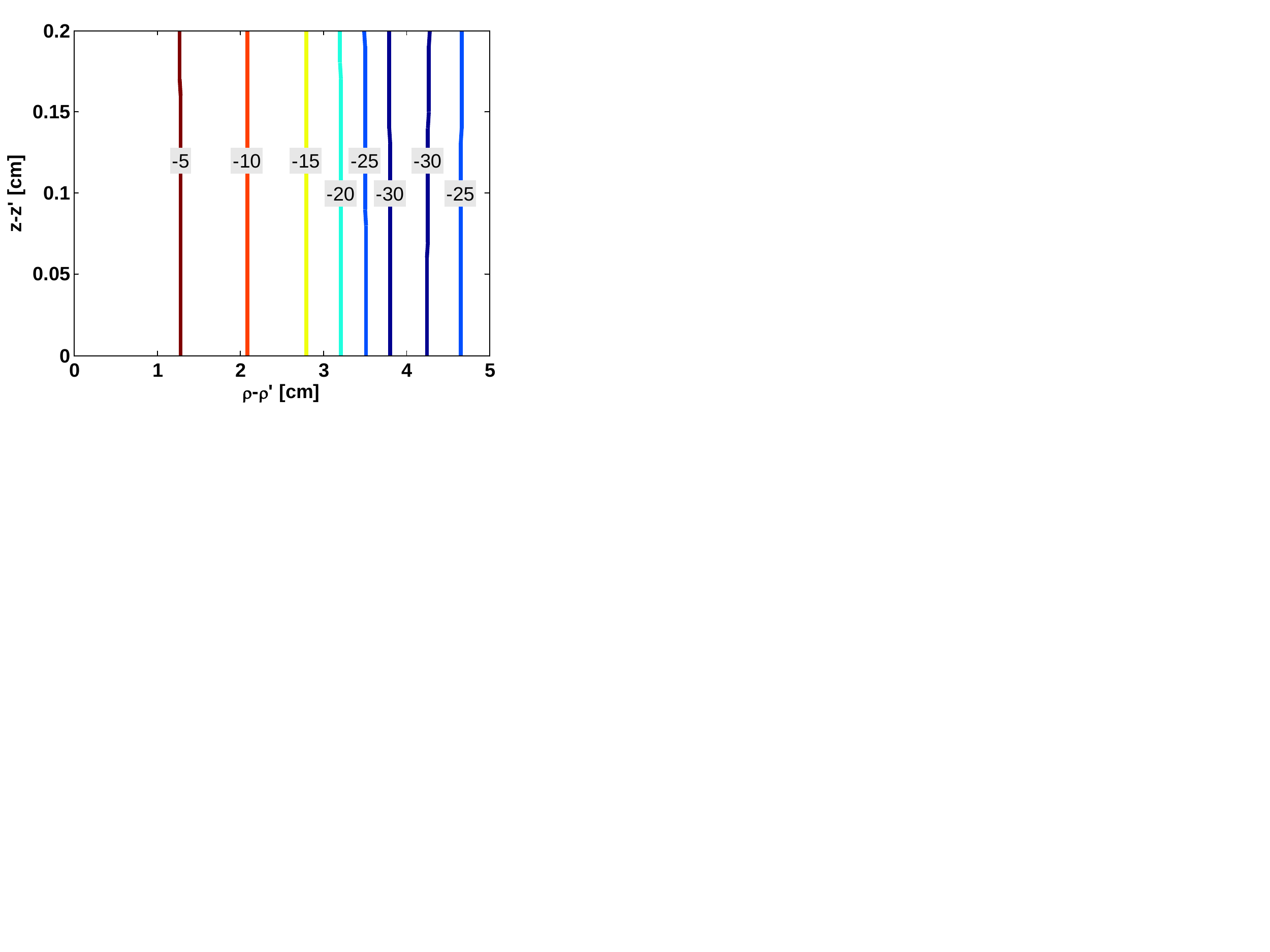}
    }
    \caption{Comparison of relative error distributions between the DSIP and SIP for a maximum order $n_{max}=30$ and integration points $n_{int}=4000$ when $z\approx z'$: (a) DSIP and (b) SIP.}
    \label{S3.F.zz.compare}
\end{figure}

\begin{figure}[!htbp]
  \centering
  \includegraphics[height=2.0in]{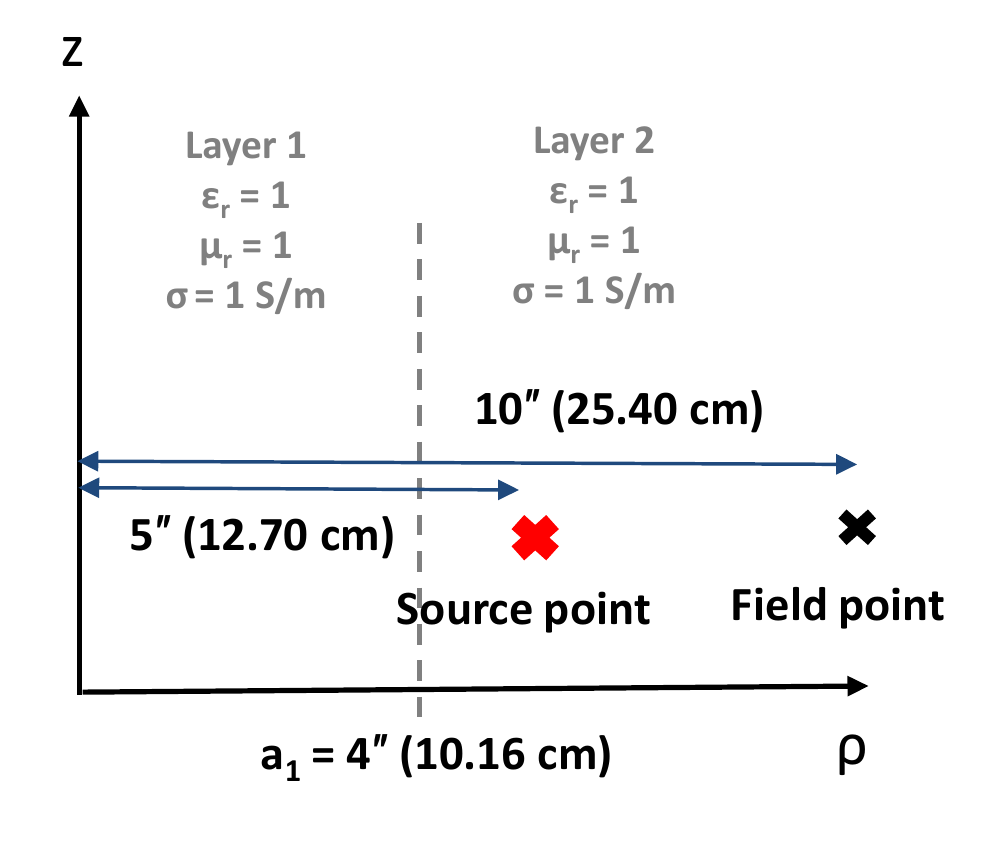}
  \caption{Two cylindrical layers with the same medium properties and position parameters.}
  \label{S3.F.first.type}
\end{figure}

When $z \approx z'$, the relative errors from the DSIP and SIP are shown in Figures \ref{S3.F.DSIP.zz.30.4000} and \ref{S3.F.SIP.zz.30.4000}, where a much smaller spatial scale is used. In the limit $z \rightarrow z'$, $\delta_4$ diverges and numerical integration along the DSIP is not feasible. This is illustrated by the large error visible near the horizontal axis in Figure \ref{S3.F.DSIP.zz.30.4000}. On the other hand, as shown in Figure \ref{S3.F.SIP.zz.30.4000}, the SIP is applicable even for $z=z'$. To determine the threshold for choosing between the SIP or DSIP, we recall from Section \ref{sec.3.1} and \ref{sec.3.2} that $\delta_{3,SIP}$ is the dominant path segment for the SIP and $\delta_4$ for the DSIP. Hence, assuming equal sampling rates for similar accuracy, the smaller value between $\delta_{3,SIP}$ and $\delta_4$ can be adopted. In other words, when $z\approx z'$, one can use the SIP if $\delta_{3,SIP} < \delta_4$, and use the DSIP otherwise. To verify the validity of this choice, we consider the example provided in Figure \ref{S3.F.first.type}. The operating frequency is 36 kHz. The modal index $n$ ranges from $-30$ to $30$ and the number of integration points is the same. The two cylindrical layers have the same electromagnetic properties so that there are no reflections at the interface and only direct (primary) field terms are present. Table \ref{S3.T.first.type} compares SIP and DSIP results, showing that this criterion is indeed a good one.

\begin{table}[!htbp]
\begin{center}
\renewcommand{\arraystretch}{1.4}
\setlength{\tabcolsep}{7pt}
\caption{Comparison of accuracy between the SIP and DSIP for $z\approx z'$.}
    \begin{tabular}{ccccc}
        \hline
        $z-z'$& $\delta_{3,SIP}$ & Relative error of $H_\phi$ (SIP) & $\delta4$ &
            Relative error of $H_\phi$ (DSIP) \\
        \hline
            1 m & 617.6628 & 0.1393 & 50.6568 & 5.3080$\times10^{-3}$ \\
          0.1 m & 617.6628 & 3.8746$\times10^{-5}$ & 506.568 & 1.3774$\times10^{-4}$ \\
         0.01 m & 617.6628 & 1.6291$\times10^{-5}$ & 5065.68 & 4.4262$\times10^{-3}$ \\
        0.001 m & 617.6628 & 1.6113$\times10^{-5}$ & 50656.8 & 212.8221 \\
        \hline
    \end{tabular}
    \label{S3.T.first.type}
\end{center}
\end{table}

\subsubsection{Small transverse separation: $\rho\approx\rho'$}
\label{sec.3.3.2}

When $\rho\approx\rho'$, the azimuthal summation over $n$ has slow convergence. This stems from the behavior of the direct field terms. The magnitudes of each term of the azimuth series representing the direct field term decrease very slowly with the mode index $n$. It should be emphasized that a direct fields contribution is only present when the field point and source point are in the same layer. In other words, this scenario only happens in cases \eqref {S2.eq.Fn.case1} and \eqref{S2.eq.Fn.case2}. Table \ref{S3.T.nonconvergence} illustrates this. The numbers in Table \ref{S3.T.nonconvergence} are $|\Hnd(k_{j\rho}\rho)\Jn(k_{j\rho}\rho')|$, which corresponds to the $\rho$-component of direct field terms when the source is $z$-oriented. It is assumed that $k_{j\rho}=0.25147 +\iu 0.79122$ and $\rho'=0.1270$.
\begin{table}[!htbp]
\begin{center}
\renewcommand{\arraystretch}{1.2}
\setlength{\tabcolsep}{7pt}
\caption{Convergence behavior of direct field terms. Values in the table are $|\Hnd(k_{j\rho}\rho)\Jn(k_{j\rho}\rho')|$.}
    \begin{tabular}{ccccc}
        \hline
        order ($n$) & $\rho=\rho'$ & $\rho=1.1\rho'$ & $\rho=2.0\rho'$ & $\rho=5.0\rho'$ \\
        \hline
         0 & 5.9670 & 5.4113 & 2.8948 & 1.0064 \\
        10 & 3.0189 & 1.0581 & 1.4732$\times10^{-3}$ & 6.1529$\times10^{-8}$ \\
        20 & 3.0189 & 0.4079 & 1.4390$\times10^{-6}$ & 6.3148$\times10^{-15}$ \\
        30 & 3.0189 & 0.1573 & 1.4054$\times10^{-9}$ & 6.4717$\times10^{-22}$ \\
        50 & 3.0189 & 0.0234 & 1.3404$\times10^{-15}$ & 6.7907$\times10^{-36}$ \\
        \hline
    \end{tabular}
    \label{S3.T.nonconvergence}
\end{center}
\end{table}

When $\rho=\rho'$, the magnitude does not decrease at all as the mode number (order) increases. For $\rho=1.1\rho'$, the magnitude decreases but only very slowly. Mathematically, the non-convergent behavior of this series can be examined using, for example, the small argument approximations in \eqref{S2.eq.small.approx.Jn} -- \eqref{S2.eq.small.approx.Hnd}. The possible direct field terms for Case 1 (refer to \eqref{S2.eq.Fn.case1.orig}) then become
\begin{subequations}
\begin{flalign}
\Hn(k_{j\rho}\rho)\Jn(k_{j\rho}\rho')&=
    G_i^{-1}\rho^{-n}\left(-\frac{\iu}{n\pi}\right)\cdot G_i(\rho')^n=
    -\frac{\iu}{n\pi}\left(\frac{\rho'}{\rho}\right)^n, \label{S3.eq.direct.HnJn}\\
\Hnd(k_{j\rho}\rho)\Jn(k_{j\rho}\rho')&=
    G_i^{-1}\rho^{-n}\left(\frac{\iu}{\pi k_{j\rho}\rho}\right)\cdot G_i(\rho')^n=
    \frac{\iu}{\pi k_{j\rho}\rho}\left(\frac{\rho'}{\rho}\right)^n, \label{S3.eq.direct.HndJn}\\
\Hn(k_{j\rho}\rho)\Jnd(k_{j\rho}\rho')&=
    G_i^{-1}\rho^{-n}\left(-\frac{\iu}{n\pi}\right)\cdot G_i(\rho')^n\left(\frac{n}{k_{j\rho}\rho'}\right)=
    -\frac{\iu}{\pi k_{j\rho}\rho'}\left(\frac{\rho'}{\rho}\right)^n, \label{S3.eq.direct.HnJnd}\\
\Hnd(k_{j\rho}\rho)\Jnd(k_{j\rho}\rho')&=
    G_i^{-1}\rho^{-n}\left(\frac{\iu}{\pi k_{j\rho}\rho}\right)\cdot
    G_i(\rho')^n\left(\frac{n}{k_{j\rho}\rho'}\right)=
    \frac{\iu n}{\pi k^2_{j\rho}\rho\rho'}\left(\frac{\rho'}{\rho}\right)^n, \label{S3.eq.direct.HndJnd}
\end{flalign}
\end{subequations}
which shows degraded convergence as $\rho \rightarrow \rho'$. 
Note that \eqref{S3.eq.direct.HnJn} refers to the $z$-component produced by the $z$-directed source but similar conclusions can be made in other cases. Therefore, in this scenario, the direct field (which can be evaluated analytically in a closed-form)
should be subtracted from the integrand for accurate computations. For example, 
\eqref{S2.eq.EHz.modified.integrand}
can be rewritten as
\begin{flalign}
    \begin{bmatrix}
    E_z \\ H_z
    \end{bmatrix}
=\frac{\iu Il}{4\pi\omega\epsilon_j}\intmp dk_z e^{\iu k_z(z-z')}
\left[
	\suma e^{\iu n(\phi-\phi')}\left\{\Fn-\bF^o_n(\rho,\rho')\right\}\cdot\Dj
\right] +
    \begin{bmatrix}
    E^o_z \\ H^o_z
    \end{bmatrix},
    \label{S3.eq.EHz.modified}
\end{flalign}
where the superscript $o$ in $\bF^o_n(\rho,\rho')$ and in the the last term indicates the direct field term. $\bF^o_n(\rho,\rho')$ is taken as $\Fn$ with all generalized reflection coefficients set to zero. The last terms can be writen in a closed-form. To determine whether direct field subtraction should be applied when $\rho\approx\rho'$, we compare the the magnitude of the direct field contribution for the maximum mode considered with the one for the lowest mode, $n=0$. If the ratio of the two magnitudes are below $10^{-20}$, the direct field subtraction is not needed, otherwise we apply the subtraction.

\subsection{Further convergence study}
\label{sec.3.4}

\begin{figure}[!htbp]
 \centering
    \includegraphics[width=2.5in]{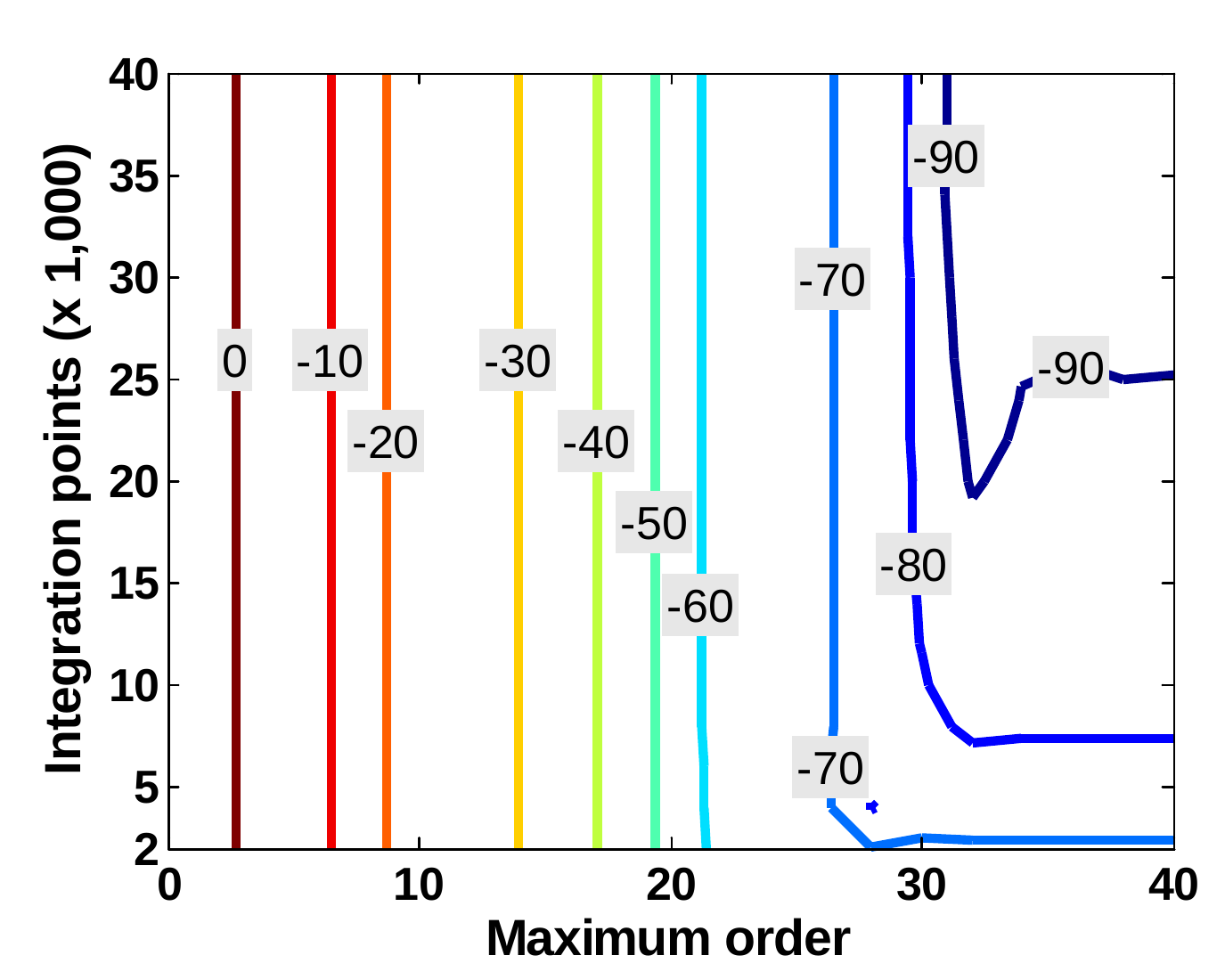}
  \caption{Relative error distribution using the DSIP in terms of various maximum orders and integration points when $\rho-\rho'=10$ cm, $\phi-\phi'=105^\circ$, and $z-z'=10$ cm. An operating frequency is 36 kHz and resistivity of the medium is 1 $\Omega\cdot m$.}
  \label{S3.F.orders.points.phi105}
\end{figure}

Figure \ref{S3.F.orders.points.phi105} shows the relative error (from DSIP integration) for the
$|H_{\phi}|$ component due to a azimuth-oriented magnetic dipole
as a function of the
number of azimuth modes (orders) and quadrature points, for an example with $\rho-\rho'=10$ cm, $\phi-\phi'=105^\circ$, and $z-z'=10$ cm. The operating frequency is 36 kHz and resistivity of the medium is 1 $\Omega\cdot m$.
This result shows good convergence as both these parameters increase. Note that in this case a large number of integration points ($> 10,000$) becomes necessary only if an error level below $-70$ $dB$ is required, with a maximum azimuth order beyond about 25 being used.

\begin{figure}[!htbp]
	\centering
	\subfloat[\label{S3.F.orders.points.f1R1}]{%
      \includegraphics[width=2.7in]{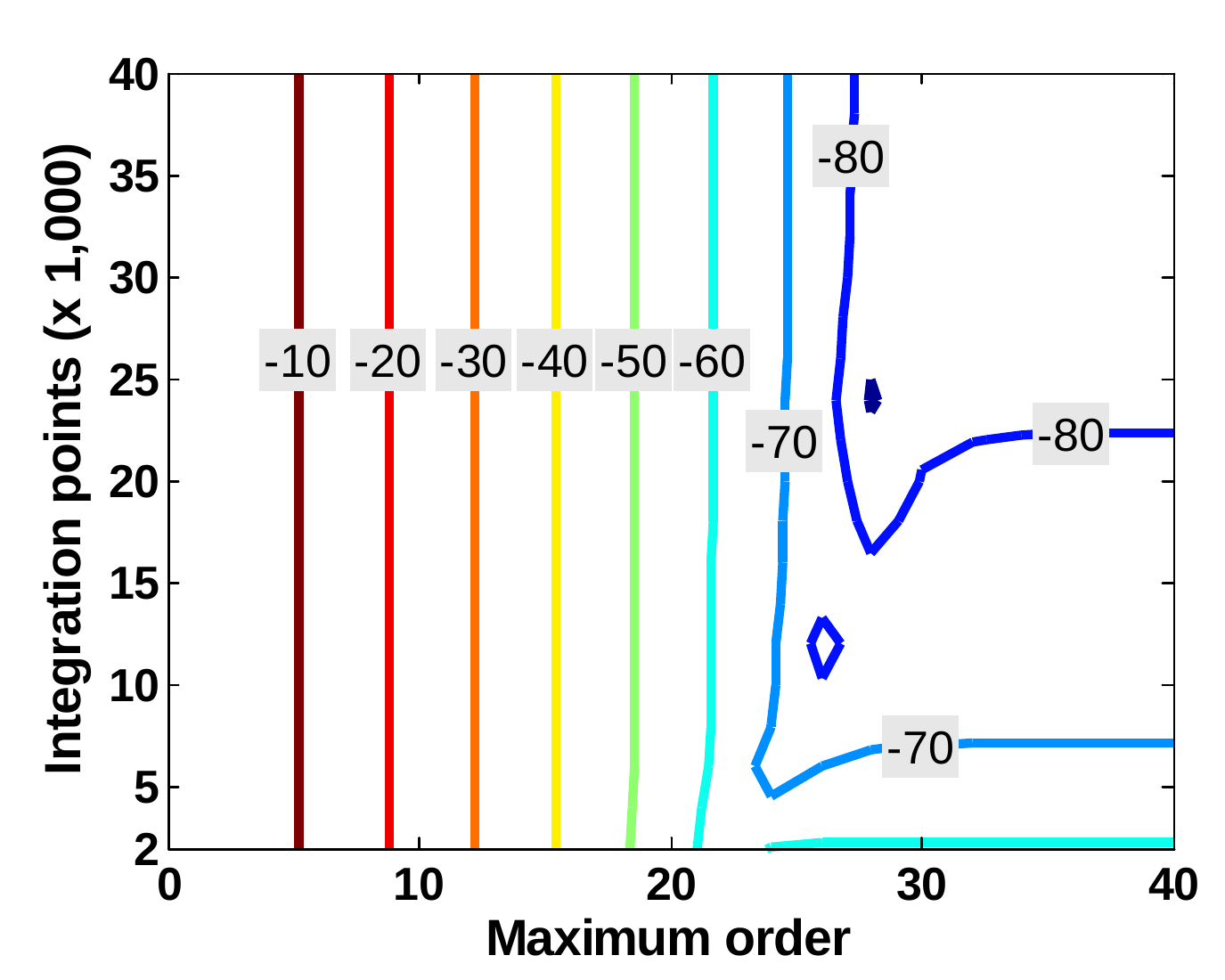}
    }
    \hfill
    \subfloat[\label{S3.F.orders.points.f1R2}]{%
      \includegraphics[width=2.7in]{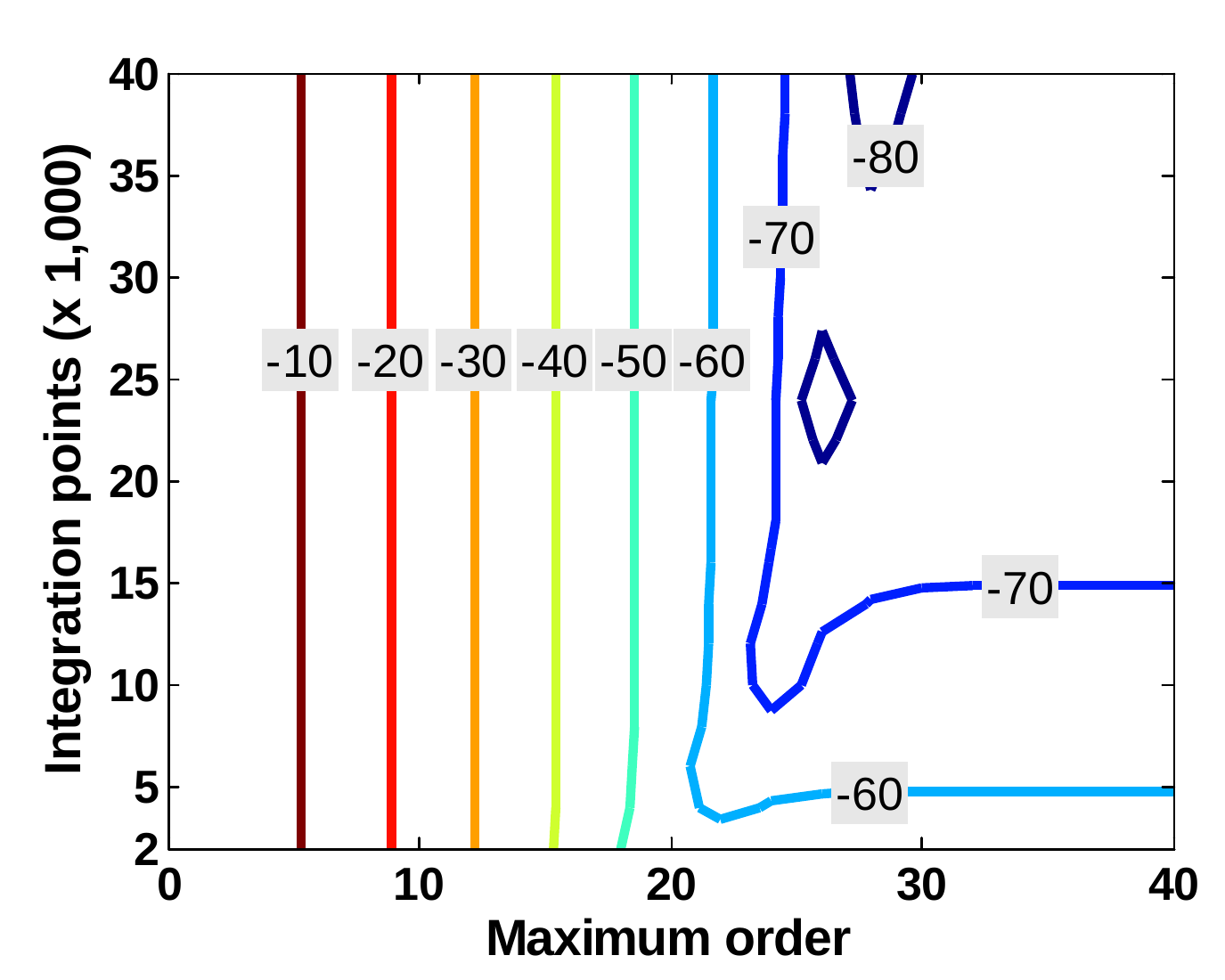}
    }
	\hfill
    \subfloat[\label{S3.F.orders.points.f1R3}]{%
      \includegraphics[width=2.7in]{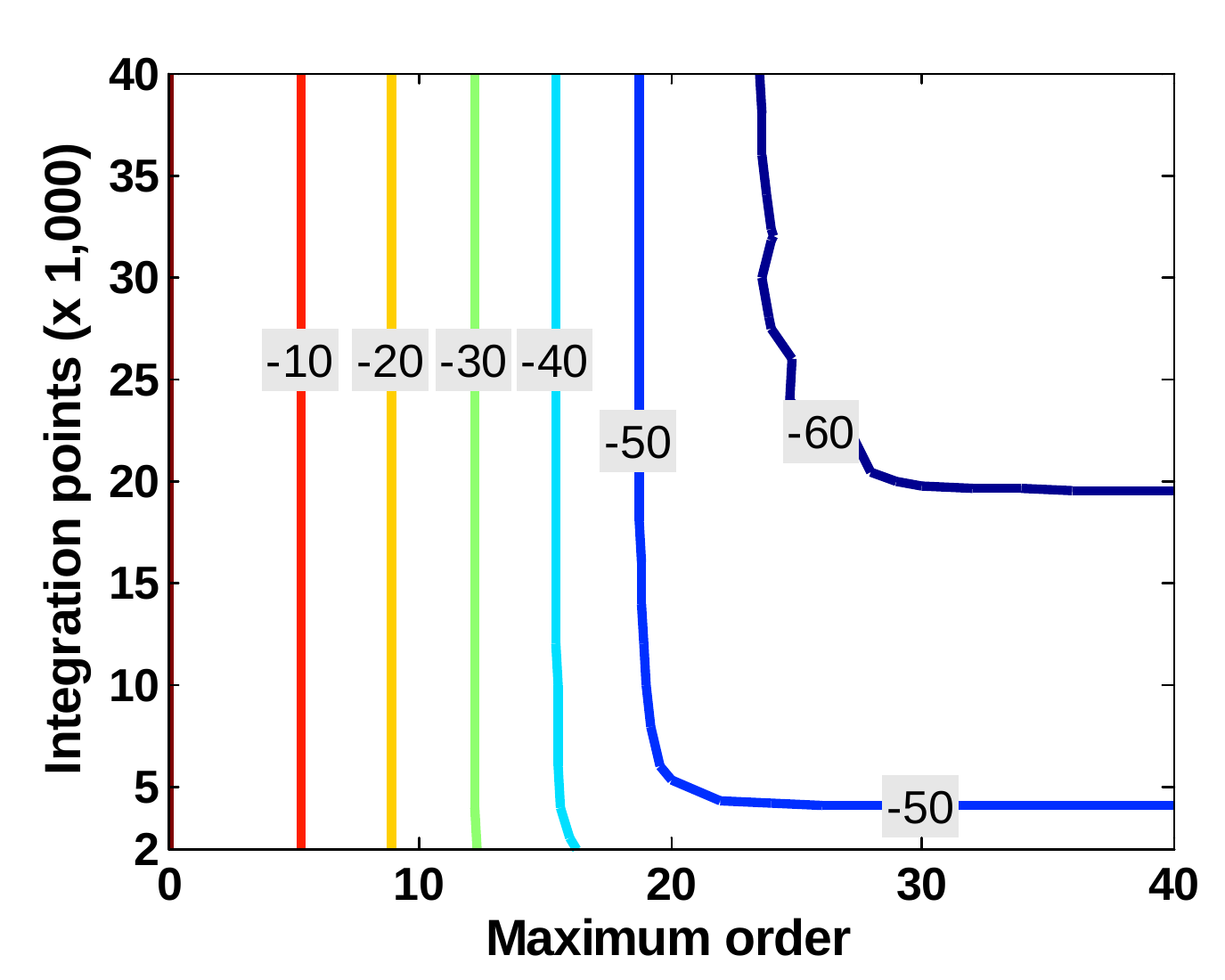}
    }
    \caption{Relative error distributions using the DSIP in terms of various maximum orders and integration points when $\rho-\rho'=10$ cm, $\phi-\phi'=0^\circ$, and $z-z'=10$ cm. An operating frequency is 0.01 Hz: (a) resistivity of the medium ($R_m$) is $10^{-8}$ $\Omega\cdot m$, (b) $R_m$ = 1 $\Omega\cdot m$, and (c) $R_m$ = $10^{3}$ $\Omega\cdot m$.}
    \label{S3.F.orders.points.f1}
\end{figure}

\begin{figure}[!htbp]
	\centering
	\subfloat[\label{S3.F.orders.points.f2R1}]{%
      \includegraphics[width=2.7in]{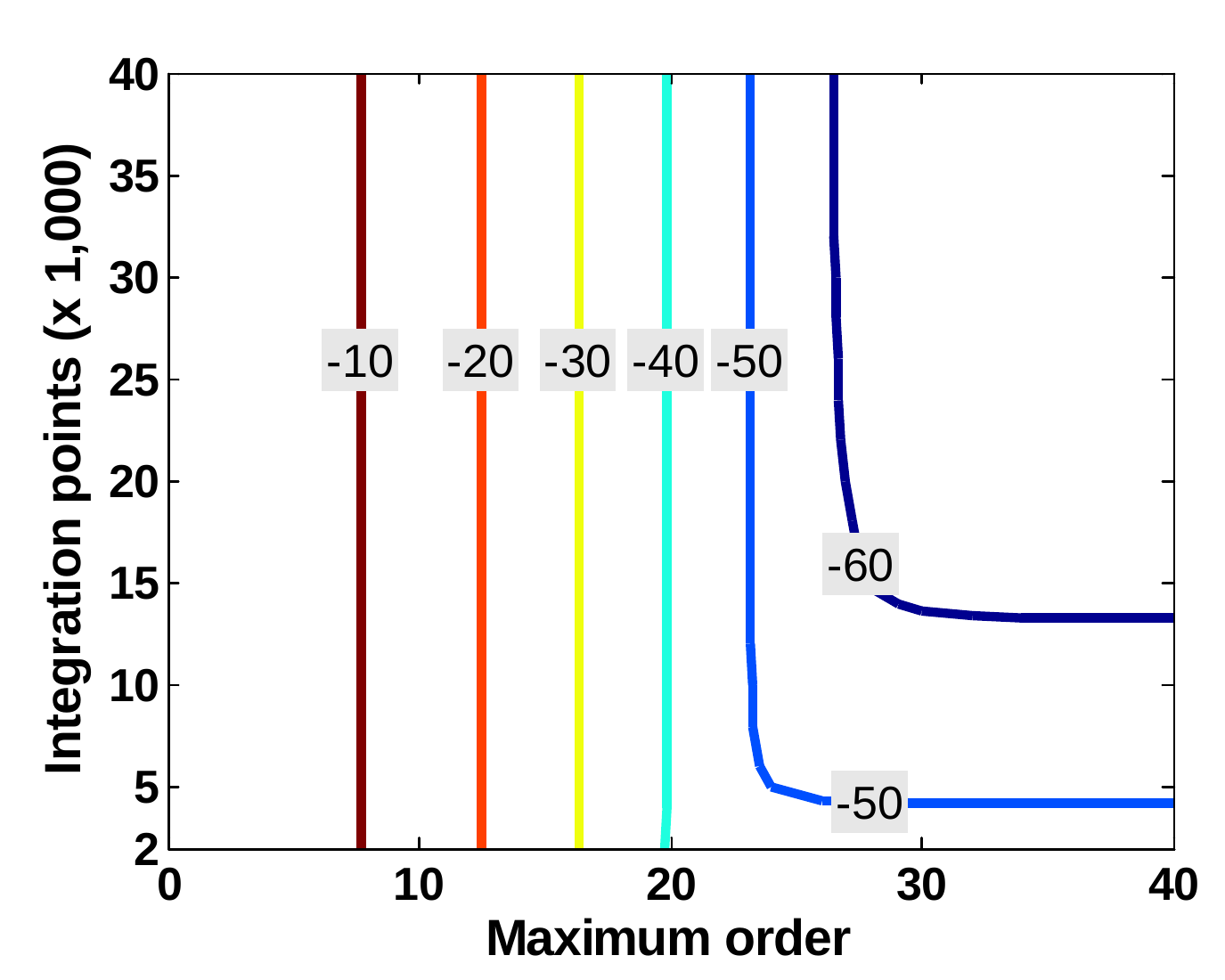}
    }
    \hfill
    \subfloat[\label{S3.F.orders.points.f2R2}]{%
      \includegraphics[width=2.7in]{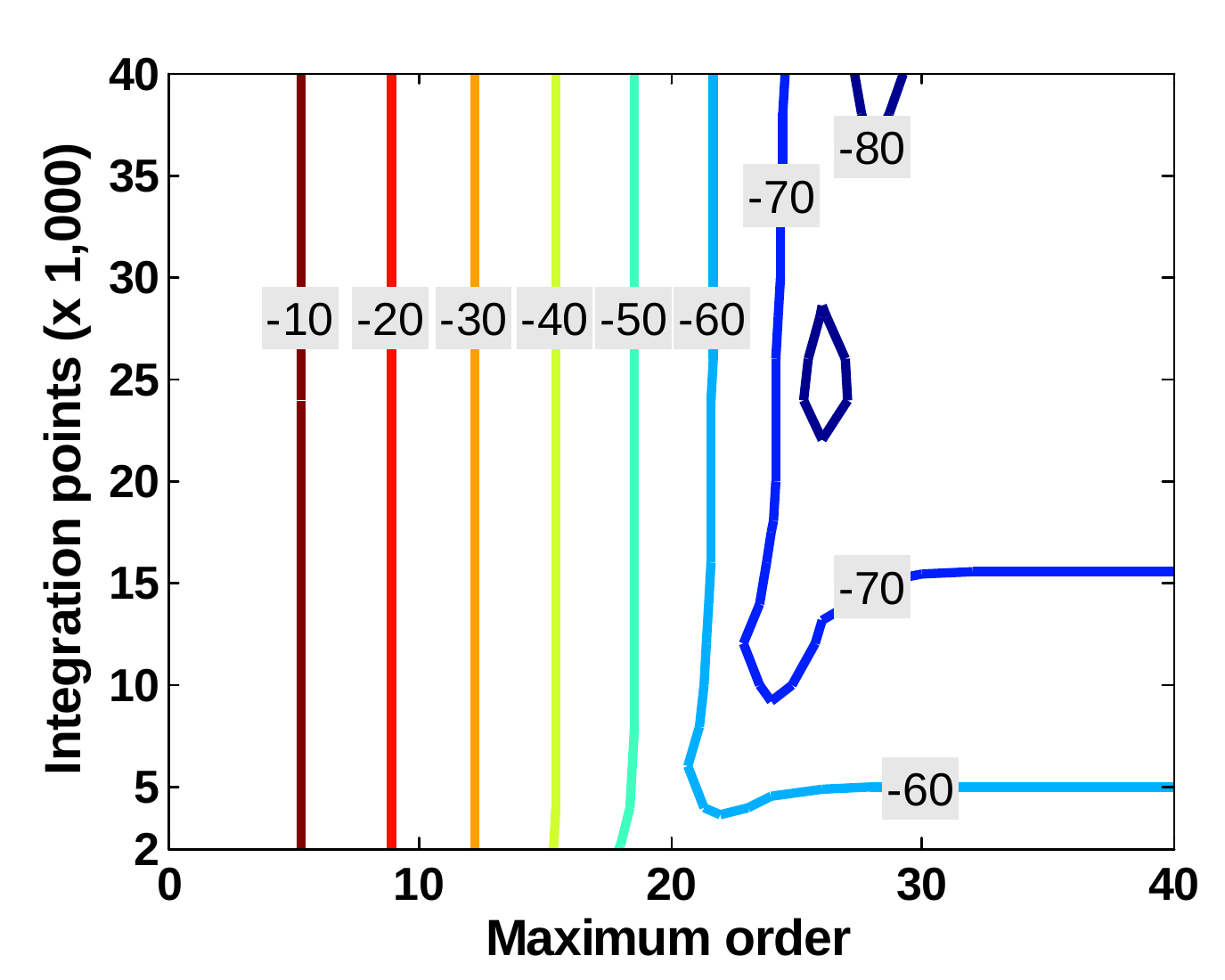}
    }
	\hfill
    \subfloat[\label{S3.F.orders.points.f2R3}]{%
      \includegraphics[width=2.7in]{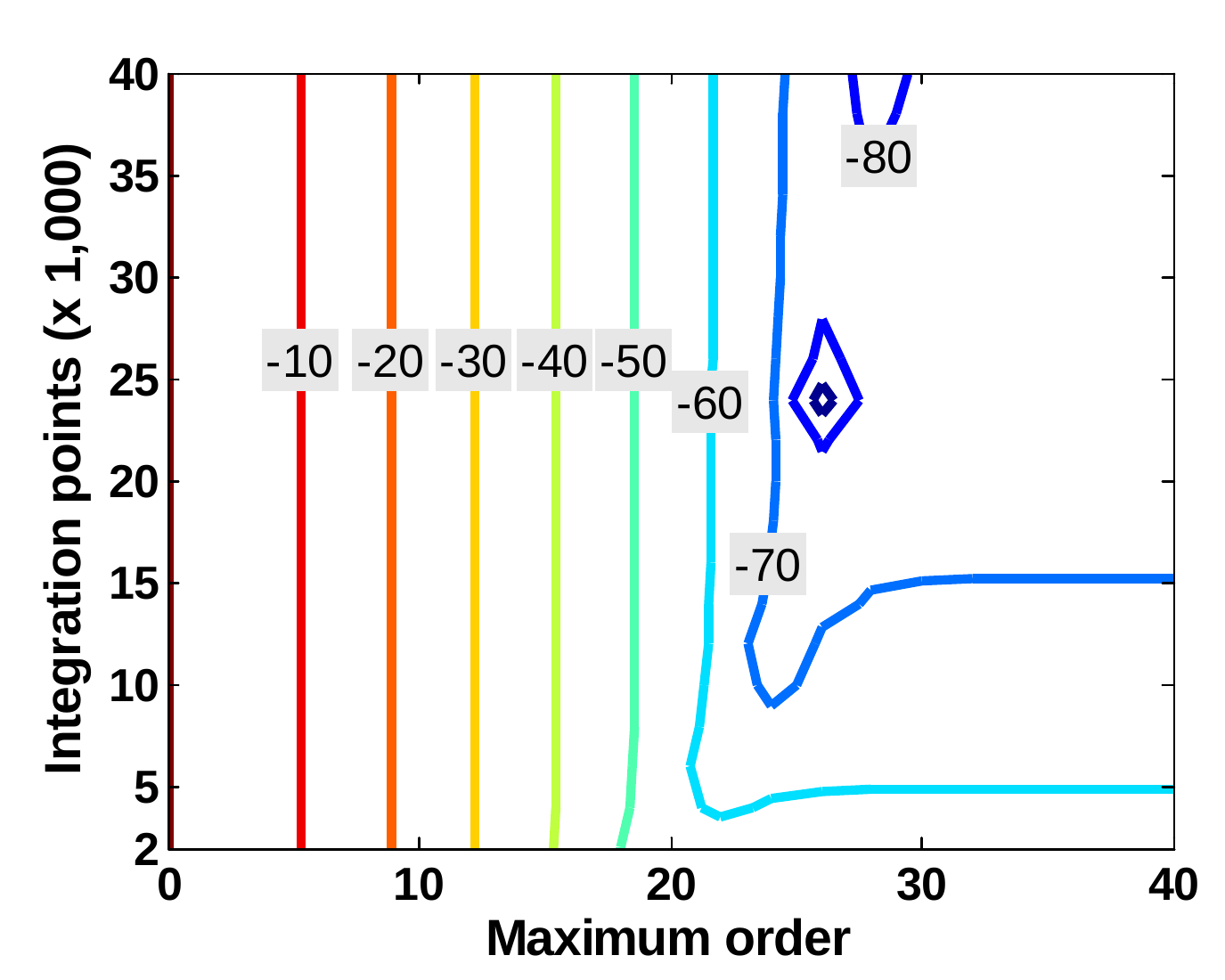}
    }
    \caption{Relative error distributions using the DSIP in terms of various maximum orders and integration points when $\rho-\rho'=10$ cm, $\phi-\phi'=0^\circ$, and $z-z'=10$ cm. An operating frequency is 10 kHz: (a) resistivity of the medium ($R_m$) is $10^{-5}$ $\Omega\cdot m$, (b) $R_m$ = 1 $\Omega\cdot m$, and (c) $R_m$ = $10^{3}$ $\Omega\cdot m$.}
    \label{S3.F.orders.points.f2}
\end{figure}

\begin{figure}[!htbp]
	\centering
	\subfloat[\label{S3.F.orders.points.f3R1}]{%
      \includegraphics[width=2.7in]{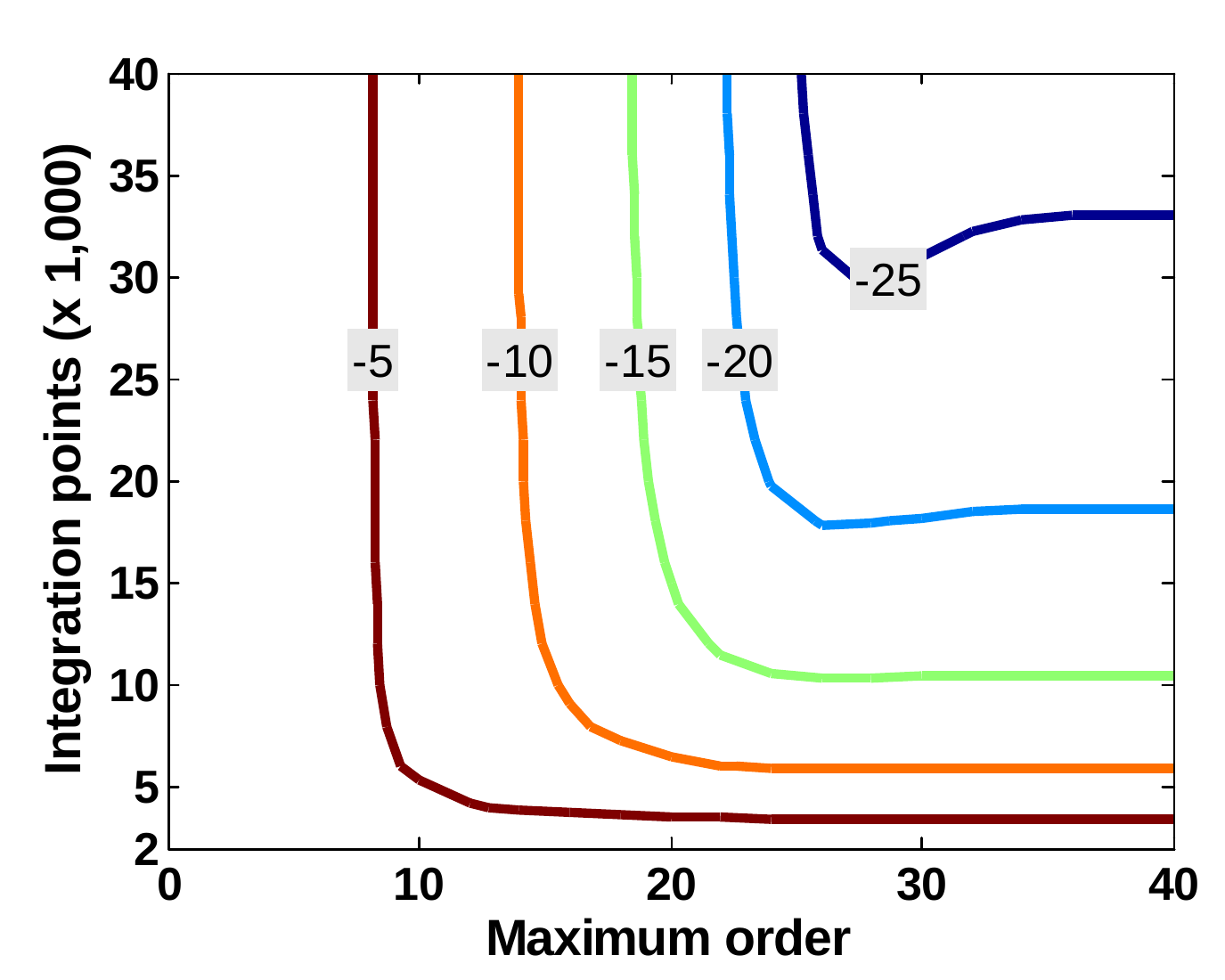}
    }
    \hfill
    \subfloat[\label{S3.F.orders.points.f3R2}]{%
      \includegraphics[width=2.7in]{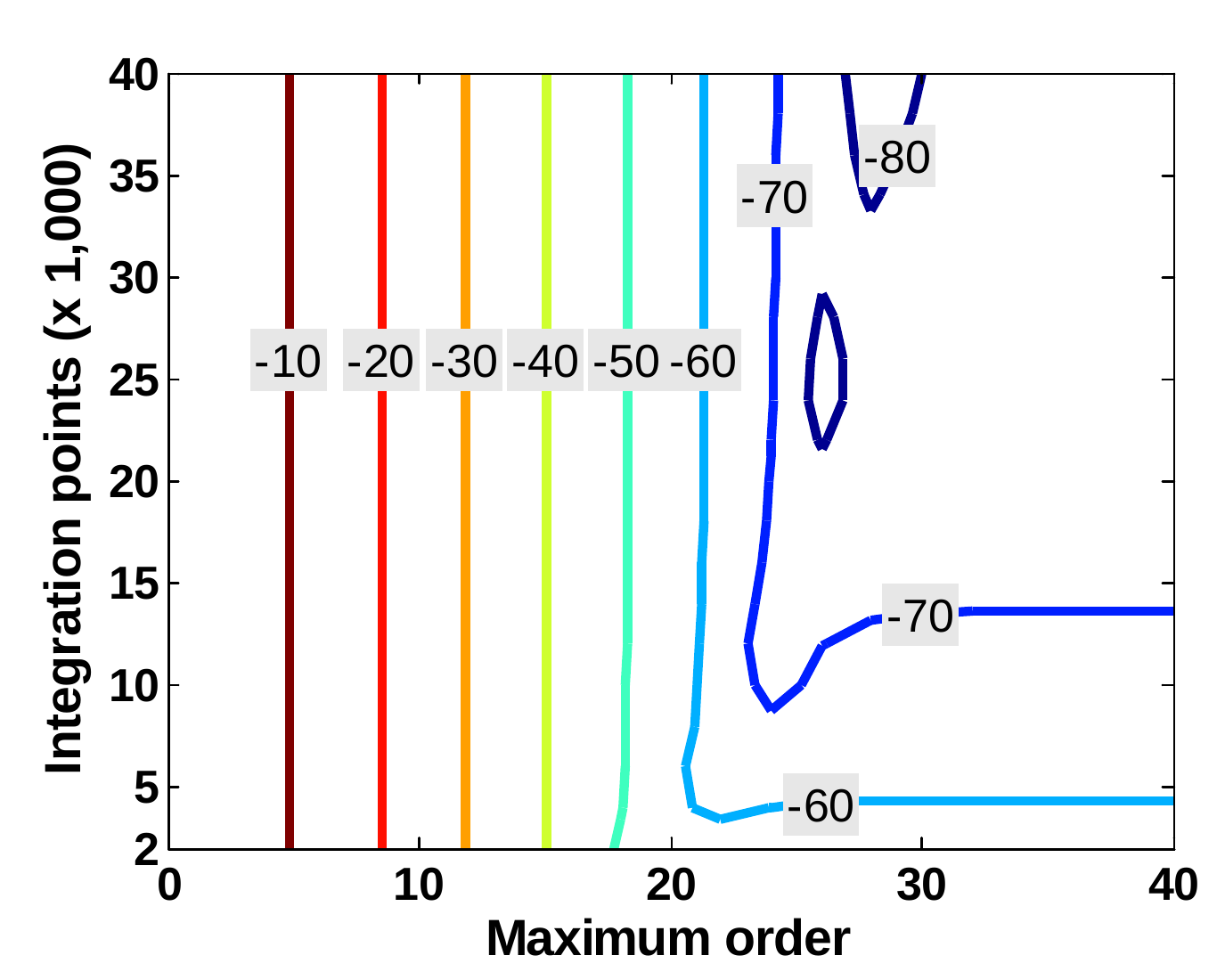}
    }
	\hfill
    \subfloat[\label{S3.F.orders.points.f3R3}]{%
      \includegraphics[width=2.7in]{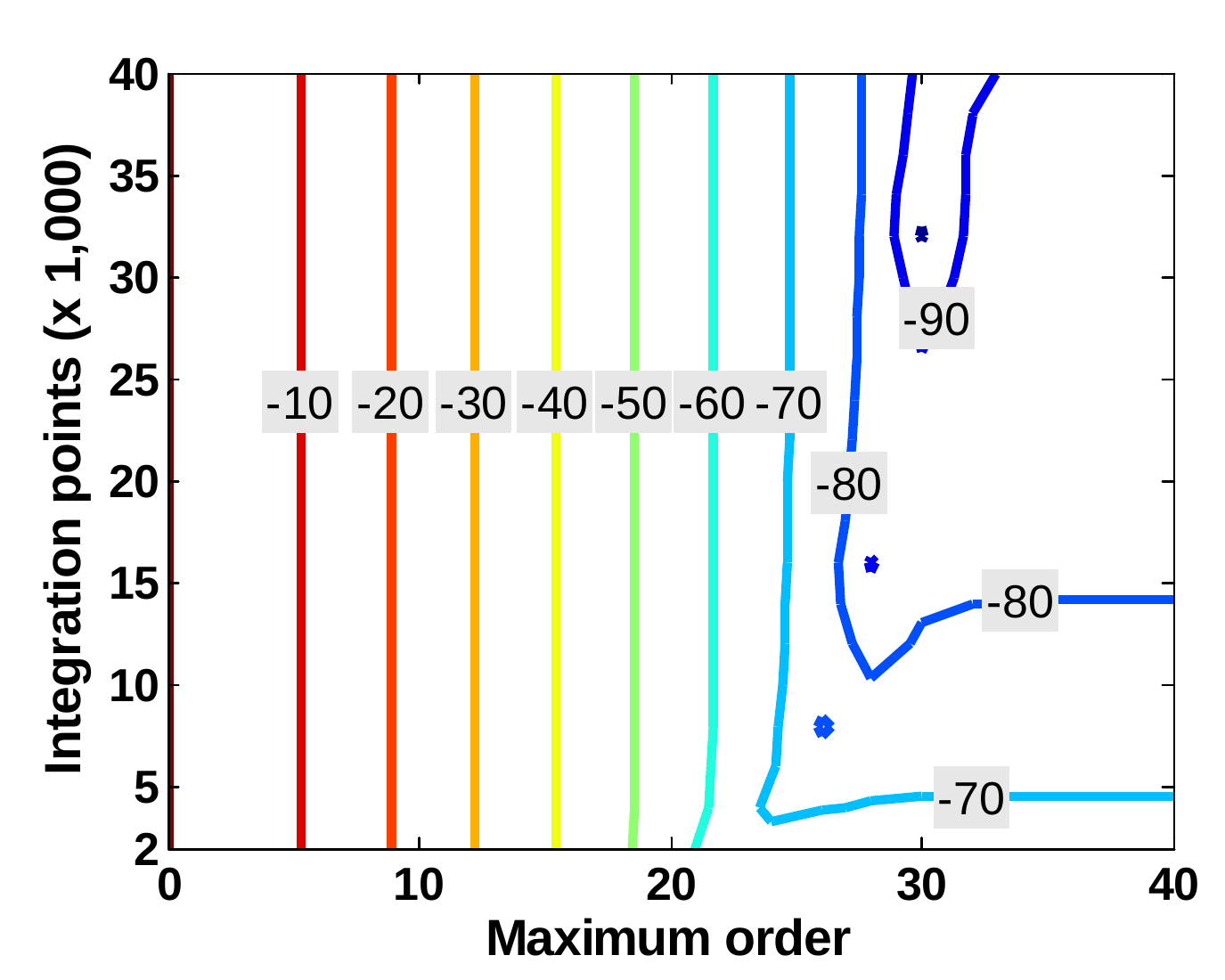}
    }
    \caption{Relative error distributions using the DSIP in terms of various maximum orders and integration points when $\rho-\rho'=10$ cm, $\phi-\phi'=0^\circ$, and $z-z'=10$ cm. An operating frequency is 10 MHz: (a) resistivity of the medium ($R_m$) is $10^{-3}$ $\Omega\cdot m$, (b) $R_m$ = 1 $\Omega\cdot m$, and (c) $R_m$ = $10^{3}$ $\Omega\cdot m$.}
    \label{S3.F.orders.points.f3}
\end{figure}

In order to further illustrate the robustness of the algorithm, we next show similar plots for a wide range of frequencies and resistivities. All figures below assume $\rho-\rho'=10$ cm, $\phi-\phi'=0^\circ$, and $z-z'=10$ cm. The operating frequency for Figure \ref{S3.F.orders.points.f1R1}, \ref{S3.F.orders.points.f1R2}, and \ref{S3.F.orders.points.f1R3} is 0.01 Hz; for Figure \ref{S3.F.orders.points.f2R1}, \ref{S3.F.orders.points.f2R2}, and \ref{S3.F.orders.points.f2R3} is 10 kHz; and for Figure \ref{S3.F.orders.points.f3R1}, \ref{S3.F.orders.points.f3R2}, and \ref{S3.F.orders.points.f3R3} is 10 MHz. All these cases show good convergence. It should be noted that Figure \ref{S3.F.orders.points.f3R1} shows a larger relative error because the field value  itself is very small (a high-frequency field in a highly conductive medium have a small skin-depth with very fast exponential decay away from the source),  $|H_{\phi}| \approx 3.5864 \times10^{-10}$ $A/m$, whereas $|H_{\phi}|$ is greater than one for the other cases.

\pagebreak

\section{Results}
\label{sec.4.results}
This section provides results of some scenarios with practical significance in borehole geophysics.
In all cases, both relative permittivity $\epsilon_r$ and relative permeability $\mu_r$ are equal to one, whereas the resistivities can exhibit large variations among the layers. Both the transmitter and receiver are small coil antennas represented by magnetic dipoles with magnetic moments normalized to one. The results of the present algorithm are compared against results produced by the finite element method (FEM). For more details about the relevant FEM algorithms, refer to \cite{Pardo06:Simulation, Pardo06:Two}. The results below are produced using a double-precision code running on a PC with 2.6 GHz Opterons and 8 cores. The code is designed to automatically increase the number of integration points and azimuth modes until the relative error between two successive iterations is below some given threshold. The relative error is defined as
$|F_i-F_{i-1}|/|F_i|$,
where $F$ is the relevant electric or magnetic field component, and the subscript $i$ indicates the iteration number. In the following, an error threshold of $10^{-4}$ is chosen.

Table \ref{S4.T.cases.geometry} provides the geometrical and electromagnetic properties of all the cases considered. $R_i$ denotes the resistivity of layer $i$ expressed in [$\Omega\cdot m$], and $a_i$ represents the cylindrical interfaces expressed in inches [$^{\prime\prime}$], see also Figure \ref{S2.F.basic.geometry}. Table \ref{S4.T.cases.Tx.Rx.} shows the operating frequency and positions of the transmitter and receiver. The $\phi$-coordinates of the transmitter and receiver are set to coincide.
Table \ref{S4.T.cases.comparison} provides a comparison of the computed magnetic field component. As seen, all cases show a good agreement between the present algorithm and FEM. It should be noted that only Cases 1 and 2 (homogeneous media) have analytical results available for reference.

\begin{table}[!htbp]
\begin{center}
\renewcommand{\arraystretch}{1.2}
\setlength{\tabcolsep}{6pt}
\caption{Geometric and electromagnetic properties sorted by the number of layers. The resistivities $R_i$ are expressed in $\Omega\cdot m$, and the interface radii $a_i$ are expressed in inches [$^{\prime\prime}$]. For Case 6, Case 9, Case 10, refer to Figure \ref{S4.F.case6}, \ref{S4.F.case9}, \ref{S4.F.case10}, respectively.}
\begin{tabular}{ccccccccc}
    \hline
     & number of layers & $R_1$ & $R_2$ & $R_3$ & $R_4$ & $a_1$ & $a_2$ & $a_3$ \\
    \hline
	Case 1 & 2 & 1 & 1 & $-$ & $-$ &
		4.0$^{\prime\prime}$ & $-$ & $-$ \\
	Case 2 & 2 & 1 & 1 & $-$ & $-$ &
		4.0$^{\prime\prime}$ & $-$ & $-$ \\
	Case 3 & 2 & 1000 & 1 & $-$ & $-$ &
		4.0$^{\prime\prime}$ & $-$ & $-$ \\
	Case 4 & 2 & 2.7$\times10^{-8}$ & 1 & $-$ & $-$ &
		4.0$^{\prime\prime}$ & $-$ & $-$ \\
	Case 5 & 2 & 2.7$\times10^{-8}$ & 1 & $-$ & $-$ &
		4.0$^{\prime\prime}$ & $-$ & $-$ \\
	Case 6 & 3 & 1.0$\times10^{-5}$ & 1 & 5 & $-$ &
		4.0$^{\prime\prime}$ & 5.5$^{\prime\prime}$ & $-$ \\
	Case 7 & 3 & 1.0$\times10^{-5}$ & 1 & 1.0$\times10^{-5}$ & $-$ &
		4.0$^{\prime\prime}$ & 5.5$^{\prime\prime}$ & $-$ \\
	Case 8 & 3 & 1.0$\times10^{-5}$ & 1 & 5 & $-$ &
		4.0$^{\prime\prime}$ & 16.0$^{\prime\prime}$ & $-$ \\
	Case 9 & 3 & 1.0$\times10^{-5}$ & 1 & 5 & $-$ &
		4.0$^{\prime\prime}$ & 5.5$^{\prime\prime}$ & $-$ \\		
	Case 10 & 4 & 1.0$\times10^{-5}$ & 1 & 1.0$\times10^{-5}$ & 5 &
		4.0$^{\prime\prime}$ & 5.5$^{\prime\prime}$ & 5.625$^{\prime\prime}$ \\
	Case 11 & 4 & 1.0$\times10^{-5}$ & 1 & 1.0$\times10^{-5}$ & 5 &
		4.0$^{\prime\prime}$ & 5.5$^{\prime\prime}$ & 5.625$^{\prime\prime}$ \\
	Case 12 & 4 & 1.0$\times10^{-5}$ & 1 & 1.0$\times10^{-5}$ & 5 &
		4.0$^{\prime\prime}$ & 5.5$^{\prime\prime}$ & 5.625$^{\prime\prime}$ \\
	Case 13 & 4 & 1.0$\times10^{-5}$ & 1 & 1.0$\times10^{-5}$ & 5 &
		4.0$^{\prime\prime}$ & 5.5$^{\prime\prime}$ & 5.625$^{\prime\prime}$ \\
	Case 14 & 4 & 1.0$\times10^{-5}$ & 1 & 1.0$\times10^{-5}$ & 5 &
		4.0$^{\prime\prime}$ & 5.5$^{\prime\prime}$ & 5.625$^{\prime\prime}$ \\
	Case 15 & 4 & 1.0$\times10^{-5}$ & 1 & 1.0$\times10^{-5}$ & 5 &
		4.0$^{\prime\prime}$ & 5.5$^{\prime\prime}$ & 5.625$^{\prime\prime}$ \\
    \hline
\end{tabular}
\label{S4.T.cases.geometry}
\end{center}
\end{table}

\begin{table}[!htbp]
\begin{center}
\renewcommand{\arraystretch}{1.2}
\setlength{\tabcolsep}{6pt}
\caption{Transmitter and receiver positions and polarizations, as well as frequency of operation. For Case 6, Case 9, Case 10, refer to Figure \ref{S4.F.case6}, \ref{S4.F.case9}, \ref{S4.F.case10}, respectively.}
\begin{tabular}{cccccccc}
    \hline
     & \multirow{2}{*}{frequency} & \multicolumn{3}{c}{Transmitter} & \multicolumn{3}{c}{Receiver} \\
     & & \multicolumn{2}{c}{position ($\rho, z$)} & polarization &
		\multicolumn{2}{c}{position ($\rho', z'$)} & polarization \\
    \hline
	Case 1 & 36 kHz & 5$^{\prime\prime}$ & 0$^{\prime\prime}$ & $\phi$
		& 5$^{\prime\prime}$ & 16$^{\prime\prime}$ & $\phi$ \\
	Case 2 & 36 kHz & 5$^{\prime\prime}$ & 0$^{\prime\prime}$ & $z$
		& 5$^{\prime\prime}$ & 16$^{\prime\prime}$ & $z$ \\		
	Case 3 & 36 kHz & 5$^{\prime\prime}$ & 0$^{\prime\prime}$ & $\phi$
		& 5$^{\prime\prime}$ & 16$^{\prime\prime}$ & $\phi$ \\
	Case 4 & 36 kHz & 5$^{\prime\prime}$ & 0$^{\prime\prime}$ & $\phi$
		& 5$^{\prime\prime}$ & 16$^{\prime\prime}$ & $\phi$ \\
	Case 5 & 36 kHz & 5$^{\prime\prime}$ & 0$^{\prime\prime}$ & $z$
		& 5$^{\prime\prime}$ & 16$^{\prime\prime}$ & $z$ \\
	Case 6 & 36 kHz & 5$^{\prime\prime}$ & 0$^{\prime\prime}$ & $\phi$
		& 5$^{\prime\prime}$ & 16$^{\prime\prime}$ & $\phi$ \\
	Case 7 & 36 kHz & 5$^{\prime\prime}$ & 0$^{\prime\prime}$ & $\phi$
		& 5$^{\prime\prime}$ & 16$^{\prime\prime}$ & $\phi$ \\
	Case 8 & 36 kHz & 5$^{\prime\prime}$ & 0$^{\prime\prime}$ & $\phi$
		& 5$^{\prime\prime}$ & 16$^{\prime\prime}$ & $\phi$ \\
	Case 9 & 36 kHz & 6$^{\prime\prime}$ & 0$^{\prime\prime}$ & $\phi$
		& 6$^{\prime\prime}$ & 16$^{\prime\prime}$ & $\phi$ \\
	Case 10 & 36 kHz & 5$^{\prime\prime}$ & 0$^{\prime\prime}$ & $\phi$
		& 5$^{\prime\prime}$ & 16$^{\prime\prime}$ & $\phi$ \\
	Case 11 & 1 kHz & 5$^{\prime\prime}$ & 0$^{\prime\prime}$ & $\phi$
		& 5$^{\prime\prime}$ & 16$^{\prime\prime}$ & $\phi$ \\
	Case 12 & 125 kHz & 5$^{\prime\prime}$ & 0$^{\prime\prime}$ & $\phi$
		& 5$^{\prime\prime}$ & 16$^{\prime\prime}$ & $\phi$ \\
	Case 13 & 36 kHz & 5$^{\prime\prime}$ & 0$^{\prime\prime}$ & $z$
		& 5$^{\prime\prime}$ & 16$^{\prime\prime}$ & $\rho$ \\
	Case 14 & 36 kHz & 5$^{\prime\prime}$ & 0$^{\prime\prime}$ & $\phi$
		& 5$^{\prime\prime}$ & 4$^{\prime\prime}$ & $\phi$ \\
	Case 15 & 36 kHz & 5$^{\prime\prime}$ & 0$^{\prime\prime}$ & $\phi$
		& 5$^{\prime\prime}$ & 64$^{\prime\prime}$ & $\phi$ \\
    \hline
\end{tabular}
\label{S4.T.cases.Tx.Rx.}
\end{center}
\end{table}

\begin{figure}[!htbp]
	\centering
	\subfloat[\label{S4.F.case6}]{%
      \includegraphics[height=2.5in]{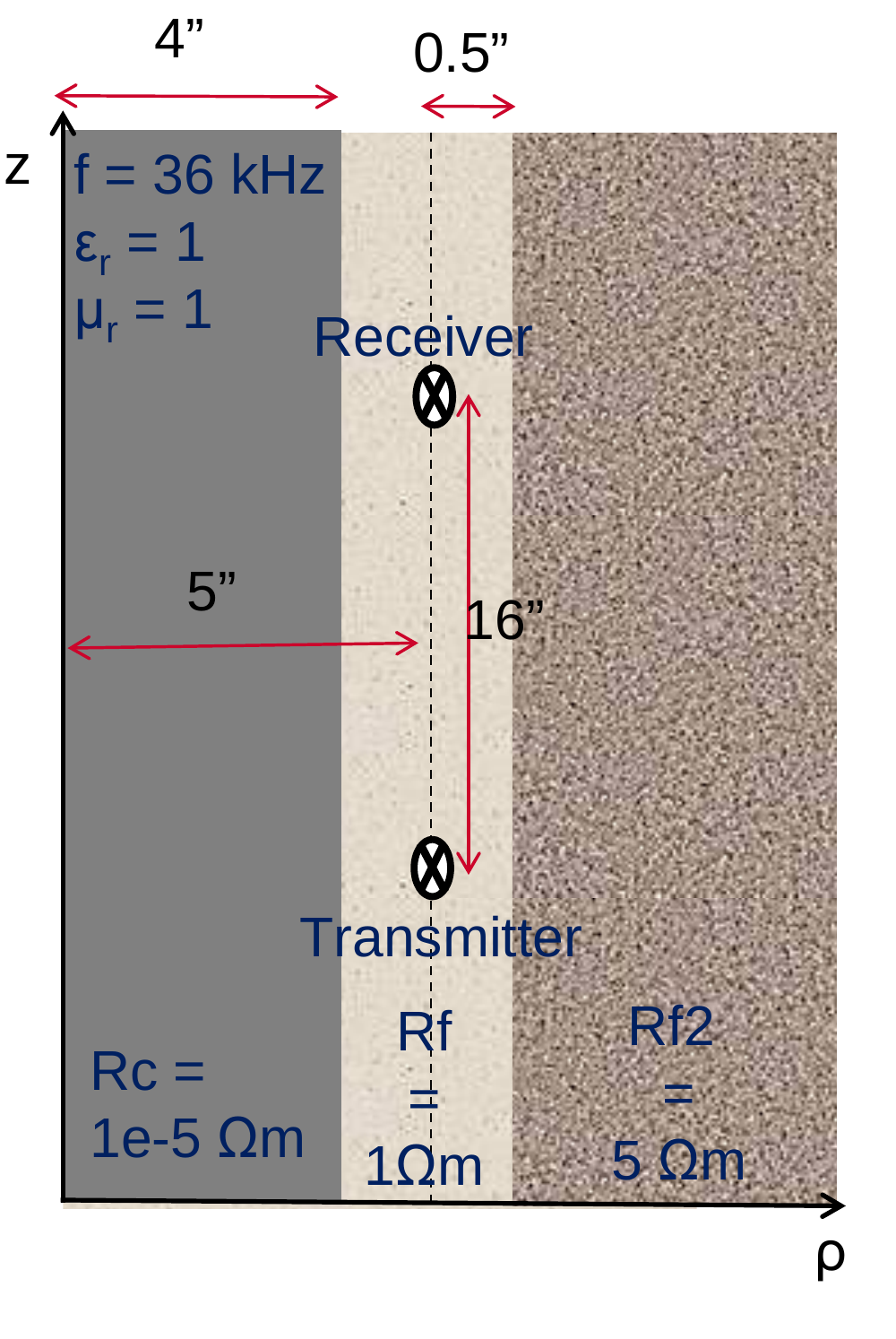}
    }
    \hfill
    \subfloat[\label{S4.F.case9}]{%
      \includegraphics[height=2.5in]{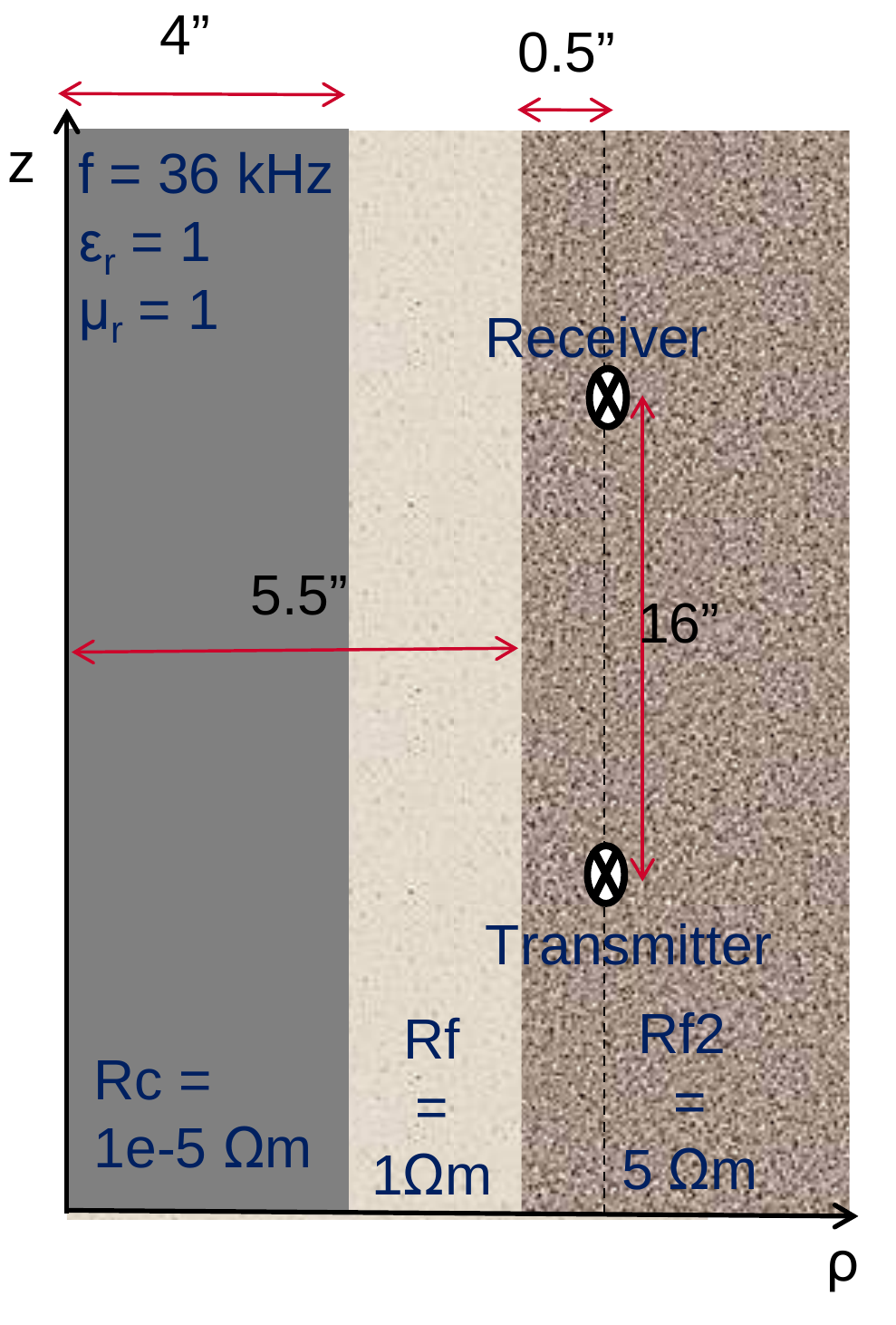}
    }
    \hfill
    \subfloat[\label{S4.F.case10}]{%
      \includegraphics[height=2.5in]{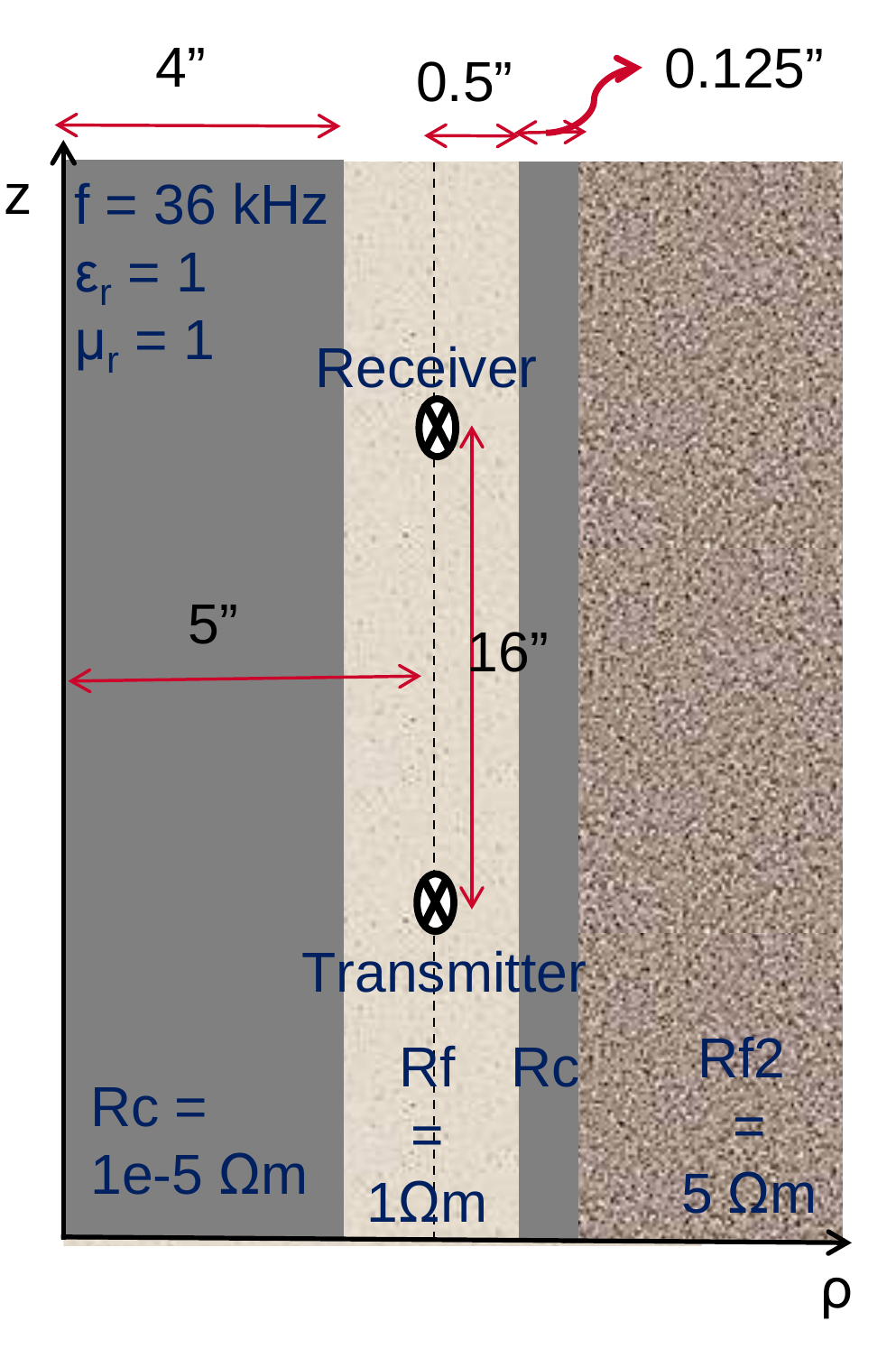}
    }
    \caption{Description of three selected cases in the $\rho z$-plane: (a) Case 6, (b) Case 9, and (c) Case 10.}
    \label{S3.F.delat3.case12}
\end{figure}

\begin{table}[!htbp]
\begin{center}
\renewcommand{\arraystretch}{1.2}
\setlength{\tabcolsep}{2pt}
\caption{Computed results. The designated component of a magnetic field in the second column is compared. There are analytical answers for Case 1 and Case 2 because the medium surrounding the transmitter is homogeneous. Field values are expressed in a phasor form with $e^{-\iu\omega t}$ convention.}
\begin{tabular}{cccccc}
    \hline
     & \multirow{2}{*}{component} & \multirow{2}{*}{analytical} & \multirow{2}{*}{FEM}
	 & \multirow{2}{*}{present algorithm} & \multirow{2}{*}{computing time} \\
     &  & & & &  \\
    \hline
	Case 1 & $H_\phi$ & 4.1884 $\angle$-91.0681$^\circ$ &
		4.0476 $\angle$-91.1087$^\circ$ & 4.1884 $\angle$-91.0681$^\circ$ & 4 sec.\\
	Case 2 & $H_z$ & 8.3259 $\angle$91.2105$^\circ$ &
		8.2662 $\angle$91.2178$^\circ$ & 8.3259 $\angle$91.2105$^\circ$ & 4 sec.\\
	Case 3 & $H_\phi$ & $-$ & 4.0473 $\angle$-91.2619$^\circ$ & 4.1881 $\angle$-91.2172$^\circ$ & 4 sec.\\
	Case 4 & $H_\phi$ &$-$& 12.1745 $\angle$-100.9810$^\circ$ & 12.4300 $\angle$-100.7265$^\circ$ & 17 sec.\\
	Case 5 & $H_z$ &$-$& 11.2415 $\angle$91.0181$^\circ$ & 11.3623 $\angle$90.9977$^\circ$ & 4 sec.\\
	Case 6 & $H_\phi$ &$-$& 10.3116 $\angle$-98.1898$^\circ$ & 10.5471 $\angle$-98.1354$^\circ$ & 31 sec.\\
	Case 7 & $H_\phi$ &$-$& 46.4257 $\angle$-114.2428$^\circ$ & 46.4265 $\angle$-114.2420$^\circ$ & 30 sec.\\
	Case 8 & $H_\phi$ &$-$& 10.7855 $\angle$-100.0572$^\circ$ & 10.7856 $\angle$-100.0573$^\circ$ & 29 sec.\\
	Case 9 & $H_\phi$ &$-$& 8.1258 $\angle$-97.0379$^\circ$ & 8.1326 $\angle$-97.0322$^\circ$ & 30 sec.\\
	Case 10 & $H_\phi$ &$-$& 46.6309 $\angle$-118.4329$^\circ$ & 46.6307 $\angle$-118.4320$^\circ$ & 41 sec.\\
	Case 11 & $H_\phi$ &$-$& 290.2144 $\angle$-127.4332$^\circ$ & 290.2652 $\angle$-127.4151$^\circ$ & 9 sec.\\
	Case 12 & $H_\phi$ &$-$& 18.7957 $\angle$-110.9753$^\circ$ & 18.8069 $\angle$-110.9191$^\circ$ & 36 sec.\\
	Case 13 & $H_\rho$ &$-$& 1.2589 $\angle$-67.6502$^\circ$ & 1.2589 $\angle$-67.5950$^\circ$ & 41 sec.\\
	Case 14 & $H_\phi$&$-$& 1007.5854 $\angle$-107.9553$^\circ$ & 1045.5344 $\angle$-107.8155$^\circ$ & 20 sec.\\
	Case 15 & $H_\phi$ &$-$& 12.1271 $\angle$-111.9689$^\circ$ & 12.1266 $\angle$-111.9684$^\circ$ & 40 sec.\\
    \hline
\end{tabular}
\label{S4.T.cases.comparison}
\end{center}
\end{table}

\pagebreak

The spatial field distributions of the magnetic field around the transmitter dipole in Cases 6, 10, and 11 are plotted in Figures 19, 20, and 21. Because of the large variation in magnitude, we use a decibel scale for these plots; i.e., we plot the distribution of $10\log_{10}|\mathbf{H}|$. In the following figures, thick black lines represent the layers' interfaces and thinner black lines are field contours. It should be noted that the third layer of Cases 10 and 11 is too thin to be discerned in such Figures. The variability of the skin effect on the field strength according to the frequency of operation is clearly visible. The field distribution seen in Case 10 is more confined within the source layer (second layer) than Case 6 due to the presence of an extra (outer) conductive layer (third layer).

\begin{figure}[!htbp]
	\centering
	\subfloat[\label{S4.F.case6.z0}]{%
      \includegraphics[height=2.5in]{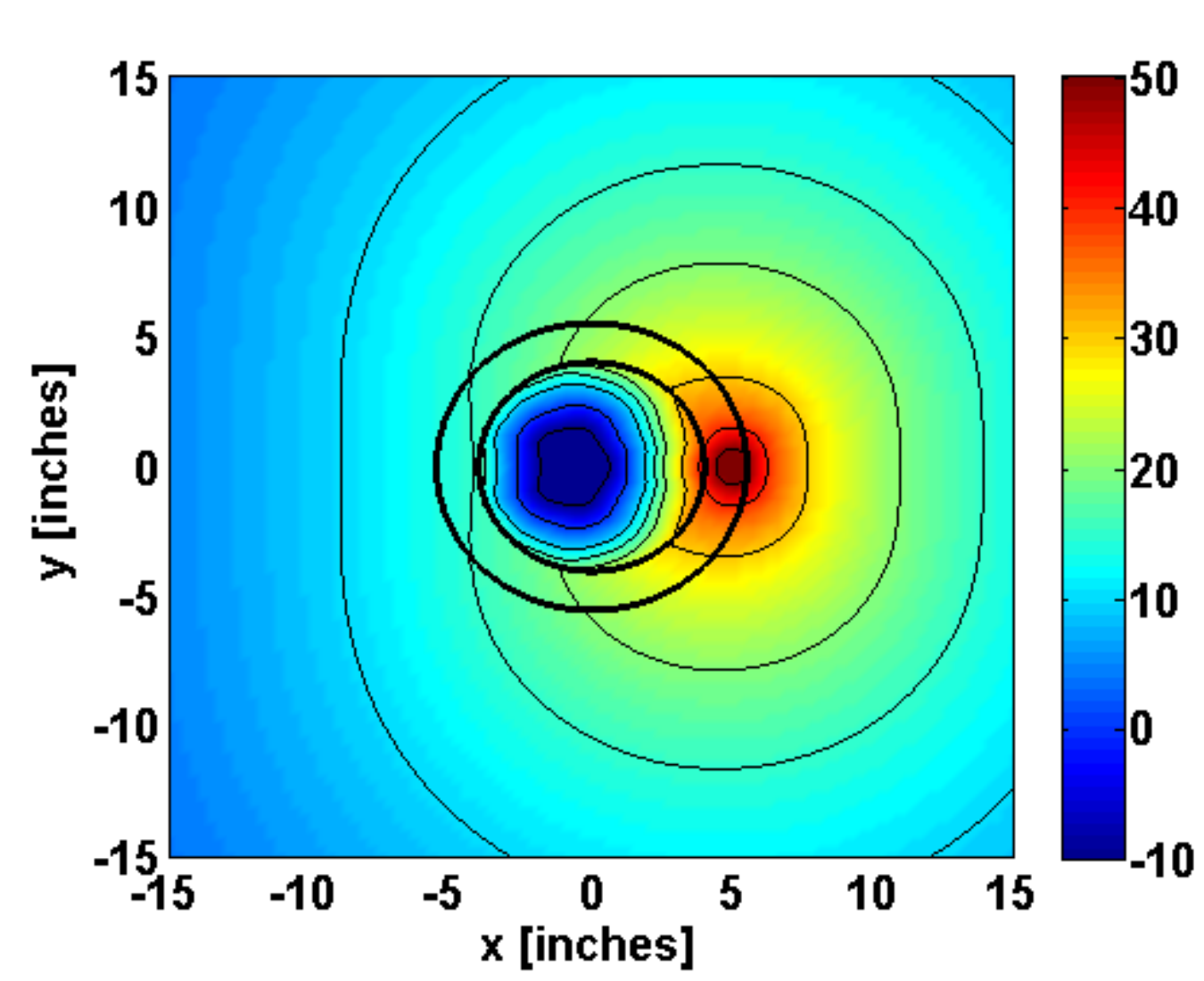}
    }
    \hfill
    \subfloat[\label{S4.F.case10.z0}]{%
      \includegraphics[height=2.5in]{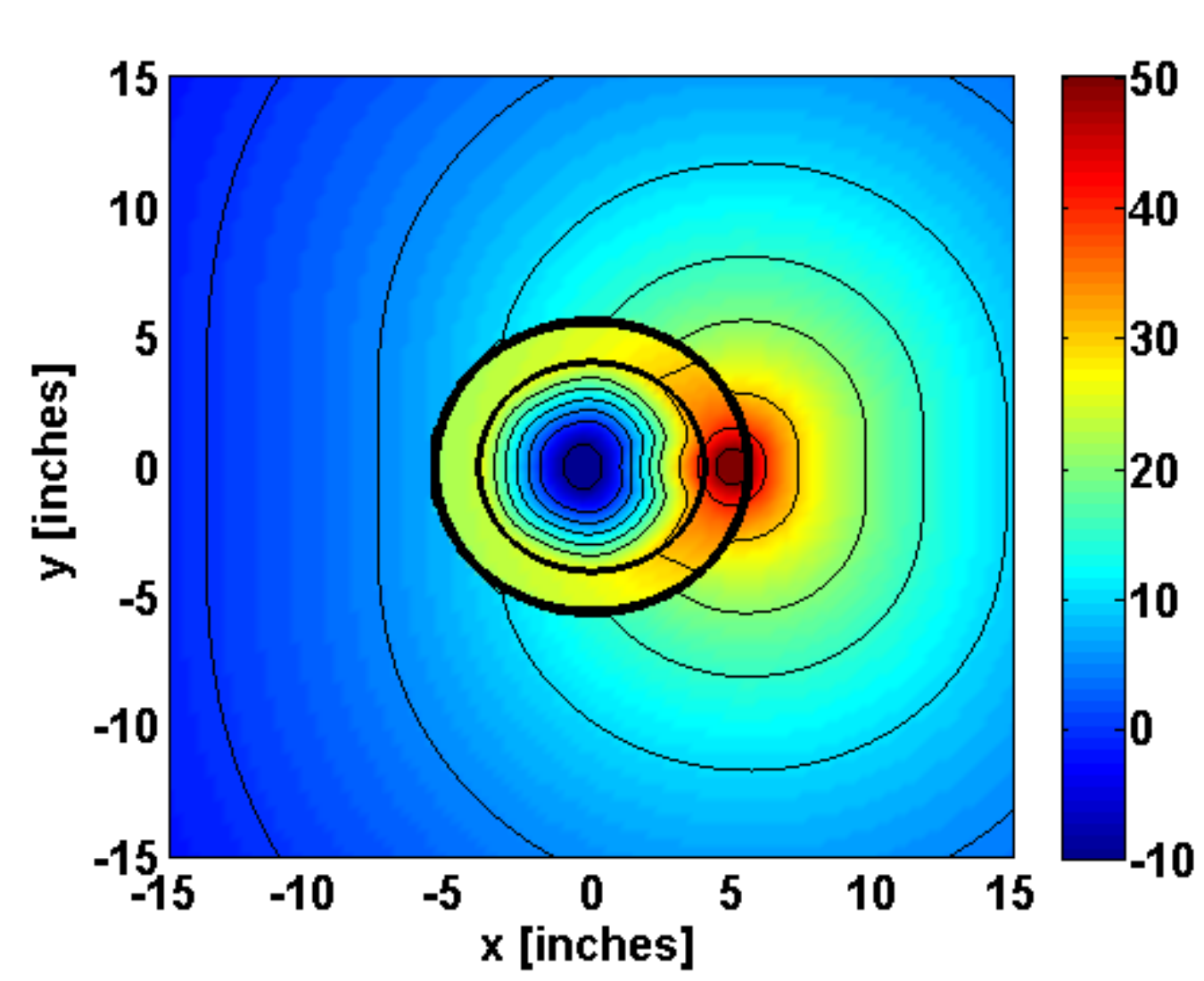}
    }
    \caption{Magnitude of the magnetic field at the $z=0^{\prime\prime}$ plane at 36 kHz: (a) Case 6 and (b) Case 10.}
    \label{S4.F.case6.and.10.z0}
\end{figure}

\begin{figure}[!htbp]
	\centering
	\subfloat[\label{S4.F.case6.y0}]{%
      \includegraphics[height=2.5in]{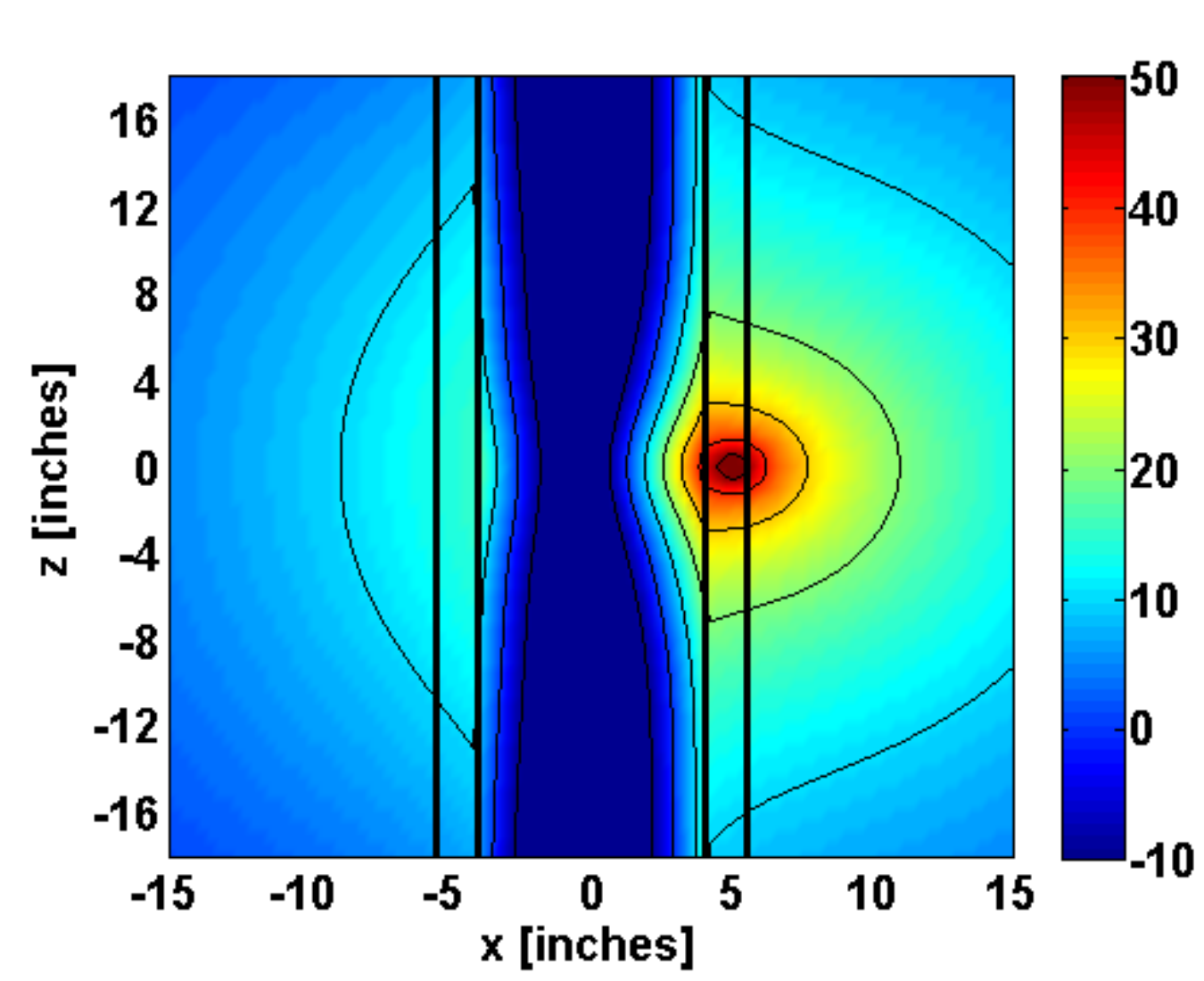}
    }
    \hfill
    \subfloat[\label{S4.F.case10.y0}]{%
      \includegraphics[height=2.5in]{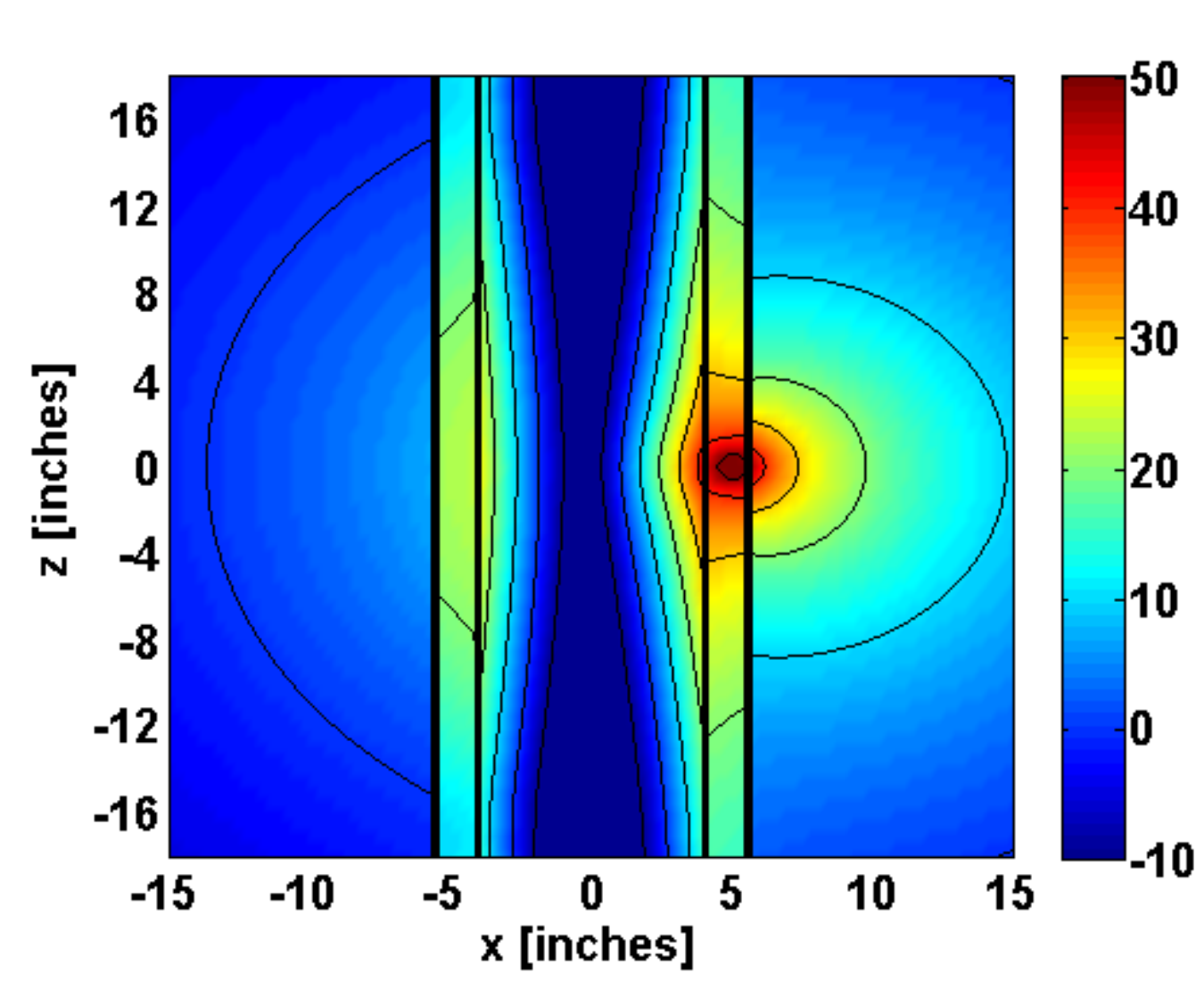}
    }
    \caption{Magnitude of the magnetic field at the $y=0^{\prime\prime}$ plane at 36 kHz: (a) Case 6 and (b) Case 10.}
    \label{S4.F.case6.and.10.y0}
\end{figure}

\begin{figure}[!htbp]
	\centering
	\subfloat[\label{S4.F.case10.3D.1kHz}]{%
      \includegraphics[height=2.5in]{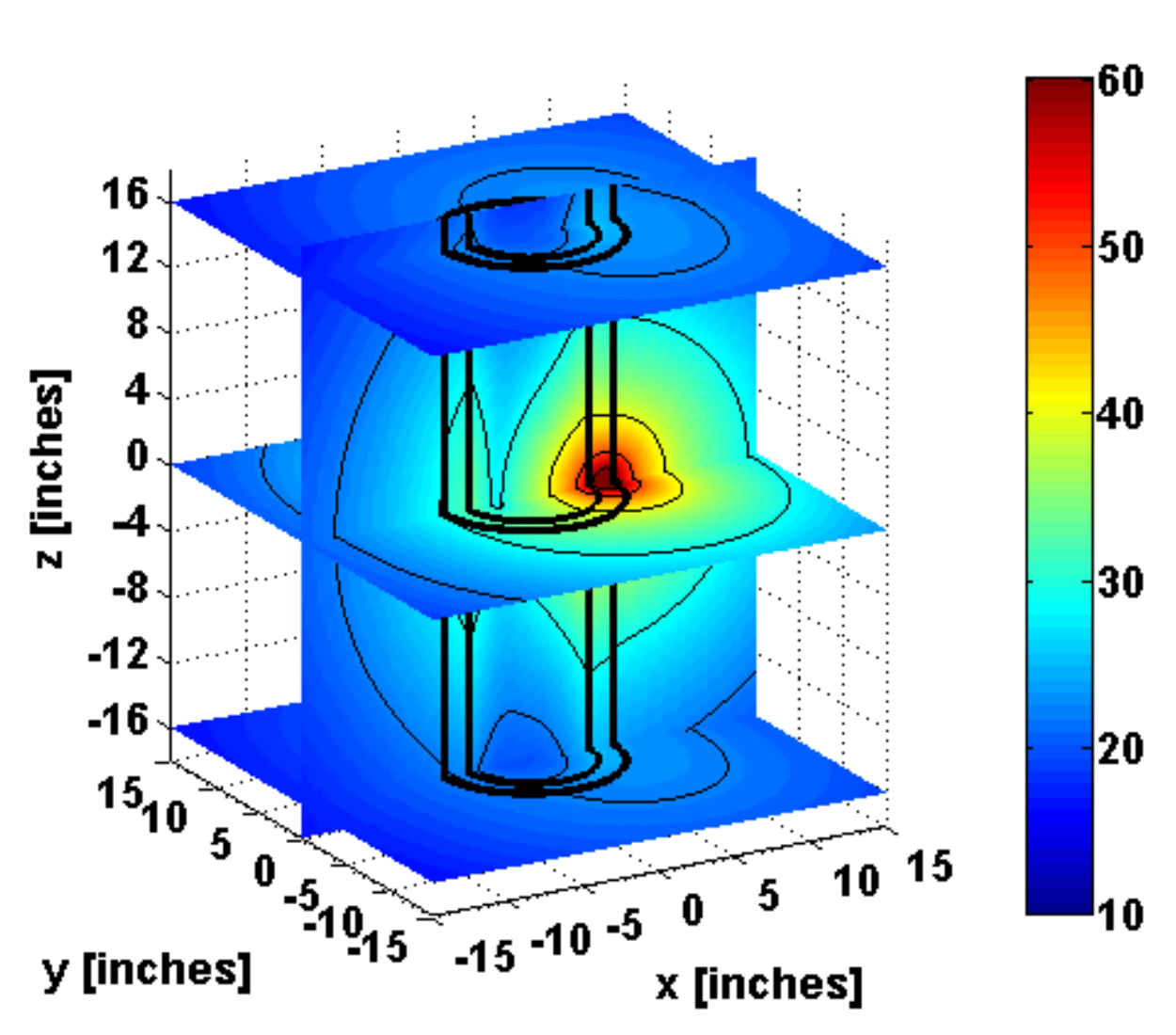}
    }
    \hfill
    \subfloat[\label{S4.F.case10.y0.1kHz}]{%
      \includegraphics[height=2.5in]{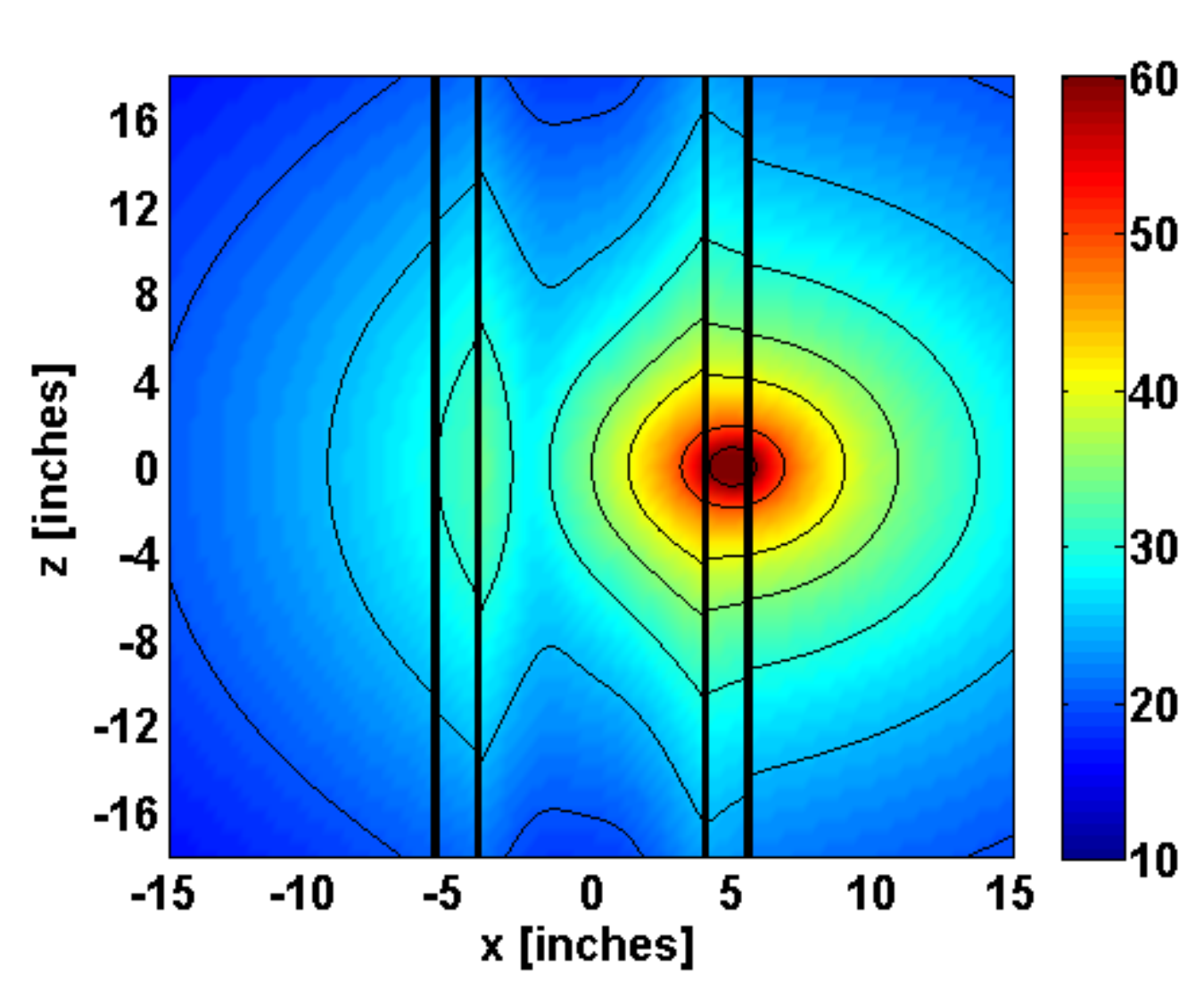}
    }
	
	\subfloat[\label{S4.F.case10.z16.1kHz}]{%
      \includegraphics[height=2.5in]{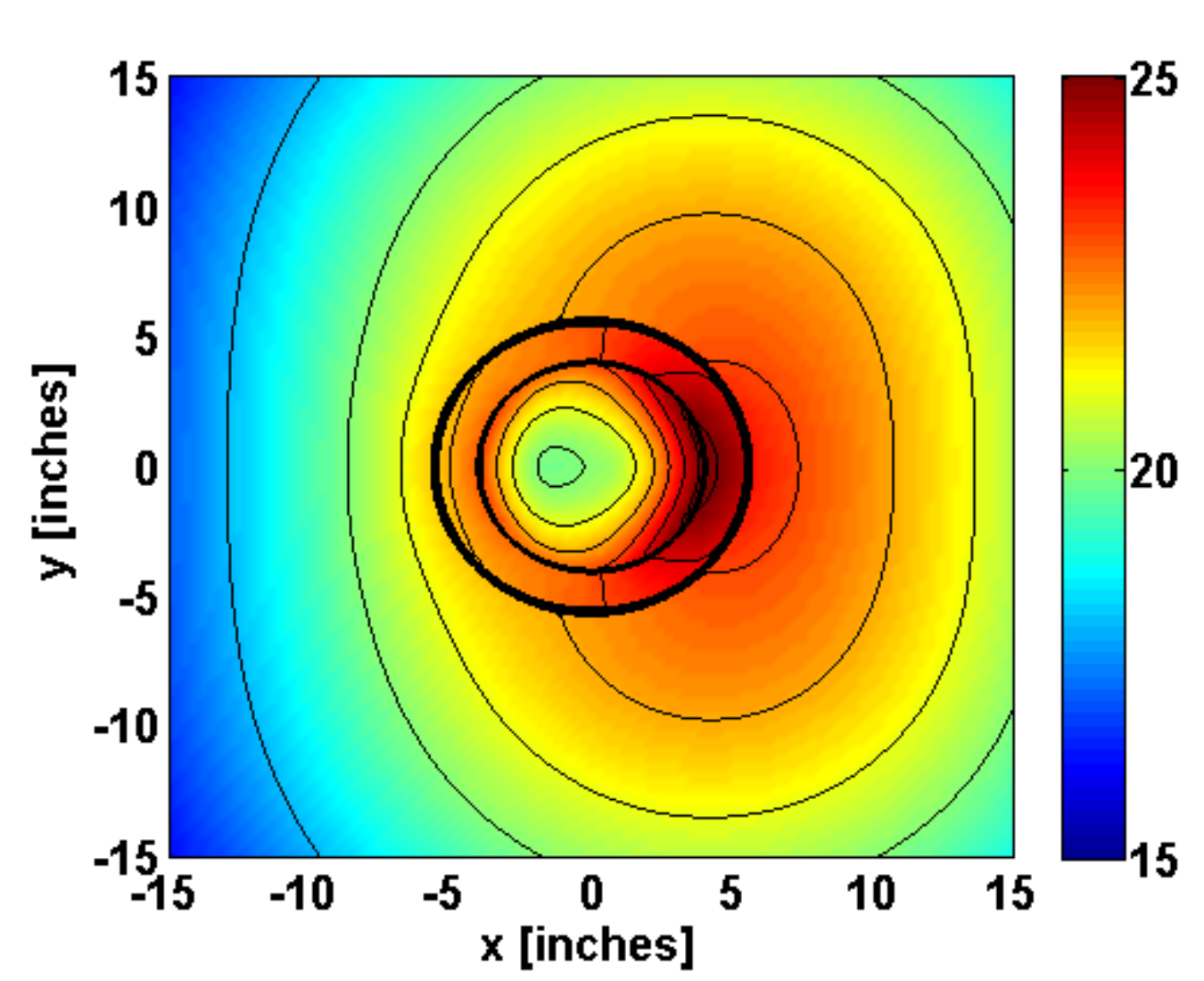}
    }
    \hfill
    \subfloat[\label{S4.F.case10.z0.1kHz}]{%
      \includegraphics[height=2.5in]{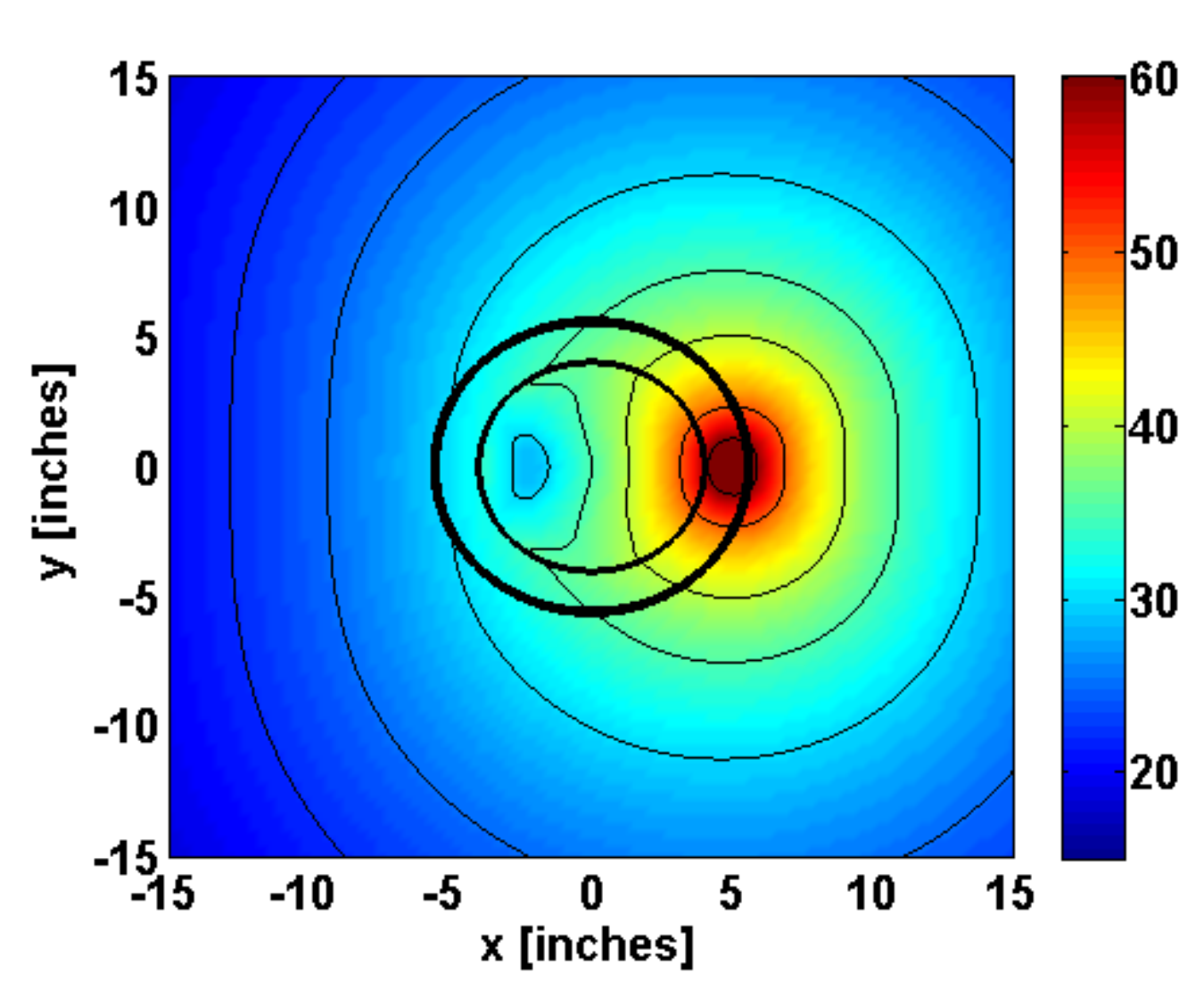}
    }
    \caption{Magnitude of the magnetic field of Case 11: (a) Three-dimensional perspective view, (b) $y=0^{\prime\prime}$ plane, (c) $z=16^{\prime\prime}$ plane, and (d) $z=0^{\prime\prime}$ plane.}
    \label{S4.F.case11.views}
\end{figure}

\pagebreak 
\section{Conclusions}
\label{sec.5.con}
We have developed a stable pseudoanalytical methodology for the accurate computation of electromagnetic fields due to point sources (tensor Green's function) in cylindrically stratified media. Existing canonical formulations for this problem are complete but are unstable under double-precision arithmetics for extreme variations on the physical parameters such as layer thicknesses, conductivities, source and field observation locations, and frequencies of operation. The use of the range-conditioned cylindrical functions in conjunction with carefully selected (sub)domains of evaluation set by their argument and order, as well as adaptively-chosen integration paths on the complex spectral plane allow for a robust computation that is always stable. The algorithm has been illustrated in a number of scenarios relevant to borehole geophysics.

\section*{Acknowledgement}
 We thank Halliburton Energy Services for the permission to publish this work, and Dr. Baris Guner for providing some validation data and suggesting comparison scenarios.

\appendix
\section{Reflection and transmission coefficients for two cylindrical layers}
\label{app.a}
\begin{figure}[!htbp]
  \centering
  \includegraphics[height=2.0in]{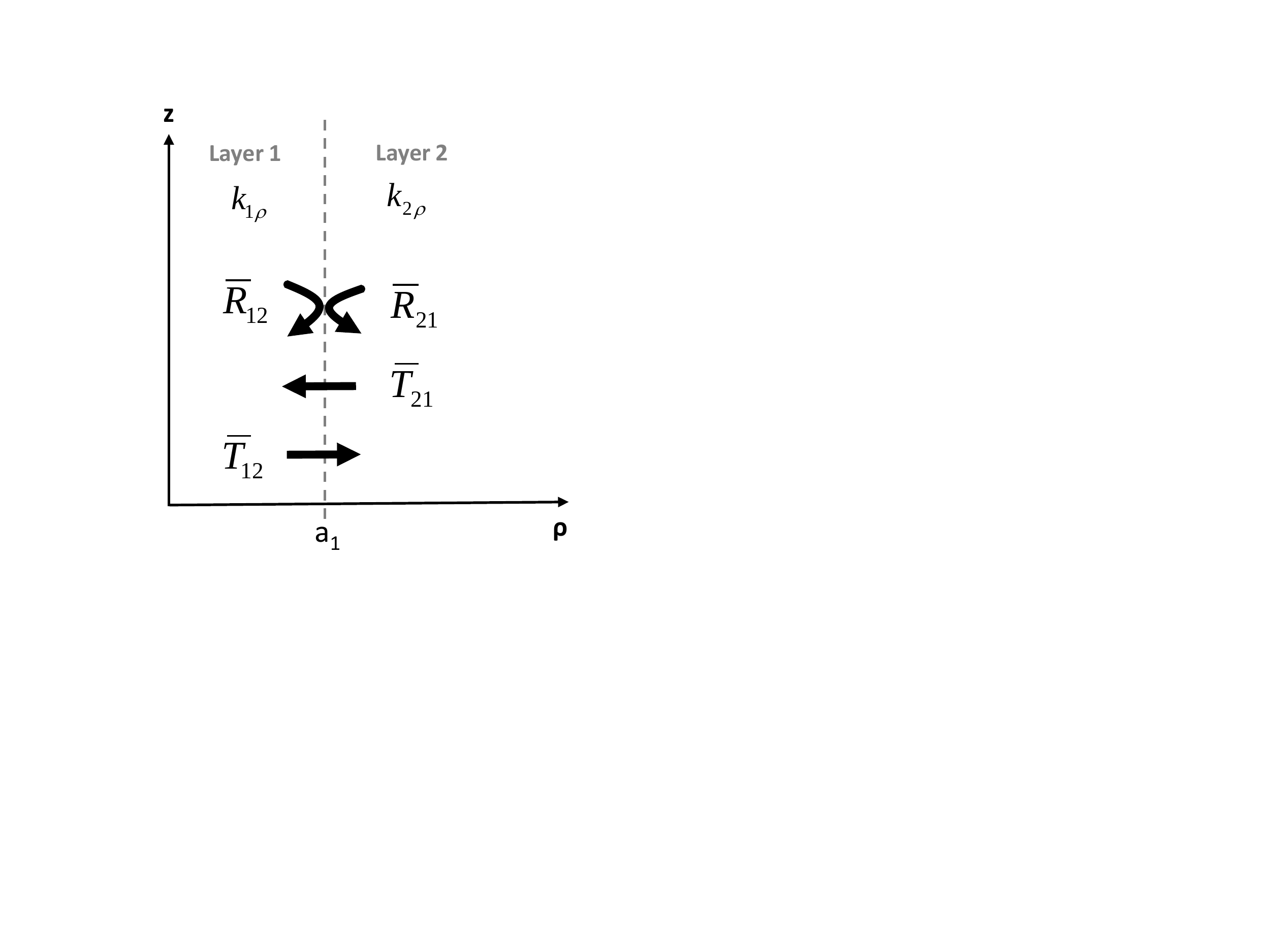}
  \caption{Reflection and transmission coefficients for two cylindrical layers in the $\rho z$-plane.}
  \label{A.F.local.region12}
\end{figure}

The (local) reflection and transmission coefficients between two cylindrical layers are written as~\cite[ch. 3]{Chew:Inhomogeneous}
\begin{flalign}
\bR_{12}&=\bD^{-1}\cdot
	\left[\Hn(k_{1\rho}a_1)\bHn(k_{2\rho}a_1)-\Hn(k_{2\rho}a_1)\bHn(k_{1\rho}a_1)\right],
	\label{A.eq.R12matrix}\\
\bR_{21}&=\bD^{-1}\cdot
    \left[\Jn(k_{1\rho}a_1)\bJn(k_{2\rho}a_1)-\Jn(k_{2\rho}a_1)\bJn(k_{1\rho}a_1)\right],
	\label{A.eq.R21matrix}\\
\bT_{12}&=\frac{2\omega}{\pi k^2_{1\rho}a_1}\bD^{-1}\cdot
    \begin{bmatrix}
    \epsilon_1 & 0 \\ 0 & -\mu_1
    \end{bmatrix},
	\label{A.eq.T12matrix}\\
\bT_{21}&=\frac{2\omega}{\pi k^2_{2\rho}a_1}\bD^{-1}\cdot
    \begin{bmatrix}
    \epsilon_2 & 0 \\ 0 & -\mu_2
    \end{bmatrix},
	\label{A.eq.T21matrix}\\
\bD&=\Hn(k_{2\rho}a_1)\bJn(k_{1\rho}a_1)-\Jn(k_{1\rho}a_1)\bHn(k_{2\rho}a_1),
	\label{A.eq.Dmatrix}
\end{flalign}
where
\begin{subequations}
\begin{flalign}
\bJn(k_{j\rho}\rho)&=\frac{1}{k^2_{j\rho}\rho}
    \begin{bmatrix}
    \iu\omega\epsilon_j k_{j\rho}\rho\Jnd(k_{j\rho}\rho) & -nk_z\Jn(k_{j\rho}\rho)\\
    -nk_z\Jn(k_{j\rho}\rho) & -\iu\omega\mu_j k_{j\rho}\rho\Jnd(k_{j\rho}\rho)
    \end{bmatrix},
\label{A.eq.small.matrixJ}\\
\bHn(k_{j\rho}\rho)&=\frac{1}{k^2_{j\rho}\rho}
    \begin{bmatrix}
    \iu\omega\epsilon_j k_{j\rho}\rho\Hnd(k_{j\rho}\rho) & -nk_z\Hn(k_{j\rho}\rho)\\
    -nk_z\Hn(k_{j\rho}\rho) & -\iu\omega\mu_{j}k_{j\rho}\rho\Hnd(k_{j\rho}\rho)
    \end{bmatrix}.
\label{A.eq.small.matrixH}
\end{flalign}
\end{subequations}
Figure \ref{A.F.local.region12} depicts the meaning of such reflection and transmission coefficients for two cylindrical layers in the $\rho z$-plane.

On the other hand, $\hbR_{12}$, $\hbR_{21}$, $\hbT_{12}$, and $\hbT_{21}$ are defined below in such a way that they are comprised of the range-conditioned cylindrical functions.
\begin{flalign}
\hbR_{12}&=\hbD^{-1}\cdot
	\left[\hHn(k_{1\rho}a_1)\hbHn(k_{2\rho}a_1)-\hHn(k_{2\rho}a_1)\hbHn(k_{1\rho}a_1)\right],
	\label{A.eq.R12matrix}\\
\hbR_{21}&=\hbD^{-1}\cdot
    \left[\hJn(k_{1\rho}a_1)\hbJn(k_{2\rho}a_1)-\hJn(k_{2\rho}a_1)\hbJn(k_{1\rho}a_1)\right],
	\label{A.eq.R21matrix}\\
\hbT_{12}&=\frac{2\omega}{\pi k^2_{1\rho}a_1}\hbD^{-1}\cdot
    \begin{bmatrix}
    \epsilon_1 & 0 \\ 0 & -\mu_1
    \end{bmatrix},
	\label{A.eq.T12matrix}\\
\hbT_{21}&=\frac{2\omega}{\pi k^2_{2\rho}a_1}\hbD^{-1}\cdot
    \begin{bmatrix}
    \epsilon_2 & 0 \\ 0 & -\mu_2
    \end{bmatrix},
	\label{A.eq.T21matrix}\\
\hbD&=\hHn(k_{2\rho}a_1)\hbJn(k_{1\rho}a_1)-\hJn(k_{1\rho}a_1)\hbHn(k_{2\rho}a_1),
	\label{A.eq.Dmatrix}
\end{flalign}
where the expressions for $\hbJn$ and $\hbHn$ under the various argument types are provided in the main text. 
\section{Thresholds for small, moderate, and large argument types}
\label{app.b}

In this Appendix, we determine numerical threshold values that determine small, moderate, and large arguments assuming standard double-precision arithmetic.

To determine an adequate small-argument threshold, we briefly examine the error incurred by the
small argument approximations considered in Section \ref{sec.2.1.1} using the relative error $(|F_{e}(z)-F_{a}(z)|)/|F_{e}(z)|$,
where $F(z)$ stands for either $\Jn(z)$ or $\Hn(z)$, with $F_{e}(z)$ and $F_{a}(z)$ the exact and
small argument approximate values, respectively. Table \ref{S2.T.threshold.small.J} and Table
\ref{S2.T.threshold.small.H} show the relative errors for different argument magnitudes and orders by
considering $z=|z|e^{\iu\frac{\pi}{4}}$. The threshold for small arguments can be chosen according to
the desired accuracy, but it cannot be too small otherwise overflow might still occur for modes with
very high order and arguments with magnitude just above the threshold. In addition, the threshold
should not be applied for the lowest order modes because the small argument approximation becomes
relatively poorer. With this in mind, a threshold value between $10^{-3}$ and $10^{-5}$ for $|z|$ with
$n\geq5$ is recommended for small arguments.
\begin{table}[!htbp]
\begin{center}
\renewcommand{\arraystretch}{1.4}
\setlength{\tabcolsep}{7pt}
\caption{Relative errors of $\Jn(z)$ using the small argument approximation.}
\begin{tabular}{cccccc}
	\hline
	 & $n=1$ & $n=5$ & $n=10$ & $n=20$ & $n=30$ \\
	\hline
	$|z|=10^{-1}$ & 1.2491$\times10^{-3}$ & 4.1658$\times10^{-4}$ & 2.2724$\times10^{-4}$ &
			1.1904$\times10^{-4}$ & 8.0642$\times10^{-5}$ \\
	$|z|=10^{-3}$ & 1.2499$\times10^{-7}$ & 4.1667$\times10^{-8}$ & 2.2727$\times10^{-8}$ &
			1.1905$\times10^{-8}$ & 8.0645$\times10^{-9}$ \\
	$|z|=10^{-5}$ & 1.2499$\times10^{-11}$ & 4.1636$\times10^{-12}$ & 2.2833$\times10^{-12}$ &
			1.1860$\times10^{-12}$ & 8.3446$\times10^{-13}$ \\
	$|z|=10^{-7}$ & 2.3161$\times10^{-14}$ & 7.0166$\times10^{-15}$ & 3.3853$\times10^{-14}$ &
			3.5975$\times10^{-14}$ & 1.7873$\times10^{-14}$ \\
	\hline
\end{tabular}
\label{S2.T.threshold.small.J}
\end{center}
\end{table}

\begin{table}[!htbp]
\begin{center}
\renewcommand{\arraystretch}{1.4}
\setlength{\tabcolsep}{7pt}
\caption{Relative errors of $\Hn(z)$ using the small argument approximation.}
\begin{tabular}{cccccc}
	\hline
	 & $n=1$ & $n=5$ & $n=10$ & $n=20$ & $n=30$ \\
	\hline
	$|z|=10^{-1}$ & 1.4879$\times10^{-2}$ & 6.2510$\times10^{-4}$ & 2.7780$\times10^{-4}$ &
			1.3158$\times10^{-4}$ & 8.6209$\times10^{-5}$ \\
	$|z|=10^{-3}$ & 3.7647$\times10^{-6}$ & 6.2500$\times10^{-8}$ & 2.7778$\times10^{-8}$ &
			1.3158$\times10^{-8}$ & 8.6207$\times10^{-9}$ \\
	$|z|=10^{-5}$ & 6.0662$\times10^{-10}$ & 6.2500$\times10^{-12}$ & 2.7784$\times10^{-12}$ &
			1.3190$\times10^{-12}$ & 8.6646$\times10^{-13}$ \\
	$|z|=10^{-7}$ & 8.3734$\times10^{-14}$ & 4.3195$\times10^{-16}$ & 2.6175$\times10^{-16}$ &
			4.4286$\times10^{-16}$ & 4.0776$\times10^{-15}$ \\
	\hline
\end{tabular}
\label{S2.T.threshold.small.H}
\end{center}
\end{table}

To determine the large-argument threshold, we examine \eqref{S2.eq.large.approx.Jn}. The two
trigonometric functions present there mostly determine the magnitude of $\Jn(z)$ and can be decomposed
into two terms as shown in \eqref{S2.eq.cos.chi} and \eqref{S2.eq.sin.chi}. For example, when
$\chi''=k''_{i\rho}a_i=30$, the ratio of the two terms is
$e^{30}/e^{-30}=1.1420\times10^{+26}$.
If double precision is assumed, such a threshold for large arguments is sufficient since double
precision supports up to 16 or 17 digits.

The two threshold values defined above are sufficient to provide stable and accurate evaluations under
all circumstances except when it becomes necessary to include very high-order modes with
(moderate) arguments slightly larger than the small-argument threshold. In this case, even though
it can still be possible to numerically evaluate $\Jn(z)$ and $\Hn(z)$  (and their derivatives), the evaluation of some integrand factors that involve products of these functions and their
derivatives (such as reflection coefficients) might yield overflows. To illustrate this, Table
\ref{S2.T.threshold.moderate} shows the expressible range
of $\Jn(z)$ and $\Hnd(z)$, with negative values representing $\log_{10}|\Jn(z)|$ and positive values
representing $\log_{10}|\Hnd(z)|$. The lemniscate symbols indicate instances of underflow and overflow
(note that in this case,
the functions could still be expressed using either small or large argument approximations, but at the
cost of accuracy).
Let us consider $z=10^{-3}$ with $n=50$, for example, and assume that the adopted small argument
threshold is smaller than $10^{-3}$. In this case, both $\Jn(z)$ and $\Hnd(z)$ can still be expressed
in double precision; however, $\bR_{12}$ is roughly proportional to the square of $\Hnd(z)$ (see \ref{app.a}), which is about $10^{+460}$, leading to overflow.
Consequently, an additional type of threshold, this time of function {\it magnitude}, not argument,
is introduced within the moderate-argument region to avoid such numerical overflow. Since the
supported range of values under double-precision arithmetic is from about $10^{-300}$ to $10^{+300}$,
we choose this magnitude threshold to be $T_{m}=10^{+100}$
where the exponent of +100 instead of +150 is used to provide a sufficient margin for calculation of
product factors (e.g., reflection coefficients). Using $T_{m}$, the multiplicative factor $P_{ii}$
associated with the range-conditioned cylindrical functions for moderate arguments is defined as
\begin{subequations}
\begin{flalign}
&\text{If } |\Jn(k_{i\rho}a_i)|^{-1} < T_{m}, \quad P_{ii} = 1. \label{S2.eq.Pii.a}\\
&\text{If } |\Jn(k_{i\rho}a_i)|^{-1} \geq T_{m},
	\quad P_{ii} = |\Jn(k_{i\rho}a_i)|. \label{S2.eq.Pii.b}
\end{flalign}
\end{subequations}
It should be noted that $P_{ii}$ is defined using $\Jn(k_{i\rho}a_i)$, and not $\Hn(k_{i\rho}a_i)$, to
be consistent with the multiplicative factor chosen for small arguments. As noted before,
multiplicative factors such as $P_{ii}$ are to be subsequently manipulated (reduced) algebraically
before numerical computations. This particular choice facilitates such manipulations.

\begin{table}[!htbp]
\begin{center}
\renewcommand{\arraystretch}{1.4}
\setlength{\tabcolsep}{7pt}
\caption{Expressible range of $\Jn(z)$ and $\Hnd(z)$ for the double-precision arithmetic. The negative value for each
$z$ and $n$ is $\log_{10}|\Jn(z)|$ and positive value is $\log_{10}|\Hnd(z)|$.}\vspace{0.5em}
\begin{tabular}{cccccccc}
	\hline
	 & $n=40$ & $n=50$ & $n=60$ & $n=70$ & $n=80$ & $n=90$ & $n=100$ \\
	\hline
	\multirow{2}{*}{$z=10^{-1}$}
		& -100 & -130 & -160 & -192 & -223 & -255 & -288 \\
	    & +100 & +130 & +160 & +191 & +223 & +255 & +288 \\

	\multirow{2}{*}{$z=10^{-2}$}
		& -140 & -180 & -220 & -261 & -303 & $-\infty$ & $-\infty$ \\
	    & +141 & +181 & +221 & +262 & +304 & $+\infty$ & $+\infty$ \\

	\multirow{2}{*}{$z=10^{-3}$}
		& -180 & -230 & -280 & $-\infty$ & $-\infty$ & $-\infty$ & $-\infty$ \\
	    & +182 & +232 & +282 & $+\infty$ & $+\infty$ & $+\infty$ & $+\infty$ \\

	\multirow{2}{*}{$z=10^{-4}$}
		& -220 & -280 & $-\infty$ & $-\infty$ & $-\infty$ & $-\infty$ & $-\infty$ \\
	    & +223 & +283 & $+\infty$ & $+\infty$ & $+\infty$ & $+\infty$ & $+\infty$ \\

	\multirow{2}{*}{$z=10^{-5}$}
		& -260 & $-\infty$ & $-\infty$ & $-\infty$ & $-\infty$ & $-\infty$ & $-\infty$ \\
	    & +264 & $+\infty$ & $+\infty$ & $+\infty$ & $+\infty$ & $+\infty$ & $+\infty$ \\
	\hline
\end{tabular}
\label{S2.T.threshold.moderate}
\end{center}
\end{table}

\begin{figure}[!htbp]
    \centering
    \includegraphics[width=5.0in]{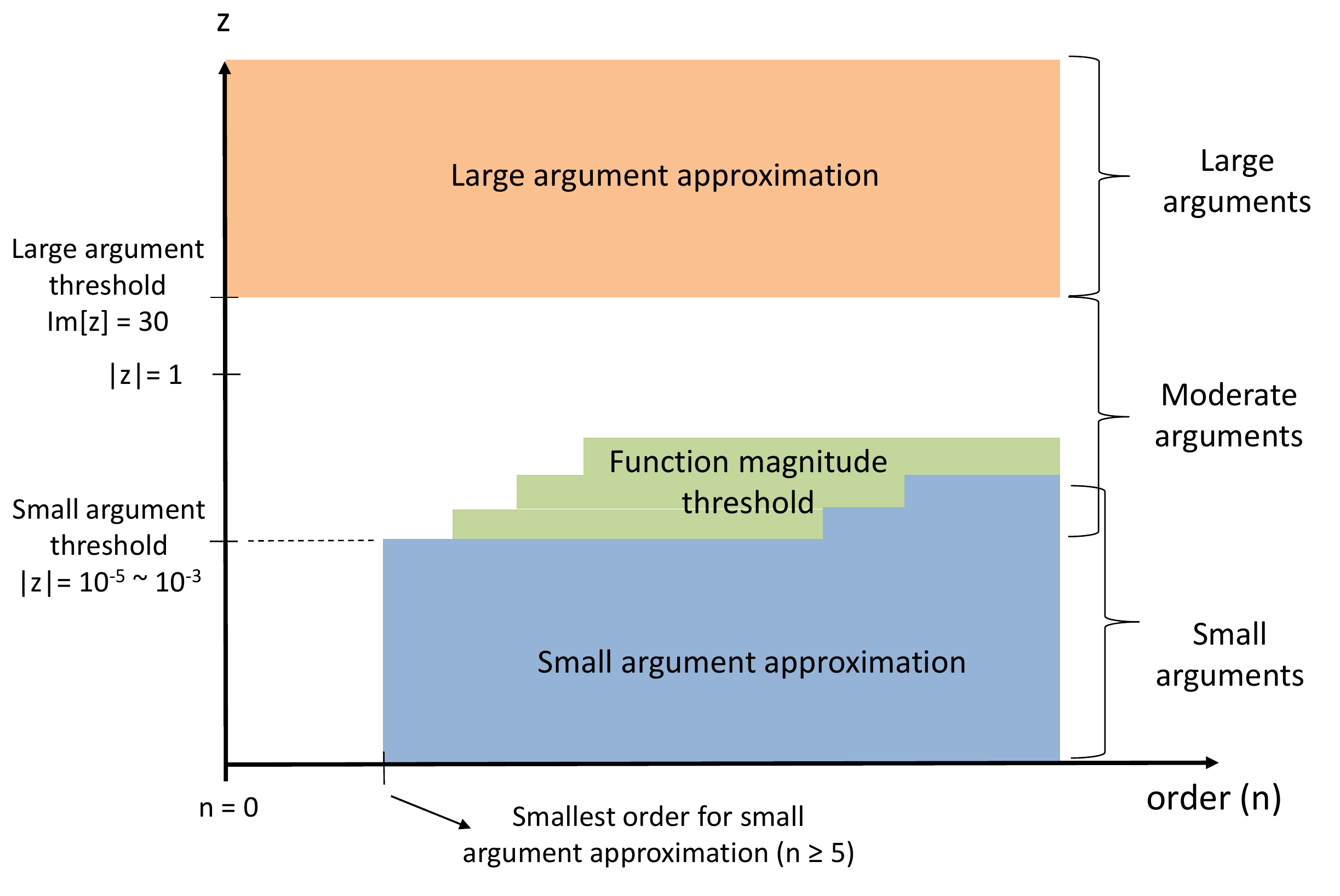}\\
    \caption{Schematic description of the three types of arguments.}
    \label{S2.F.arguments}
\end{figure}

Figure \ref{S2.F.arguments} schematically
shows the associated regions versus argument and order. In the white-space region,
no conditioning is necessary.

\section{Conditioned integral expressions for transverse field components}
\label{app.c}

The transverse electric and magnetic fields can be expressed as \cite{Kong72:Electromagnetic}
\begin{subequations}
\begin{flalign}
\mathbf{E}_s(\rr)&=\intmp dk_z \widetilde{\mathbf{E}}_s(k_z,\rr), \label{S2.eq.spectral.E}\\
\mathbf{H}_s(\rr)&=\intmp dk_z \widetilde{\mathbf{H}}_s(k_z,\rr), \label{S2.eq.spectral.H}
\end{flalign}
\end{subequations}
where $\widetilde{\mathbf{E}}_s(k_z,\rr)$ and $\widetilde{\mathbf{H}}_s(k_z,\rr)$ are the
corresponding spectral components and the subscript $s$ indicates transverse to the $z$ direction. For
each $k_z$, Maxwell's equations can be used to express these spectral components in terms of the
longitudinal components $E_z$ and $H_z$ as
\begin{flalign}
\renewcommand{\arraystretch}{1.2}
    \begin{bmatrix}
    E_\rho \\ H_\rho
    \end{bmatrix}
&=\frac{1}{k^2_\rho}
    \begin{bmatrix}
    \iu k_z\frac{\pa}{\pa\rho} & -\frac{n\omega\mu}{\rho} \\
    \frac{n\omega\epsilon}{\rho} & \iu k_z\frac{\pa}{\pa\rho}
    \end{bmatrix}
    \begin{bmatrix}
    E_z \\ H_z
    \end{bmatrix}
=\frac{1}{k^2_\rho}\bBn
    \begin{bmatrix}
    E_z \\ H_z
    \end{bmatrix}, \label{S2.eq.EHrho}\\
\renewcommand{\arraystretch}{1.2}
    \begin{bmatrix}
    E_\phi \\ H_\phi
    \end{bmatrix}
&=\frac{1}{k^2_\rho}
    \begin{bmatrix}
    -\frac{nk_z}{\rho} & -\iu\omega\mu\frac{\pa}{\pa\rho} \\
    \iu\omega\epsilon\frac{\pa}{\pa\rho} & -\frac{nk_z}{\rho} \\
    \end{bmatrix}
    \begin{bmatrix}
    E_z \\ H_z
    \end{bmatrix}
=\frac{1}{k^2_\rho}\bCn
    \begin{bmatrix}
    E_z \\ H_z
    \end{bmatrix}. \label{S2.eq.EHphi}
\end{flalign}
Since partial derivatives of the $z$-components with respect to $\rho$ are required for transverse
components, the integrands shown in Section \ref{sec.2.4} (see \eqref{S2.eq.Fn.case1}, \eqref{S2.eq.Fn.case2},
\eqref{S2.eq.Fn.case3}, and \eqref{S2.eq.Fn.case4}) should be decomposed into three matrices for
convenient computation as follows
\begin{flalign}
\Fn=\bLn(\rho)\cdot\bMn\cdot\bRn(\rho'), \label{S2.eq.Fn.decompose}
\end{flalign}
where $\bLn(\rho)$ is the function of $\rho$ only, $\bMn$ is function of cylindrical layers interfaces
(but neither $\rho$ nor $\rho'$), and $\bRn(\rho')$ is the function of $\rho'$ only. In this fashion,
the transverse field components can be expressed as
\begin{flalign}
    \begin{bmatrix}
    E_\rho \\ H_\rho
    \end{bmatrix}
&=\frac{\iu Il}{4\pi\omega\epsilon_j}\intmp dk_z e^{\iu k_z(z-z')}\frac{1}{k^2_{\rho}}
    \left[\suma e^{\iu n(\phi-\phi')}\bBn\cdot\bLn(\rho)\cdot\bMn\cdot
        \bRn(\rho')\cdot\Dj\right],
\label{S2.eq.EHrho.modified.integrand}\\
    \begin{bmatrix}
    E_\phi \\ H_\phi
    \end{bmatrix}
&=\frac{\iu Il}{4\pi\omega\epsilon_j}\intmp dk_z e^{\iu k_z(z-z')}\frac{1}{k^2_{\rho}}
    \left[\suma e^{\iu n(\phi-\phi')}\bCn\cdot\bLn(\rho)\cdot\bMn\cdot
        \bRn(\rho')\cdot\Dj\right].
\label{S2.eq.EHphi.modified.integrand}
\end{flalign}
It should be noted that $\bBn$ and $\bCn$ only acts upon $\bLn(\rho)$ and $\Dj$ only acts upon $\bRn$.
Therefore, $\bBn\cdot\bLn(\rho)$, $\bCn\cdot\bLn(\rho)$, and $\bRn(\rho')\cdot\Dj$ can be calculated
separately.
A set of coefficients can be defined for convenience as
\begin{subequations}
\begin{flalign}
\bBn\cdot\bLn(\rho)\cdot\bMn=\bW_{\rho,n}, \label{S2.eq.W.rhon}\\
\bCn\cdot\bLn(\rho)\cdot\bMn=\bW_{\phi,n}. \label{S2.eq.W.phin}
\end{flalign}
\end{subequations}
Furthermore, since the source factor consists of three terms
\begin{flalign}
\bRn(\rho')\cdot\Dj
    =\frac{\iu}{2}\bRn(\rho')\cdot
    \left(\Dja+\Djb+\Djc\frac{\pa}{\pa\rho'}\right),
\label{S2.eq.RnDj}
\end{flalign}
it follows that the squared bracket factor for the $\rho$-components shown in \eqref{S2.eq.EHrho.modified.integrand} can be
expanded as
\begin{flalign}
\left[\suma e^{\iu n(\phi-\phi')}\bW_{\rho,n}\cdot\bRn(\rho')\cdot\Dj\right]
&=\frac{\iu}{2}
    \left[\suma e^{\iu n(\phi-\phi')}
        \bW_{\rho,n}\cdot\bRn(\rho')\right]\cdot\Dja \notag\\
&\quad+\frac{\iu}{2}
    \left[\suma e^{\iu n(\phi-\phi')}
        \bW_{\rho,n}\cdot\bRn(\rho')\cdot\Djb\right] \notag\\
&\quad+\frac{\iu}{2}
    \left[\suma e^{\iu n(\phi-\phi')}\bW_{\rho,n}\cdot\frac{\pa}{\pa\rho'}
        \left[\bRn(\rho')\right]
    \right]\cdot\Djc.
\label{S2.eq.bracket.rho.expanded}
\end{flalign}
For negative integer orders, the off-diagonal elements of $\bW_{\rho,n}\cdot\bRn(\rho')$ and
$\bW_{\rho,n}\cdot\frac{\pa}{\pa\rho'}\left[\bRn(\rho')\right]$ in \eqref{S2.eq.bracket.rho.expanded}
change sign. The same folding technique used for the $z$-components in Section \ref{sec.2.6} can be applied to \eqref{S2.eq.bracket.rho.expanded}. In other words, the summations in the right hand side of \eqref{S2.eq.bracket.rho.expanded} can be
folded such that only non-negative modes are involved.
For the $\phi$-components, the squared bracket factor shown in \eqref{S2.eq.EHphi.modified.integrand}
can be likewise expanded as
\begin{flalign}
\left[\suma e^{\iu n(\phi-\phi')}\bW_{\phi,n}\cdot\bRn(\rho')\cdot\Dj\right]
&=\frac{\iu}{2}
    \left[\suma e^{\iu n(\phi-\phi')}
        \bW_{\phi,n}\cdot\bRn(\rho')\right]\cdot\Dja \notag\\
&\quad+\frac{\iu}{2}
    \left[\suma e^{\iu n(\phi-\phi')}
        \bW_{\phi,n}\cdot\bRn(\rho')\cdot\Djb\right] \notag\\
&\quad+\frac{\iu}{2}
    \left[\suma e^{\iu n(\phi-\phi')}\bW_{\phi,n}\cdot\frac{\pa}{\pa\rho'}
        \left[\bRn(\rho')\right]
    \right]\cdot\Djc.
\label{S2.eq.bracket.phi.expanded}
\end{flalign}
In contrast to the $\rho$-components, the diagonal elements of $\bW_{\phi,n}\cdot\bRn(\rho')$ and
$\bW_{\phi,n}\cdot\frac{\pa}{\pa\rho'}\left[\bRn(\rho')\right]$ in \eqref{S2.eq.bracket.phi.expanded}
change sign. Again, the summations in the right hand side of \eqref{S2.eq.bracket.phi.expanded} can be
easily folded in such a way that only non-negative modes are involved.




\bibliographystyle{model1-num-names}
\bibliography{biblioJCP}







\end{document}